\documentclass{article}
\usepackage[T1]{fontenc}
\usepackage[utf8]{inputenc}
\usepackage{authblk}
\usepackage{amsmath}
\usepackage{booktabs} 
\usepackage{dcolumn}
\usepackage{subfigure}
\usepackage{numprint}
\usepackage{mathtools}
\usepackage{amssymb}
\usepackage{color}
\usepackage{stmaryrd}
\usepackage{geometry}
\usepackage{tablefootnote}
\usepackage{float}
\usepackage{tikz}

\newcommand{\revision}[1]{{#1}}

\newcommand{\bX}{\boldsymbol{X}}
\newcommand{\bx}{\boldsymbol{x}}
\newcommand{\by}{\boldsymbol{y}}
\newcommand{\bY}{\boldsymbol{Y}}
\newcommand{\bbeta}{\boldsymbol{\beta}}

\newcommand{\btheta}{\boldsymbol{\theta}}
\newcommand{\bepsilon}{\boldsymbol{\varepsilon}}
\newcommand{\bz}{\boldsymbol{z}}

\newcommand{\ba}{\boldsymbol{a}}
\newcommand{\bb}{\boldsymbol{b}}

\newcommand{\bs}{\boldsymbol{s}}

\usepackage{chicago}

\newcommand{\transpose}{^{\text{\sffamily T}}}
\newcommand{\inverse}{^{-1}}

\newcommand{\Esp}{\mathbb{E}}    
\newcommand{\Var}{\text{Var}}    
   
\newcommand{\esp}{\mathbb{E}}    
\newcommand{\var}{\text{Var}}    

\renewcommand{\thefootnote}{\arabic{footnote}}

\title{\bf \'Econom\'etrie \& {\em Machine Learning}}% \\ Toi aussi, rejoins le c\^ot\'e obscur de la Force}--> :) 

\author[1]{Arthur Charpentier}
\author[2]{\\Emmanuel Flachaire}
\author[3]{\\J\'er\'emie Jakubowicz}
\author[4]{Antoine Ly}
\affil[1]{Universit\'e de Rennes 1 \& CREM}
\affil[2]{Aix-Marseille Universit\'e \& AMSE}
\affil[3]{T\'el\'ecom SudParis}
\affil[4]{Universit\'e Paris-Est}

\date{Juillet 2017}

\begin{document}

%\maketitle

\begin{titlepage}

\begin{center}

\mbox{\,}
\bigskip\bigskip
\bigskip\bigskip

{\bf\LARGE 
\'Econom\'etrie \& {\em Machine Learning}
}

%Median-based Inequality Measures:\\[.5ex]
%{\Large\normalfont Why should we no longer use the Gini index}
%{\Large\normalfont Why should we prefer the MLD  to  the Gini index}
%{\Large\normalfont Use the mean logarithmic deviation index, not the Gini}
%{\Large\normalfont Variations of  inequality are underestimated}
%{\LARGE\normalfont Inequality increased  more than we thought}
\par\end{center}

\bigskip

\renewcommand*{\thefootnote}{\fnsymbol{footnote}}
\begin{center}
\Large
\textbf{Arthur Charpentier}\\[1ex]
\large
Universit\'e de Rennes 1 \& CREM\\
{\small 7 Place Hoche, 35065 Rennes Cedex, France}\\
{\small arthur.charpentier@univ-rennes1.fr}\\

\bigskip

%{\Large and}

\bigskip

\Large
\textbf{Emmanuel Flachaire}\\[1ex]
\large
Aix-Marseille Universit\'e, AMSE, CNRS \& EHESS\\
{\small 5 bd Maurice Bourdet, CS 50498, 13205 Marseille Cedex 01, France}\\
{\small emmanuel.flachaire@univ-amu.fr}\\

%{\Large and}

\bigskip

%{\Large and}

%\bigskip

%\Large
%\textbf{J\'er\'emie Jakubowicz}\\[1ex]
%\large
%T\'el\'ecom SudParis\\

%\bigskip

{\Large et}

\bigskip

\Large
\textbf{Antoine Ly}\\[1ex]
\large
Universit\'e Paris-Est\\
{\small 5, boulevard Descartes, 77454 Marne-la-Vallée cedex, France}\\
{\small antoine.ly@gmail.com }\\

\bigskip

\par\end{center}

\setcounter{footnote}{0}

\bigskip
\bigskip
\begin{abstract}
L'\'econom\'etrie et l'apprentissage machine semblent avoir une finalit\'e en commun: construire un mod\`ele pr\'edictif, pour une variable d'int\'er\^et, \`a l'aide de variables explicatives (ou {\em features}). Pourtant, ces deux champs se sont d\'evelopp\'es en parall\`ele, cr\'eant ainsi deux cultures diff\'erentes, pour paraphraser \citeNP{Breiman}. Le premier visait \`a construire des mod\`eles probabilistes permettant de d\'ecrire des ph\'enom\`emes \'economiques. Le second utilise des algorithmes qui vont apprendre de leurs erreurs, dans le but, le plus souvent de classer (des sons, des images, etc). Or r\'ecemment, les mod\`eles d'apprentissage se sont montr\'es plus efficaces que les techniques \'econom\'etriques traditionnelles (avec comme prix \`a payer un moindre pouvoir explicatif), et surtout, ils arrivent \`a g\'erer des donn\'ees beaucoup plus volumineuses. Dans ce contexte, il devient n\'ecessaire que les \'econom\`etres comprennent ce que sont ces deux cultures, ce qui les oppose et surtout ce qui les rapproche, afin de s'approprier des outils d\'evelopp\'es par la communaut\'e de l'apprentissage statistique, pour les int\'egrer dans des mod\`eles \'econom\'etriques.

\bigskip



\end{abstract}

\vfill
\begin{center}
\Large 
Mars  2018
\par\end{center}
\vfill

\end{titlepage}

%\newpage
\mbox{\,}
\vfill
\tableofcontents
\vfill
\newpage

\section{Introduction}

L'utilisation de techniques quantitatives en \'economie remonte probablement au 16\`eme si\`ecle, comme le montre \citeNP{Morgan}. Mais il faudra attendre le d\'ebut du XXi\`eme si\`ecle pour que le terme « \'econom\'etrie » soit utilis\'e pour la premi\`ere fois, donnant naissance \`a l'{\em Econometric Society} en 1933. Les techniques de {\em machine learning} (apprentissage machine) sont plus r\'ecentes. C'est \`a Arthur Samuel, consid\'er\'e comme le p\`ere du premier programme d'auto-apprentissage, que l'on doit le terme « {\em machine learning} » qu'il d\'efinit comme « {\em a field of study that gives computer the ability without being explicitly programmed} ». Parmi les premi\`eres techniques, on peut penser \`a la th\'eorie des assembl\'ees de neurones propos\'ee dans \citeNP{Hebb} (qui donnera naissance au {\em perceptron} dans les ann\'ees 1950, puis aux r\'eseaux de neurones) dont \citeNP{Widrow} montreront quinze ans plus tard les liens avec les m\'ethodes des moindres carr\'es, aux SVM ({\em support vector machine}) et plus r\'ecemment aux m\'ethodes de {\em boosting}. Si les deux communaut\'es ont grandi en parall\`ele, les donn\'ees massives imposent de cr\'eer des passerelles entre les deux approches, en rapprochant les « deux cultures » \'evoqu\'ees par \citeNP{Breiman}, opposant la statistique math\'ematique (que l'on peut rapprocher de l'\'econom\'etrie traditionnelle, comme le note \citeNP{Aldrich}) \`a la statistique computationnelle, et \`a l'apprentissage machine de mani\`ere g\'en\'erale.

\subsection{La Mod\'elisation \'econom\'etrique}

L'\'econom\'etrie et les techniques d'apprentissage statistique supervis\'e sont proches, tout en \'etant tr\`es diff\'erentes. Proches au d\'epart, car toutes les deux utilisent une base (ou un tableau) de donn\'ees, c'est \`a dire des observations $\lbrace (y_i,\bx_i) \rbrace$, avec $i=1,\cdots,n$, $\bx_i\in\mathcal{X}\subset\mathbb{R}^p$ et $y_i\in\mathcal{Y}$. Si $y_i$ est qualitative, on parlera d'un probl\`eme de classification\footnote{\revision{Nous utiliserons ici le terme « classification » lorsque $\mathcal{Y}$ est un ensemble de classes, typiquement une classification binaire, $\mathcal{Y}=\{0,1\}$, ce cas correspondant à la réalisation d'une variable indicatrice, $\boldsymbol{1}_{Y_t\leq 0}$, ou $\boldsymbol{1}_{Y\in\mathcal{A}}$, par exemple. Ce terme est moins dat\'e que « discrimination » par exemple, et plus général que la constitution d'un « score » (qui est souvent une étape intermédiaire). Il ne doit pas être confondu avec la classification non-supervisée (comme la « classification ascendante hiérarchique ») qui est la constitution de classe homogène à partir d'une mesure de similarité (on utilisera parfois, dans ce cas, le terme de « constitution de classes », ou de « clusters »).}}, et dans le cas contraire, d'un probl\`eme de r\'egression. Proches \`a l'arriv\'ee, car dans les deux cas, on cherche \`a construire un « mod\`ele », c'est \`a dire une fonction $m:\mathcal{X}\mapsto\mathcal{Y}$ qui sera interpr\'et\'ee comme une pr\'evision.

Mais entre le d\'epart et l'arriv\'ee, il existe de r\'eelles diff\'erences. Historiquement, les mod\`eles \'econom\'etriques s'appuient sur une th\'eorie \'economique, avec le plus souvent des mod\`eles param\'etriques. On a alors recours aux outils classiques de l'inf\'erence statistique (comme le maximum de vraisemblance, ou la m\'ethode des moments) pour estimer les valeurs d'un vecteur de param\`etres $\btheta$, dans un mod\`ele param\'etrique $m_{\btheta}(\cdot)$, par une valeur $\widehat{\btheta}$. Comme en statistique, avoir des estimateurs sans biais est important car on peut quantifier une borne inf\'erieure pour la variance (borne de Cram\'er-Rao). La th\'eorie asymptotique joue alors un r\^ole important (d\'eveloppements de Taylor, loi des grands nombres, et th\'eor\`eme central limite). En apprentissage statistique, en revanche, on construit souvent des mod\`eles non-param\'etriques, reposant presque exclusivement sur les donn\'ees (i.e. sans hypothèse de distribution), et les m\'eta-param\`etres utilis\'es (profondeur de l'arbre, param\`etre de p\'enalisation, etc) sont optimis\'es par validation crois\'ee.

Au del\`a des fondements, si l'\'econom\'etrie \'etudie abondamment les propri\'et\'es (souvent asymptotiques) de $\widehat{\btheta}$ (vu comme une variable al\'eatoire, gr\^ace \`a la repr\'esentation stochastique sous-jacente), l'apprentissage statistique s'int\'eresse davantage aux propri\'et\'es du mod\`ele optimal  $m^\star(\cdot)$ (suivant un crit\`ere qui reste \`a d\'efinir), voire simplement $m^\star(\boldsymbol{x}_i)$ pour quelques observations $i$ jug\'ees d'int\'er\^et (par exemple dans une population de test). Le probl\`eme de choix de mod\`ele est aussi vu sous un angle assez diff\'erent. Suivant la loi de Goodhart (« {\em si une mesure devient un objectif, elle cesse d'\^etre une mesure} »), les \'econom\`etres utilisent des crit\`eres de type AIC ou BIC pour choisir un mod\`ele optimal (p\'enalisant la qualit\'e d'ajustement d'un mod\`ele par sa complexit\'e, ex-post, lors de la phase de validation ou de choix), alors qu'en apprentissage statistique, c'est la fonction objectif qui tiendra compte d'une p\'enalisation, comme pour le {\sc lasso}, ressemblant \`a une forme de p\'enalisation ex-ante.

\subsection{Applications}

Avant de revenir sommairement sur l'\'evolution des mod\`eles \'econom\'etriques, c'est \`a Francis Galton que l'on doit le terme « r\'egression », comme le rappelle \citeNP{KoenkerGalton}. Si le terme est parfois devenu synonyme de « mod\`ele \'e\-co\-no\-m\'e\-tri\-que », il avait \'et\'e introduit dans le contexte de « {\em regression towards mediocraty inhereditary stature} », pour reprendre le titre de l'article paru en 1886. Galton utilisait un mod\`ele lin\'eaire pour mod\'eliser la taille moyenne d'un gar\c{c}on (\`a l'\^age adulte) en fonction de la taille de son p\`ere. Si cette technique de r\'egression \'etait connue par les \'economistes, il a fallu attendre les ann\'ees 1930 pour voir surgir le concept de « mod\`ele » \'economique. Comme le note \citeNP{Debreu}, la premi\`ere \'etape a \'et\'e de formuler des affirmations \'economiques dans un language math\'ematique. Les diff\'erentes grandeurs sont vues comme des variables, et dans les ann\'ees 1930, on verra appara\^itre \revision{les} « {\em statistical demand curves} », pour reprendre la terminologie d'Henry Schultz. Cette approche statistique permettra d'aller plus loin que les travaux pionners de \citeNP{Engel} qui \'etudiait empiriquement la relation entre la consommation et le revenu des m\'enages, par exemple, dans une approche uniquement descriptive.

Les mod\`eles \'econom\'etriques se sont d\'evelopp\'es en parall\`ele des mod\`eles macro-\'economiques. Les premiers travaux de la Commission Cowles ont port\'e sur l'identification  des mod\`eles \'economiques, et l'estimation de mod\`eles \`a \'equations simultan\'ees. Ces d\'eveloppements vont aboutir \`a un \^age d'or de l'\'econom\'etrie, dans les ann\'ees 1960, o\`u les mod\`eles \'econom\'etriques seront utilis\'es afin d'am\'eliorer les pr\'evisions macro\'economiques. On va alors voir appara\^itre tout un ensemble de « lois » qui sont souvent traduites comme des relations lin\'eaires entre des grandeurs agr\'eg\'ees, telle que la « loi de Okun » introduite dans \citeNP{Okun} qui postule une relation lin\'eaire entre le variation du nombre de demandeurs d'emploi et de la croissance du PIB,
$$
\Delta\text{Ch\^omage}_t=\beta_0+\beta_1\text{Croissance}_t+\varepsilon_t,
$$
quand on \'etudie ces grandeurs au cours du temps ($t$), ou la loi de « Feldstein-Horioka » introduite dans \citeNP{FH} qui suppose une relation lin\'eaire entre les taux d'investissement et d'\'epargne, relativement au revenu national,
$$
\frac{\text{investissement}_i}{\text{revenu national}_i}=\beta_0+\beta_1
\frac{\text{\'epargne}_i}{\text{revenu national}_i}+\varepsilon_i
$$
quand on mod\`elise les liens entre les allocations investissement-\'epargne pour plusieurs pays ($i$). Cet \^age d'or correspond aussi \`a un questionnement profond, suite \`a la critique de \citeNP{Lucas}, s'interrogeant sur l'inefficacit\'e de ces outils \`a expliquer et \`a pr\'evoir des crises. La principale explication \'etait alors le manque de fondement micro-\'economiques de ces mod\`eles, ce qui donnera un second souffle aux mod\`eles micro-\'econom\'etriques. On pourra rappeler que cette critique dite « de Lucas » avait \'et\'e formul\'ee dans \citeNP{Orcutt}, qui avan\c{c}ait l'id\'ee que les donn\'ees macro\'economiques posaient des probl\`emes insolubles d'identification. La solution passait forc\'ement par de l'\'econom\'etrie sur donn\'ees individuelles (au lieu de donn\'ees obtenues par aggr\'egation), ce qui sera reformul\'e quelques ann\'ees plus tard par \citeNP{Koopmans}. 

Malheureusement, les mod\`eles micro-\'econom\'etriques sont g\'en\'eralement plus complexes, car ils se doivent de tenir compte d'une \'eventuelle censure dans les donn\'ees, avec par exemple le mod\`ele introduit par \citeNP{Tobin}, d'erreurs sur les variables (qui pourront \^etre corrig\'ees par des instruments avec une technique initi\'ee par \citeNP{Reiersol}) ou avoir \'et\'e collect\'ee avec un biais de s\'election, avec les techniques propos\'ees par \citeNP{Heckman}. On notera que les \'econom\`etres se sont beaucoup interrog\'es sur la mani\`ere dont les donn\'ees \'etaient construites, et ne se sont jamais content\'es de « construire des mod\`eles ». Un exemple peut \^etre l'\'evaluation des politiques publiques, largement d\'etaill\'e dans \citeNP{Givord}. Dans ce cas, en effet, deux \'ecoles se sont oppos\'ees (initiant un d\'ebat que l'on verra resurgir tout au long de l'article sur les m\'ethodes d'apprentissage statistique). La premi\`ere, dite « structuraliste », cherchera \`a construire un mod\`ele complet afin de d\'ecrire le comportement des agents \'e\-co\-no\-mi\-ques. La seconde, souvent qualifi\'ee d'« empiriste », vise \`a tester l'effet d'une mesure sans pour autant expliciter les m\'ecanismes sous-jacents. C'est ce qu'explique \citeNP{AngristKrueger}, en affirmant « {\em research
in a structuralist style relies heavily on economic theory to guide
empirical work $[\cdots]$ An alternative to structural modeling, $[\cdots]$ the
`experimentalist' approach, $[\cdots]$ puts front and center the problem
of identifying causal effects from specific events or situations} ».

\revision{On peut aussi souligner que si l'approche de la Commission Cowles était très exigeante, en supposant le modèle connu, toute une littérature s'est développée en allégeant cette hypothèse, soit en essayant de choisir le bon modèle (avec les travaux de \citeNP{HendryK} par exemple) ou en proposant de faire des moyennes de modèles (comme développé récemment par \citeNP{Lili})}. Et plus généralement, alors que l'analyse \'econom\'etrique (en particulier \`a des fins de politique \'economique) s'est d\'evelopp\'ee plus r\'ecemment autour de l'inf\'erence causale, les techniques d'apprentissage machine ont \'et\'e vues, traditionnellement, autour de la pr\'ediction (o\`u la recherche de corr\'elations suffisamment fortes entre variables suffit) d'o\`u leur popularit\'e dans des usages plus industriels de classification, comme la reconnaissance de caract\`eres, de signature, d'images, ou de traduction, comme le rappelle \citeNP{Bishop}. En biologie, ces techniques ont \'et\'e appliqu\'ees pour cr\'eer des classifications d'esp\`eces animales en fonction d'analyse d'ADN, ou dans le domaine militaire et s\'ecuritaire pour l'identification de cibles ou de terroristes (potentiels). Il faudra attendre les ann\'ees 1990 pour voir des applications en finance avec \citeNP{Altman} par exemple, ou \citeNP{Herbrich} pour une revue de litt\'erature sur les applications potentielles en \'economie. Si des applications sont aujourd'hui nombreuses, et si ces techniques concurrencent les mod\`eles de micro-\'econom\'etrie (on pourra penser au scoring bancaire, \`a la d\'etection de fraude fiscale ou assurantielle, ou \`a l'identification de prospects en marketing), les algorithmes d'apprentissage sont devenus tr\`es populaires en reconnaissance de parole, puis d'images, et plus r\'ecemment avec les applications en ligne et les applications aux jeux (d'\'echec, et plus r\'ecemment de go). Si l'\'econom\'etrie s'est d\'evelopp\'ee au confluent des math\'ematiques et de l'\'economie, l'apprentissage machine (que l'on pourrait avoir tendance \`a rapprocher de l'intelligence artificelle) s'est d\'evelopp\'e \`a la fronti\`ere des math\'ematiques et de l'informatique (avec des r\'esultats fondamentaux en optimisation - en particulier autour des m\'ethodes de gradient stochastique - et sur les espaces « \revision{sparse} » ou « \revision{parcimonieux} »).

\revision{\subsection{De la grande dimension aux donn\'ees massives}}

Dans cet article, une variable sera un vecteur de $\mathbb{R}^n$, de telle sorte qu'en concat\'enant les variables ensemble, on puisse les stocker dans une matrice $\bX$, de taille $n\times p$, avec $n$ et $p$ potentiellement grands\footnote{L\`a encore, des extensions sont possibles, en particulier dans les donn\'ees m\'edicales avec des images de type IRM comme variables pr\'edictives, ou des donn\'ees climatiques avec des cartes en variables pr\'edictives, ou plus g\'en\'eralement une variable tensorielle en dimension plus ou moins grande. Comme le montre \citeNP{Kolda} il est toutefois possible de se ramener dans le cas usuel (de donn\'ees sous formes de vecteurs) en utilisant la d\'ecomposition de Tucker.}. Le fait que $n$ soit grand n'est, a priori, pas un probl\`eme en soi, au contraire. De nombreux th\'eor\`emes en \'econom\'etrie et en statistique sont obtenus lorsque $n\rightarrow\infty$ (c'est la th\'eorie asymptotique). En revanche, le fait que $p$ soit grand est probl\'ematique, en particulier si $p>n$. Les deux dimensions sont \`a distinguer, car elles vont engendrer des probl\`emes relativement diff\'erents.

\citeNP{Portnoy} a montr\'e que l'estimateur du maximum de vraisemblance conserve la propri\'et\'e de normalit\'e asymptotique si $p$ reste petit devant $n$, ou plus pr\'ecis\'ement, si $p^2/n\rightarrow 0$ lorsque $n,p\rightarrow\infty$. Aussi, il n'est pas rare de parler de \revision{grande dimension} d\`es lors que $p>\sqrt{n}$. Un autre concept important est celui de sparcit\'e, qui repose non pas sur la dimension $p$ mais sur la dimension effective, autrement dit le nombre de variables effectivement significatives. Il est alors possible d'avoir $p>n$ tout en ayant des estimateurs convergents.

La grande dimension en terme de nombre de variables, $p$, peut faire peur \`a cause de la mal\'ediction de la dimension, introduit par \citeNP{Bellman}. L'ex\-pli\-ca\-tion de cette mal\'ediction est que le volume de la sph\`ere unit\'e, en dimension $p$, tend vers $0$ lorsque $p\rightarrow\infty$. On dit alors que l'espace est « \revision{parcimonieux} » - c'est \`a dire que la probabilit\'e de trouver un point proche d'un autre devient de plus en plus faible (on pourrait parler d'espace « \revision{clairsem\'e} »). Ou de mani\`ere duale, pour reprendre la formulation de \citeNP{HastieEtal}, le volume qu'il convient de consid\'erer pour avoir une proportion donn\'ee d'observations augmente avec $p$. L'id\'ee de r\'eduire la dimension en consid\'erant une analyse en composante principale peut para\^itre s\'eduisante, mais l'analyse souffre d'un certain nombre de d\'efauts en grande dimension. La solution est alors souvent la s\'election de variables, qui pose le probl\`eme des tests multiples, ou du temps de calcul, pour s\'electionner $k$ variables parmi $p$, lorsque $p$ est grand.

\revision{Pour reprendre la terminologie de \citeNP{Buhlmann}, les probl\`emes que nous \'evoquons ici correspondent \`a ceux observ\'es en grande dimension, qui est un probl\`eme essentiellement statistique. D'un point de vue informatique, on peut aller un peu plus loin, avec des donn\'ees r\'eellement massives (qui occupent \'enorm\'ement de place en m\'emoire). Dans ce qui pr\'ec\`ede, les donn\'ees \'etaient stock\'ees dans une matrice $\boldsymbol{X}$, de taille $n\times p$. Si cet objet formel est toujours bien d\'efini, il peut y avoir des soucis \`a stocker une telle matrice, voire manipuler une matrice abondament utilis\'ee en \'econom\'etrie, $\boldsymbol{X}^{\text{\sffamily T}}\boldsymbol{X}$ (matrice $n\times n$). La condition du premier ordre (dans le mod\`ele lin\'eaire) est associ\'ee \`a la r\'esolution de $\boldsymbol{X}^{\text{\sffamily T}}(\boldsymbol{X}\boldsymbol{\beta}-\boldsymbol{y})=\boldsymbol{0}$. En dimension raisonnable, on utilise la d\'ecomposition QR (c'est \`a dire la d\'ecomposition de Gram-Schmidt). En grande dimension, on peut utiliser des m\'ethodes num\'eriques de descente de gradient, o\`u le gradient est approch\'e sur un sous-\'echantillon de donn\'ees (comme d\'ecrit par exemple dans \citeNP{Zinkevichetal}) Cet aspect informatique est souvent oubli\'e alors qu'il a \'et\'e \`a la base de bon nombre d'avanc\'ees m\'ethodologiques, en \'econom\'etrie. Par exemple, \citeNP{HoerlKennard} reviennent sur l'origine de l'utilisation de la r\'egression Ridge: « {\em Nous facturions  90\$ par jour pour notre temps, mais avons dû facturer 450\$ par heure d'ordinateur sur un Univac I} ($ \cdots $) {\em Avec cette machine, il a fallu 75 minutes de traitement pour inverser une matrice  $40 \times 40 $ en passant par une partition $ 4 \times 4 $ de sous-matrices $10\times 10$, en utilisant des bandes magnétiques pour le stockage temporaire. Nous avons not\'e que les coefficients d'un régression linéaire calculés en utilisant les moindres carrés n'avaient pas toujours de sens. Les coefficients avaient tendance à être trop grands en valeur absolue, certains avaient même le mauvais signe, et ils pouvaient être instables avec de très petits changements dans les données} ($ \cdots $) {\em Comme la méthode que nous proposions attaquait l'une des vaches sacrées de la régression linéaire - les moindres carrés - nous avons fait face \`a une une résistance considérable}».} 

%%%% EN ANGLAIS {\em We were charging \$90/day for our time, but had to charge \$450/hour for computer time on a Univac I that had} ($\cdots$) {\em With this machine, it took 75 processing minutes to invert a $40 \times 40$ matrix through a $4 \times 4$ partition of $10 \times 10$ submatrices, using magnetic tapes for temporary storage. We found that multiple linear regression coefficients computed using least squares didn’t always make sense when put into the context of the process generating the data. The coefficients tended to be too large in absolute value, some would even have the wrong sign, and they could be unstable with very small changes in the data} ($\cdots$) {\em Since the method proposed attacked one of the sacred cows of linear regression - least squares - there was considerable resistance }».} \

\

\subsection{Statistique computationnelle et non-param\'etrique}

L'objet de ce papier est d'expliquer les diff\'erences majeures entre l'\'econom\'etrie et l'apprentissage statistique, correspondant aux deux cultures mentionn\'ees par \citeNP{Breiman}, lorsqu'il \'evoque en mod\'elisation statistique la « {\em data modeling culture} » (reposant sur un mod\`ele stochastique, comme la r\'egression logistique ou le mod\`ele de Cox) et la « {\em algorithmic modeling culture} » (reposant sur la mise en \oe{}uvre d'un algorithme, comme dans les for\^ets al\'eatoires ou les supports vecteurs machines, une liste exhaustive est présenté dans \citeNP{Shalev-Shwartz}). Mais la fronti\`ere entre les deux est tr\`es poreuse. \`A l'intersection se retrouve, par exemple, l'\'econom\'etrie non-param\'etrique. Cette derni\`ere repose sur un mod\`ele probabiliste (comme l'\'e\-co\-no\-m\'e\-trie), tout en insistant davantage sur les algorithmes (et leurs performances) plut\^ot que sur des th\'eor\`emes asymptotiques.

\revision{L'\'econom\'etrie} non-param\'etrique repose sur des d\'ecompositions dans des bases fonctionnelles. L'\'eco\-no\-m\'e\-trie lin\'eaire consiste \`a approcher la fonction $m:\bx\mapsto\esp[Y\vert \bX=\bx]$ par une fonction lin\'eaire. Mais plus g\'en\'eralement, on peut consid\'erer une d\'ecomposition dans une base fonctionnelle, et s'int\'eresser \`a une approximation obtenue sur un nombre fini de termes :
$$
m(\bx)=\sum_{j=0}^\infty \omega_j g_j(\bx)  \quad\text{ et }\quad
\widehat{m}(\bx)=\sum_{j=0}^{h^\star} \widehat{\omega}_j g_j(\bx),
$$
o\`u les poids $\omega_j$ sont estim\'es, alors que le nombre de composantes $h^\star$ est optimis\'e. On retrouvera ici les mod\`eles additifs (dits {\sc gam}), par exemple, \'etudi\'es dans  \citeNP{Trevor}. Une autre solution consiste \`a consid\'erer un mod\`ele simple, mais local. Par exemple un mod\`ele constant, au voisinage de $\bx$, obtenu en consid\'erant seulement les observations proches de $\bx$ :
$$
\widehat{g}(\bx)=\sum_{i=1}^n \widehat{\omega}_{\bx} y_i\text{ \quad par exemple \quad }\widehat{g}(\bx)=\frac{1}{n_{\bx}}\sum_{i:\Vert \bx_i-\bx\Vert \leq h} y_i
$$
o\`u $n_{\bx}$ est le nombre d'observations au voisinage de $\bx$. En mettant des poids fonctions de la distance \`a $\bx$, on retrouve ici le mod\`ele obtenu par \citeNP{Nadaraya} et \citeNP{Watson}, ou les m\'ethodes de r\'egression locale.

Les diff\'erentes m\'ethodes reposent sur des m\'eta-param\`etres - correspondant param\`etres de lissage - c'est \`a dire $h$ dans les exemples pr\'ec\'edents. Pour un \'econom\`etre, le param\`etre « optimal » pour $h$ est obtenu soit \`a l'aide de th\'eor\`emes asymptotiques, soit \`a l'aide de techniques de validation, comme en {\em machine learning}.  On obtient alors une valeur num\'erique, mais on n'a pas d'interpr\'etation en lien avec la taille de l'\'echantillon, ou les variances des diff\'erentes grandeurs. \revision{Si les économistes ont toujours la culture du tableau présentant la\linebreak« sortie de régression », les méthodes non-paramétriques sont utiles pour détecter des mauvaises spécifications, des non-prises en compte de nonlinéarité, ou d'effets croisées (et les outils de « machine learning » que nous allons voir peuvent probablement jouer le même rôle).}

\subsection{Plan de l'article}

Pour reprendre le titre de \citeNP{Varian}, l'objet de cet article est de pr\'esenter les diff\'erences fondamentales entre l'\'econom\'etrie et l'apprentissage \revision{machine}, et surtout de voir comment ces deux techniques peuvent apprendre l'une de l'autre, dans un contexte o\`u les bases de donn\'ees deviennent massives. La Section \ref{sec:1:econometrie} reviendra sur la construction du mod\`ele lin\'eaire. Le mod\`ele sera introduit ici \`a partir du mod\`ele Gaussien « homosc\'edastique ». Ce mod\`ele pr\'esente l'avantage d'avoir une \'el\'egante interpr\'etation g\'eom\'etrique, en terme de projection sur le sous-espace des combinaisons lin\'eaires des variables explicatives. La premi\`ere extension que nous verrons est le passage du mod\`ele lin\'eaire \`a un mod\`ele non-lin\'eaire, tout en construisant un pr\'edicteur lin\'eaire. La seconde extension proposera de construire un mod\`ele non-gaussien, pour mod\'eliser une variable indicatrice ou un comptage Poissonnien, par exemple, donnant naissance aux mod\`eles lin\'eaires g\'en\'eralis\'es (construits pour des variables dans la famille exponentielle). %Nous cl\^oturerons sur des rappels sur les mod\`eles plus r\'ecents, proposant de tester un r\'eel effet causal, plut\^ot qu'une corr\'elation significativement non nulle, par exemple.

Une fois rappel\'e l'origine des outils \'econom\'etriques standards, dans la Section \ref{sec:2:ML} nous pr\'esenterons les outils et techniques d\'evelopp\'es dans le contexte de l'apprentissage machine. Si l'outil central des mod\`eles \'econom\'etriques est la distribution de la variable d\'ependante, $Y$, les techniques d'apprentissage reposent sur une fonction de perte, $\ell$, repr\'esentant une « distance » entre la variable d'int\'er\^et $y$, et le mod\`ele $m(\cdot)$. Nous pr\'esenterons tout d'abord l'algorithme de boosting, reposant sur l'id\'ee d'un apprentissage lent, en mod\'elisant s\'equentiellement les r\'esidus. Le danger des m\'ethodes d'ap\-pren\-tis\-sage est qu'il est ais\'e de construire un mod\`ele « parfait », dont le pouvoir pr\'edictif serait faible. Nous \'evoquerons alors les techniques de p\'enalisation, utilis\'ees pour \'eviter le sur-apprentissage. Nous \'evoquerons en particulier les notions d'{\em in-sample} et {\em out-of-sample}, et les techniques de validation crois\'ee. Pour conclure cette section, nous reviendrons sur les interpr\'etations probabilistes des outils d'apprentissage, qui permettront de faire le lien entre les diff\'erentes approches, tout en restant sur une discussion g\'en\'erale sur la philosophie de ces deux cultures.

Apr\`es cette section sur la philosophie des m\'ethodes de {\em machine learning}, nous reviendrons dans la section \ref{sec:algo} sur quelques algorithmes importants : les r\'eseaux de neurones, les supports vecteurs machine (SVM) et enfin les m\'ethodes de type arbres et for\^ets.

La Section \ref{sec:3:classif} proposera des exemples concrets de comparaison entre les diff\'erentes techniques, dans le cas de classifications (binaires) pour des variables $y\in\{0,1\}$ (achat d'assurance, non-remboursement d'un cr\'edit) et dans un contexte de r\'egression (lorsque la variable d'int\'er\^et n'est plus qualitative - ce que nous simplifierons en notant $y\in\mathbb{R}$). Nous reviendrons avant sur les courbes ROC, outils importants pour juger de la qualit\'e d'un classifieur, malheureusement peu utilis\'es en \'econom\'etrie.
%L'outil classique en \'econom\'etrie est la r\'egression logistique, alors qu'une technique populaire en apprentissage est celle des supports vecteurs machines (SVM). Si ces deux m\'ethodes sont tr\`es diff\'erentes, les pr\'edictions sont - \'e\-tran\-gement - tr\`es proches (m\^eme si la r\'egression logistique permet de pr\'edire une probabilit\'e $\widehat{y}\in[0,1]$, alors que l'algorithme SVM pr\'edit la classe $\widehat{y}\in\{0,1\}$). Nous verrons aussi les r\'esaux de neurones, les arbres de classification, et les for\^ets al\'eatoires, outils qui ont permis clairement une jonction entre les deux communaut\'es. Nous conclurons cette section en pr\'esentant les outils de s\'election de variables utilis\'ees en apprentissage statistique. Ces outils seront utilis\'es sur diff\'erents jeux de donn\'ees.
%Enfin, La Section \ref{sec:4:reg} reviendra sur l'utilisation des techniques de {\em machine learning} dans un contexte de r\'egression (lorsque la variable d'int\'er\^et n'est plus qualitative - ce que nous simplifierons en notant $y\in\mathbb{R}$).
Nous verrons en particulier les m\'ethodes de bagging, for\^ets al\'eatoires ou boosting. Nous reviendrons aussi sur les m\'ethodes de choix de mod\`eles et des m\'eta-param\`etres. \`A travers ces exemples d'application, nous verrons comment les mod\`eles de type {\em machine learning} peuvent \^etre utilis\'es pour mieux détecter la mauvaise sp\'ecification des mod\`eles de r\'egression param\'etriques, \`a cause de non-lin\'earit\'es, et/ou d'int\'eractions manqu\'ees.

%Enfin, nous conclurons dans la Section \ref{sec:4:utilisation} en pr\'esentant des techniques classiques d'apprentissage statistiques, qui peuvent \^etre utilis\'ees dans les mod\`eles \'econom\'etriques classiques. Nous reviendrons ainsi sur les techniques dites de « bagging », bas\'ees sur des simulations (par r\'e\'echantillonnage) et d'aggr\'egation. Nous verrons ensuite comment utiliser dans un contexte de r\'egression les techniques de composantes principales. L'algorithme E.M. sera ensuite pr\'esent\'e, en lien avec les mod\`eles \`a effets al\'eatoires. Nous verrons \'egalement comment passer d'un arbre de classification \`a un arbre de r\'egression, en particulier en lien avec la formule de d\'ecomposition de la variance. Nous avons \'evoqu\'e dans la section \ref{sec:1:econometrie} les liens entre l'\'econom\'etrie et les techniques de projection, et dans la section \ref{sec:2:ML} l'apprentissage s\'equentiel. Nous verrons comment combiner les deux, introduisant la notion de poursuite de projections, en introduisant au passage les m\'ethodes de r\'eseaux de neurones. Enfin, nous verrons comment les m\'ethodes de p\'enalisation peuvent \^etre utilis\'ees pour choisir des instruments, dans un contexte \'econom\'etrique, avec des donn\'ees massives (et un vaste choix possible d'instruments).

\section{\'Econom\'etrie et mod\`ele probabiliste}\label{sec:1:econometrie}

%{\color{blue} @@ $x$ = observation, $\bx$ = vecteur d'observations,  $X$ = variable al\'eatoire, $\bX$ = vecteur al\'eatoire ET la matrice des observations $\bx_i$ }

L'importance des mod\`eles probabilistes en \'economie trouve sa source dans les questionnements de \citeNP{Working} et les tentatives de r\'eponses apport\'ees dans les deux tomes de \citeNP{Tinbergen}. Ces derniers ont engendr\'e par la suite énormément de travaux, comme le rappelle \citeNP{Duo} dans son ouvrage sur les fondements de l'\'econom\'etrie, et plus particuli\`erement dans le premier chapitre « {\em The Probability Foundations of Econometrics} ». Rappelons que Trygve Haavelmo a re\c{c}u le prix Nobel d’\'economie en 1989 pour sa « {\em clarification des fondations de la th\'eorie probabiliste de l'\'econom\'etrie} ». Car comme l'a montr\'e \citeNP{Haavelmo} (initiant un changement profond dans la th\'eorie \'econom\'etrique dans les ann\'ees 1930, comme le rappelle le chapitre 8 de \citeNP{Morgan}) l'\'econom\'etrie repose fondamentalement sur un mod\`ele probabiliste, et ceci pour deux raisons essentielles. Premi\`erement, l'utilisation de grandeurs (ou « mesures »~) statistiques telles que les moyennes, les erreurs-types et les coefficients de cor\-r\'e\-la\-tion \`a des fins inf\'erentielles ne peut se justifier que si le processus g\'en\'erant les donn\'ees peut \^etre exprim\'e en termes de mod\`ele probabiliste. Deuxi\`emement, l'approche par les probabilit\'ees est relativement g\'en\'erale, et se trouve \^etre particuli\`erement adapt\'ee \`a l'analyse des observations « d\'ependantes » et « non homog\`enes », telles qu'on les trouve souvent sur des donn\'ees \'economiques. On va alors supposer qu'il existe un espace probabiliste $(\Omega,\mathcal{F},\mathbb{P})$ tel que les observations $ (y_i,\bx_i)$ sont vues comme des r\'ealisations de variables al\'eatoires $ (Y_i,\bX_i)$. En pratique, la loi jointe du couple $ (Y,\bX)$ nous int\'eresse toutefois peu : la loi de $\bX$ est inconnue, et c'est la loi de $Y$ conditionnelle \`a $\bX$ qui nous int\'eressera. Dans la suite, nous noterons $x$ une observation, $\bx$ un vecteur d'observations,  $X$ une variable al\'eatoire, et $\bX$ un vecteur al\'eatoire et, abusivement, $\bX$ pourra aussi désigner la matrice des observations individuelles (les $\bx_i$), suivant le contexte.

\revision{\subsection{Fondements de la statistique math\'ematique}}

\revision{Comme le rappelle l'introduction de \citeNP{Vapnik98}, l'inf\'erence en statistique param\'etrique est basée sur la croyance suivante:
le statisticien connaît bien le problème à analyser, en particulier, il connaît la loi physique qui génère les propriétés stochastiques des données, et la fonction à trouver s'\'ecrit via un nombre fini de paramètres\footnote{On peut rapprocher cette approche de l'\'econom\'etrie structurelle, telle que pr\'esent\'ee par exemple dans \citeNP{Keen}.}. Pour trouver ces paramètres, on adopte la méthode du maximum de vraisemblance. Le but de la théorie est de justifier cette approche (en découvrant et en décrivant ses propriétés favorables). On verra qu'en apprentissage, la philosophie est tr\`es diff\'erente, puisqu'on ne dispose pas d'informations a priori fiables sur la loi statistique sous-jacente au problème, ni-m\^eme sur la fonction que l'on voudrait approcher (on va alors proposer des méthodes pour construire une approximation à partir de donn\'ees \`a notre disposition, pour reprendre \citeNP{Vapnik98}).
Un « âge d'or » de l'inférence paramétrique, de 1930 \`a 1960, a pos\'e les bases de la statistique math\'ematique, que l'on retrouve dans tous les manuels de statistique, y compris aujourd'hui. Comme le dit \citeNP{Vapnik98}, le paradigme paramétrique classique est basé sur les trois croyances suivantes:}

\revision{\begin{enumerate}
\item Pour trouver une relation fonctionnelle à partir des données, le statisticien est capable de définir un ensemble de fonctions, linéaires dans leurs paramètres, qui contiennent une bonne approximation de la fonction souhaitée. Le nombre de paramètres décrivant cet ensemble est petit.
\item La loi statistique sous-jacente à la composante stochastique de la plupart des problèmes de la vie réelle est la loi normale.
Cette croyance a été soutenue en se référant au théorème de limite centrale, qui stipule que dans de larges conditions la somme d'un grand nombre de variables aléatoires est approximée par la loi normale. 
\item La méthode du maximum de vraisemblance est un bon outil pour estimer les paramètres.
\end{enumerate}
Nous reviendrons dans cette partie sur la construction du paradigme \'econom\'etrique, directement inspir\'e de celui de la statistique inf\'erentielle classique.
}

\subsection{Lois conditionnelles et vraisemblance}\label{sec:proba}

L'\'econom\'etrie lin\'eaire a \'et\'e construite sous l'hypoth\`ese de donn\'ees individuelles, ce qui revient \`a supposer les variables  $ (Y_i,\bX_i)$ in\-d\'e\-pen\-dan\-tes (s'il est possible d'imaginer des observations temporelles - on aurait alors un processus $ (Y_t,\bX_t)$ - mais nous n'aborderons pas les s\'eries temporelles dans cet article). Plus pr\'ecis\'ement, on va supposer que conditionnellement aux variables explicatives $\bX_i$, les variables $Y_i$ sont ind\'ependantes. On va \'egalement supposer que ces lois conditionnelles restent dans la m\^eme famille param\'etrique, mais que le param\`etre est une fonction de $\bx$. Dans le mod\`ele lin\'eaire Gaussien on suppose que :
\begin{equation}\label{eq:model_lineaire_1}
(Y\vert \bX=\bx)\overset{\mathcal{L}}{\sim}\mathcal{N}(\mu(\bx),\sigma^2)\quad\text{ avec }\quad\mu(\bx)=\beta_0+\bx\transpose\bbeta,\text{ et }\bbeta\in\mathbb{R}^{p}.
\end{equation}
On parle de mod\`ele lin\'eaire car $\Esp[Y\vert \bX=\bx]=\beta_0+\bx\transpose\bbeta$ est une combinaison lin\'eaire des variables explicatives. C'est un mod\`ele homosc\'edastique si $\Var[Y\vert \bX=\bx]=\sigma^2$, o\`u $\sigma^2$ est une constante positive. Pour estimer les param\`etres, l'approche classique consiste \`a utiliser l'estimateur du Maximum de Vraisemblance, comme l'avait sugg\'er\'e initialement Ronald Fisher. Dans le cas du mod\`ele lin\'eaire Gaussien, la log-vraisemblance s'\'ecrit :
$$
\log\mathcal{L}(\beta_0,\bbeta,\sigma^2\vert \by,\bx) = -\frac{n}{2}\log[2\pi\sigma^2] - \frac{1}{2\sigma^2}\sum_{i=1}^n (y_i-\beta_0-\bx_i\transpose\bbeta)^2.
$$
Notons que le terme de droite, mesurant une distance entre les donn\'ees et le mod\`ele, va s'interpr\'eter comme la d\'eviance, dans les mod\`eles lin\'eaires g\'en\'eralis\'es.
On va alors poser :
$$
(\widehat{\beta}_0,\widehat{\bbeta},\widehat{\sigma}^2)=\text{argmax}\big\lbrace\log\mathcal{L}(\beta_0,\bbeta,\sigma^2\vert \by,\bx)\big\rbrace.
$$
L'estimateur du maximum de vraisemblance est obtenu par minimisation de la somme des carr\'es des erreurs (estimateur dit des « moindres carr\'es »~) que nous retrouverons dans l'approche par {\em machine learning}.

Les conditions du premier ordre permettent de retrouver les \'equations normales, dont l'\'ecriture matricielle est $\bX\transpose[\by-\bX\widehat{\bbeta}]=\boldsymbol{0}$, que l'on peut aussi \'ecrire $(\bX\transpose\bX)\widehat{\bbeta}=\bX\transpose\by$. Si la matrice $\bX$ est de plein rang colonne, alors on retrouve l'estimateur classique :
\begin{equation}\label{eq:ols}
\widehat{\bbeta}=(\bX\transpose\bX)\inverse\bX\transpose\by=\bbeta + (\bX\transpose\bX)\inverse\bX\transpose\bepsilon
\end{equation}
en utilisant une \'ecriture bas\'ee sur les r\'esidus (comme souvent en \'econom\'etrie), $y=\bx\transpose\bbeta+\varepsilon$.
Le th\'eor\`eme de Gauss Markov assure que cette estimateur est l'estimateur lin\'eaire sans biais de variance minimale. On peut alors montrer que $\displaystyle{\widehat{\bbeta}\overset{\mathcal{L}}{\sim}\mathcal{N}(\bbeta,\sigma^2 [\bX\transpose\bX]\inverse)}$, et en particulier :
$$
\Esp[\widehat{\bbeta}]=\bbeta\quad\text{ et }\quad \var[\widehat{\bbeta}]=\sigma^2 [\bX\transpose\bX]\inverse.
$$

\revision{En fait, l'hypoth\`ese de normalit\'e permet de faire un lien avec la statistique math\'ematique, mais il est possible de construire cet estimateur donn\'e par l'\'equation (\ref{eq:ols}). Si on suppose que $Y\vert\bX=\bx\overset{\mathcal{L}}{\sim}\bx\transpose\bbeta+\varepsilon$, avec $\mathbb{E}[\varepsilon]=0$, $\text{Var}[\varepsilon]=\sigma^2$ , $\text{Cov}[\varepsilon,X_j]=0$ pour tout $j$, alors $\widehat{\bbeta}$ est un estimateur sans biais de $\bbeta$ ($\Esp[\widehat{\bbeta}]=\bbeta$) et de variance minimale parmi les estimateurs sans biais lin\'eaires, avec $\var[\widehat{\bbeta}]=\sigma^2 [\bX\transpose\bX]\inverse$. De plus, cet estimateur est asymptotiquement normal  
$$
\sqrt{n}(\widehat{\bbeta}-\bbeta)\overset{\mathcal{L}}{\rightarrow}
\mathcal{N}(\boldsymbol{0},\boldsymbol{\Sigma}) \quad\text{ lorsque }\quad n\rightarrow\infty
$$
}

La condition d'avoir  une matrice $\bX$ de plein rang peut être (num\'eriquement) forte en grande dimension. Si elle n'est pas v\'erifi\'ee, 
$\widehat{\bbeta}=(\bX\transpose\bX)\inverse\bX\transpose\by$ n'existe pas. Si $\mathbb{I}$ désigne la matrice identité, notons toutefois que 
$(\bX\transpose\bX+\lambda\mathbb{I})\inverse\bX\transpose\by$ existe toujours, pour $\lambda>0$. Cet estimateur est appel\'e l'estimateur Ridge de niveau $\lambda$ (introduit dans les ann\'ees 60 par \citeNP{Hoerl}, et associ\'e \`a une r\'egularisation \'etudi\'ee par \citeNP{Tikhonov}). Cette estimateur appara\^it naturellement dans un contexte d'\'econom\'etrie Bayesienne (nous le reverrons dans la section suivante pr\'esentant les techniques de {\em machine learning}). %{\color{red} @@ check}).

\subsection{Les r\'esidus}\label{sec:residus}

Il n'est pas rare d'introduire le mod\`ele lin\'eaire \`a partir de la loi des r\'esidus, comme nous l'avions mentionn\'e auparavant. Aussi, l'\'equation (\ref{eq:model_lineaire_1}) s'\'ecrit aussi souvent :
\begin{equation}\label{eq-erreur}
y_i=\beta_0+\bx_i\transpose\bbeta+\varepsilon_i
\end{equation}
o\`u les $\varepsilon_i$ sont des r\'ealisations de variables al\'eatoires i.i.d., de loi $\mathcal {N}(0,\sigma^2)$. On notera parfois $\bepsilon\overset{\mathcal{L}}{\sim}\mathcal{N}(\boldsymbol{0},\sigma^2\mathbb{I})$, sous une forme vectorielle. Les r\'esidus estim\'es sont d\'efinis par :
$$
\widehat{\varepsilon}_i=y_i - \big[\widehat{\beta}_0+\bx_i\transpose\widehat{\bbeta}\big]
$$
Ces r\'esidus sont l'outil de base pour diagnostiquer la pertinence du mod\`ele.

Une extension du mod\`ele d\'ecrit par l'\'equation (\ref{eq:model_lineaire_1}) a \'et\'e propos\'e pour tenir compte d'un \'eventuel caract\`ere h\'et\'erosc\'edastique :
$$
(Y\vert \bX=\bx)\overset{\mathcal{L}}{\sim}\mathcal{N}(\mu(\bx),\sigma^2(\bx))
$$
o\`u $\sigma^2(\bx)$ est une fonction positive des variables explicatives.
On peut r\'e\'ecrire ce mod\`ele en posant :
$$
y_i=\beta_0+\bx_i\transpose\bbeta+\sigma^2(\bx_i)\cdot\varepsilon_i
$$
o\`u les r\'esidus sont toujours i.i.d., mais de variance unitaire, 
$$
\varepsilon_i=\frac{y_i-[\beta_0+\bx_i\transpose\bbeta]}{
\sigma(\bx_i)}.
$$
Si l'\'ecriture \`a l'aide des r\'esidus est populaire en \'econom\'etrie lin\'eaire (lorsque la variable d\'ependante est continue), elle ne l'est toutefois plus dans les mod\`eles de comptage, ou la r\'egression logistique.

L'\'ecriture \`a l'aide d'un terme d'erreur (comme dans l'\'equation (\ref{eq-erreur})) pose toutefois de nombreuses questions quant \`a la repr\'esentation d'une relation \'economique entre deux grandeurs. Par exemple, on peut supposer qu'il existe une relation (lin\'eaire pour commencer) entre les quantit\'es d'un bien \'echang\'e, $q$ et son prix $p$. On peut ainsi imaginer une \'equation d'offre
$$
q_i=\beta_0+\beta_1 p_i+u_i
$$
($u_i$ désignant un terme d'erreur) o\`u la quantit\'e vendue d\'epend du prix, mais de mani\`ere tout aussi l\'egitime, on peut imaginer que le prix d\'epend de la quantit\'e produite (ce qu'on pourrait appeler une \'equation de demande),
$$
p_i=\alpha_0+\alpha_1 q_i+v_i
$$
($v_i$ désignant un autre terme d'erreur). 
Historiquement, le terme d'erreur dans l'\'equation (\ref{eq-erreur}) a pu \^etre intepr\'et\'e comme une erreur idiosyncratique sur la variable $y$, les variables dites explicatives \'etant suppos\'ees fix\'ees, mais cette interpr\'etation rend souvent le lien entre une relation \'economique et un mod\`ele \'economique compliqu\'e, la th\'eorie \'economique parlant de mani\`ere abstraite d'une relation entre grandeur, la mod\'elisation \'econom\'etrique imposant une forme sp\'ecifique (quelle grandeur est $y$ et quelle grandeur est $x$) comme le montre plus en d\'etails le chapitre 7 de \citeNP{Morgan}.

\subsection{G\'eom\'etrie du mod\`ele lin\'eaire gaussien}\label{sec:geometrie}

%{\color{blue} je pense que ca fait que c'est beau et simple... il existe des interpretations de ce genre en {\em machine learning}, disons sur les arbres ou les reseaux de neurone ? apres, c'est aussi interessant parce que - au dela de Frish Waugh que je trouve cool - ca peut introduire plus tard les SVM, cette histoire de g\'eom\'etrie, non ? Je tente un truc (cf plus loin sur les SVM)} 

D\'efinissons le produit scalaire dans $\mathbb{R}^n$, $\langle\ba,\bb\rangle=\ba\transpose\bb$, et notons $\Vert\cdot\Vert$ la norme euclidienne associ\'ee, $\Vert \ba\Vert = \sqrt{\ba\transpose\ba}$ (not\'ee $\|\cdot\|_{\ell_2}$ dans la suite). Notons $\mathcal{E}_{\bX}$ l'espace engendr\'e par l'ensemble des combinaisons lin\'eaires des composantes $\bx$ (en rajoutant la constante). Si les variables explicatives sont lin\'eairement ind\'ependantes, $\bX$ est de plein rang colonne et $\mathcal{E}_{\bX}$ est un sous-espace de dimension $p+1$ de $\mathbb{R}^n$. Supposons \`a partir de maintenant que les variables $\bx$ et la variable $y$ sont ici centr\'ees.
\revision{Notons qu'aucune hypoth\`ese de loi n'est faite dans cette section, les propriétés géométriques découlent des propriétés de l'espérance et de la variance dans l'espace des variables de variance finie.}

Avec cette notation, notons que le mod\`ele lin\'eaire s'\'ecrit $m(\bx)=\langle\bx,\boldsymbol{\beta}\rangle$. L'espace
$\mathcal{H}_z=\{\bx\in\mathbb{R}^k:m(\bx)=z\}$ est un hyperplan (affine) qui s\'epare l'espace en deux.
D\'efinissons l'op\'erateur de projection orthogonale sur $\mathcal{E}_{\mathcal{X}}=\mathcal{H}_0$, $\Pi_{\mathcal{X}}=\bX[\bX\transpose\bX]\inverse\bX\transpose$. Aussi, la pr\'evision que l'on peut faire pour $\by$ est :
\begin{equation}\label{eq:projection}
\widehat{\by}=\underbrace{\bX[\bX\transpose\bX]\inverse\bX\transpose}_
{\Pi_{\mathcal{X}}}\by=\Pi_{\mathcal{X}}\by.
\end{equation}
Comme $\widehat{\bepsilon}=\by-\widehat{\by}=(\mathbb{I}-\Pi_{\mathcal{X}})\by=\Pi_{\mathcal{X}^\perp}\by$, on note que $\widehat{\bepsilon}\perp\bx$, que l'on interpr\'etera en disant que les r\'esidus sont un terme d'innovation, impr\'evisible, au sens o\`u $\Pi_{\mathcal{X}}\widehat{\bepsilon}=\boldsymbol{0}$.

Le th\'eor\`eme de Pythagore s'\'ecrit ici :
\begin{equation}\label{eq:pyt}
\Vert \by \Vert^2=\Vert \Pi_{\mathcal{X}}\by \Vert^2+\Vert \Pi_{\mathcal{X}^\perp}\by \Vert^2=\Vert \Pi_{\mathcal{X}}\by \Vert^2+\Vert \by-\Pi_{\mathcal{X}}\by \Vert^2=\Vert\widehat{\by}\Vert^2+\Vert\widehat{\boldsymbol{\varepsilon}}\Vert^2
\end{equation}
qui se traduit classiquement en terme de somme de carr\'es :
$$
\underbrace{\sum_{i=1}^n y_i^2}_{n\times\text{variance totale}}=\underbrace{\sum_{i=1}^n \widehat{y}_i^2}_{n\times\text{variance expliqu\'ee}}
+
\underbrace{\sum_{i=1}^n (y_i-\widehat{y}_i)^2}_{n\times\text{variance r\'esiduelle}}$$
Le coefficient de d\'etermination, $R^2$ (nous reviendrons sur ce coefficient dans la section \ref{sec:goodness}) s'interpr\^ete alors comme le carr\'e du cosinus de l'angle $\theta$ entre $\by$ et $\Pi_{\mathcal{X}}\by$ :
$$
R^2=\frac{\Vert \Pi_{\mathcal{X}} \by\Vert^2}{\Vert \by\Vert^2}=1-\frac{\Vert \Pi_{\mathcal{X}^\perp} \by\Vert^2}{\Vert \by\Vert^2}=\cos^2(\theta).
$$

Une application importante a \'et\'e obtenue par \citeNP{FrishWaugh}, lorsque l'on partitionne les variables explicatives en deux groupes, $\bX=[ \bX_1 |\bX_2]$, de telle sorte que la r\'egression devient :
$$
\by =\beta_0+ \bX_1\bbeta_1+\bX_2\bbeta_2+\varepsilon
$$
\citeNP{FrishWaugh} ont montr\'e qu'on pouvait consid\'erer deux projections successives. En effet, si $\by_2^\star=\Pi_{\mathcal{X}_1^\perp}\by$ et $\bX_2^\star=\Pi_{\mathcal{X}_1^\perp}\bX_2$, on peut montrer que
$$
\widehat{\bbeta}_2=[\bX_2^\star{}\transpose\bX_2^\star]\inverse\bX_2^\star
{}\transpose\by_2^\star
$$
Autrement dit, l'estimation globale est \'equivalente \`a l'estimation ind\'ependante des deux mod\`eles si $\bX_2^\star=\bX_2$, c'est \`a dire $\bX_2\in\mathcal{E}_{\bX_1}^\perp$, que l'on peut noter $\bx_1\perp\bx_2$. On obtient ici le th\'eor\`eme de Frisch-Waugh qui garantie que si les variables explicatives entre les deux groupes sont orthogonales, alors l'estimation globale est \'equivalente \`a deux r\'egressions ind\'ependantes, sur chacun des jeux de variables explicatives. Ce qui est un th\'eor\`eme de double projection, sur des espaces orthogonaux. Beaucoup de r\'esultats et d'interpr\'etations sont obtenus par des interpr\'etations g\'eom\'etriques (li\'ees fondamentalement aux liens entre l'esp\'erance conditionnelle et la projection orthogonale dans l'espace des variables de variance finie).

Cette vision g\'eom\'etrique permet de mieux comprendre le probl\`eme de la sous-identification, c'est \`a dire le cas o\`u le vrai mod\`ele serait $y_i=\beta_0+\bx_1\transpose\bbeta_1+\bx_2\transpose\bbeta_2+\varepsilon_i$, mais le mod\`ele estim\'e est $y_i=\beta_0+\bx_1\transpose\boldsymbol{b}_1+\eta_i$. L'estimateur du maximum de vraisemblance de $\boldsymbol{b}_1$ est :
\begin{eqnarray*}
\widehat{\boldsymbol{b}}_1
&=& {(\boldsymbol{X}_1^{\text{\sffamily{T}}}\boldsymbol{X}_1)^{-1} \boldsymbol{X}_1^{\text{\sffamily{T}}}} \by \\
&=& {(\boldsymbol{X}_1^{\text{\sffamily{T}}}\boldsymbol{X}_1)^{-1} \boldsymbol{X}_1^{\text{\sffamily{T}}}}
[\boldsymbol{X}_{1,i} \boldsymbol{\beta}_1
    + \boldsymbol{X}_{2,i} \boldsymbol{\beta}_2 + \varepsilon] \\
&=& (\boldsymbol{X}_1^{\text{\sffamily{T}}}\boldsymbol{X}_1)^{-1} \boldsymbol{X}_1^{\text{\sffamily{T}}}\boldsymbol{X}_{1} \boldsymbol{\beta}_1
+ (\boldsymbol{X}_1^{\text{\sffamily{T}}}\boldsymbol{X}_1)^{-1} \boldsymbol{X}_1^{\text{\sffamily{T}}}\boldsymbol{X}_{2} \boldsymbol{\beta}_2
+ (\boldsymbol{X}_1^{\text{\sffamily{T}}}\boldsymbol{X}_1)^{-1} \boldsymbol{X}_1^{\text{\sffamily{T}}}\varepsilon \\
&=& \boldsymbol{\beta}_1 + \underbrace{ (\boldsymbol{X}_1'\boldsymbol{X}_1)^{-1} \boldsymbol{X}_1^{\text{\sffamily{T}}}\boldsymbol{X}_{2} \boldsymbol{\beta}_2}_{\boldsymbol{\beta}_{12}}
+\underbrace{(\boldsymbol{X}_1^{\text{\sffamily{T}}}\boldsymbol{X}_1)^{-1} \boldsymbol{X}_1^{\text{\sffamily{T}}}\varepsilon}_{\nu_i}
\end{eqnarray*}
de telle sorte que $\esp[\widehat{\boldsymbol{b}}_1]=\bbeta_1+\bbeta_{12}$, le biais \'etant nul uniquement dans le cas o\`u $\boldsymbol{X}_1^{\text{\sffamily{T}}}\boldsymbol{X}_{2}=\boldsymbol{0}$ (c'est \`a dire $\boldsymbol{X}_1 \perp \boldsymbol{X}_{2}$): on retrouve ici une cons\'equence du th\'eor\`eme de Frisch-Waugh.

En revanche, la sur-identification correspond au cas o\`u le vrai mod\`ele serait $y_i=\beta_0+\bx_1\transpose\bbeta_1+\varepsilon_i$, mais le mod\`ele estim\'e est $y_i=\beta_0+\bx_1\transpose\boldsymbol{b}_1+\bx_2\transpose\boldsymbol{b}_2+\eta_i$.
Dans ce cas, l'estimation est sans biais, au sens o\`u $\mathbb{E}(\widehat{\boldsymbol{b}}_1)=\boldsymbol{\beta}_1$ mais l'estimateur n'est pas efficient. Et comme nous l'avons vu dans la section pr\'ec\'edente, il n'est pas rare d'avoir des valeurs de $\widehat{\boldsymbol{b}}_2$ qui sont consid\'er\'ees comme significativement non-nulles. Nous \'evoquerons dans la section suivante une m\'ethode efficace de choix de variables (et \'eviter la sur-identification).

\subsection{Du param\'etrique au  non-param\'etrique}\label{sec:nonlineaire}

La r\'e\'ecriture de l'\'equation (\ref{eq:projection}) sous la forme
$$
\widehat{\by}=\bX\widehat{\bbeta} 
=\underbrace{\bX[\bX\transpose\bX]\inverse\bX\transpose}_{ \Pi_{\mathcal{X}}}\by
$$ 
permet de voir la pr\'evision directement comme une transformation lin\'eaire des observations. De mani\`ere plus g\'en\'erale, on peut obtenir un pr\'edicteur lin\'eaire en consid\'erant $m(\bx)=\bs_{\bx}\transpose\by$, o\`u $\bs_{\bx}$ est un vecteur de poids, qui d\'ependent de $\bx$, interpr\'et\'e comme un vecteur de lissage. En utilisant les vecteurs $\bs_{\bx_i}$, calcul\'es \`a partir des ${\bx_i}$, on obtient une matrice  $\boldsymbol{S}$ de taille $n\times n$, et $\widehat{\by}=\boldsymbol{S}\by$.
Dans le cas de la r\'egression lin\'eaire d\'ecrite auparavant, $\bs_{\bx}=\bX[\bX\transpose\bX]\inverse\bx$, et classiquement, $\text{trace}( \boldsymbol{S})$ est le nombre de colonnes de la matrice $\bX$ (le nombre de variables explicatives). Dans ce contexte de pr\'edicteurs lin\'eaires,  $\text{trace}( \boldsymbol{S})$ est souvent vu comme un \'equivalent au nombre de param\`etres (ou complexit\'e, ou dimension, du mod\`ele), et $\nu=n-\text{trace}( \boldsymbol{S})$ est alors le nombre de degr\'es de libert\'e (comme d\'efini dans \citeNP{Ruppert} et \citeNP{Simonoff}). Le principe de parcimonie\footnote{« {\em pluralitas non est
ponenda sine necessitate} » pour reprendre le principe \'enonc\'e par Guillaume d'Occam (les multiples ne
doivent pas être utilisés sans nécessité).} consiste \`a minimiser cette dimension (la trace de la matrice $\boldsymbol{S}$) autant que faire se peut. Mais dans le cas g\'en\'eral, cette dimension est plus complexe \`a d\'efinir. Notons que l'estimateur introduit par \citeNP{Nadaraya} et \citeNP{Watson}, dans le cas d'une r\'egression non-param\'etrique simple, s'\'ecrit \'egalement sous cette forme puisque
$$
\widehat{m}_h(x)=\bs_{x}\transpose\by=\sum_{i=1}^n s_{x,i}y_i \quad\text{ avec }\quad s_{x,i}=\frac{K_h(x-x_i)}{K_h(x-x_1)+\cdots+K_h(x-x_n)},
$$
o\`u $K(\cdot)$ est une fonction noyau, qui attribue une valeur d'autant plus faible que $x_i$ est proche de $x$, et $h>0$ est la fen\^etre de lissage. 
%Notons de plus que la statistique {\color{red} (que veut-on dire ici ?  )}
%$$T=\frac{\Vert \boldsymbol{S}\bY-\boldsymbol{H}\bY \Vert}{\text{trace}([\boldsymbol{S}-\boldsymbol{H}]\transpose[\boldsymbol{S}-\boldsymbol{H}])}$$
%permet d'avoir un test de lin\'earit\'e (avec comme hypoth\`ese alternative le mod\`ele nonlin\'eaire obtenu \`a l'aide de la matrice de lissage $\boldsymbol{S}$). Sous l'hypoth\`ese de mod\`ele lin\'eaire, $T$ suit en effet une loi de Fisher (voir \citeNP{Simonoff}).

\begin{figure}
\begin{center}
\includegraphics[width=0.4\textwidth]{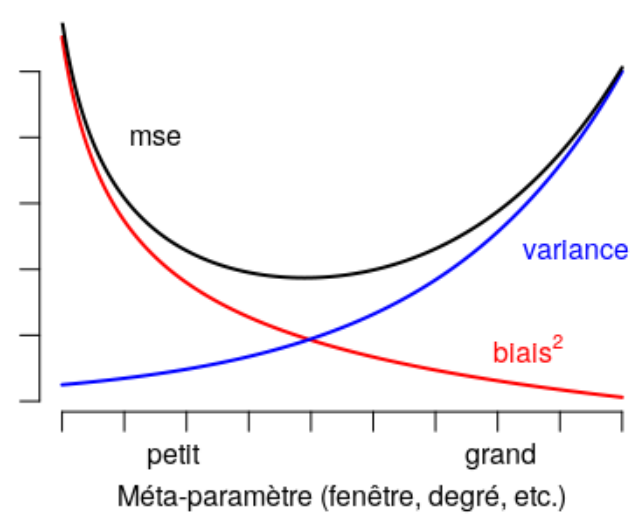}~
\includegraphics[width=0.4\textwidth]{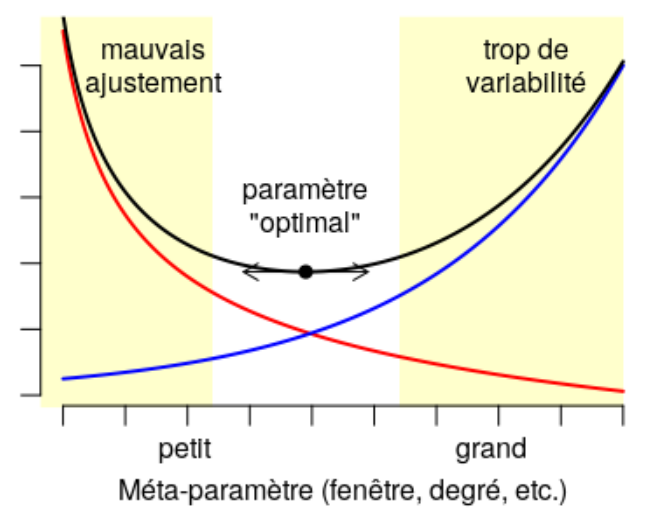}
\end{center}
\caption{Choix de $h$ et le probl\`eme de Boucle d'Or : $h$ ne doit \^etre ni trop grand (sinon il y a trop de variance), ni trop petit (sinon il y a trop de biais) .}\label{Fig:tradeoff}
\end{figure}

\revision{L'introduction de ce m\'eta-param\`etre $h$ pose un soucis, car il convient de le choisir judicieusement. En faisant des d\'eveloppement limit\'es, on peut montrer que si $X$ a pour densit\'e $f$,
$$
\text{biais}[\widehat{m}_h(x)]=\mathbb{E}[\widehat{m}_h(x)]-m(x)\sim {h^2}\left(
\frac{C_1 }{2}m''(x)+C_2 m'(x)\frac{f'(x)}{f(x)}
\right)
\text{ et }\displaystyle{{\text{Var}[\widehat{m}_h(x)]\sim\frac{C_3}{{nh}}\frac{\sigma(x)}{f(x)}}}
$$
pour des constantes que l'on peut estimer (voir \citeNP{Simonoff} par exemple). Ces deux fonctions \'evoluent inversement en fonction de $h$, comme le rappelle la Figure \ref{Fig:tradeoff}. L'id\'ee naturelle est alors de chercher \`a minimiser l'erreur quadratique moyenne, le MSE, $\text{biais}[\widehat{m}_h(x)]^2+\text{Var}[\widehat{m}_h(x)]$, ce qui donne une valeur optimale pour $h$ de la forme $h^\star=O(n^{-1/5})$, ce qui rappelle la r\`egle de \citeNP{Silverman}. En plus grande dimension, pour des variables $\bx$ continues, on peut utiliser un noyau multivari\'e, de fen\^etre matricielle $\boldsymbol{H}$, 
$$
\mathbb{E}[\widehat{m}_{\boldsymbol{H}}(\boldsymbol{x})]\sim m(\boldsymbol{x})
+\frac{C_1}{2}\text{trace}\big(\boldsymbol{H}^{\text{\sffamily T}}m''(\boldsymbol{x})\boldsymbol{H}\big)+C_2\frac{m'(\boldsymbol{x})^{\text{\sffamily T}}\boldsymbol{H}\boldsymbol{H}^{\text{\sffamily T}} \nabla f(\boldsymbol{x})}{f(\boldsymbol{x})}
\text{ et }
\text{Var}[\widehat{m}_{\boldsymbol{H}}(\boldsymbol{x})]\sim
\frac{C_3}{n~\text{det}(\boldsymbol{H})}\frac{\sigma(\boldsymbol{x})}{f(\boldsymbol{x})}.
$$
Si $\boldsymbol{H}$ est une matrice diagonale, avec le m\^eme terme $h$ sur la diagonale, alors $h^\star=\displaystyle{O(n^{-1/(4+\text{dim}(\boldsymbol{x}))})]}$. Cela dit, en pratique, on sera davantage int\'eress\'e par la version int\'egr\'ee de l'erreur quadratique,
$$
MISE(\widehat{m}_{h})=\mathbb{E}[MSE(\widehat{m}_{h}(X))]=\int MSE(\widehat{m}_{h}(x))dF(x),
$$
dont on peut montrer que
\begin{eqnarray*}
MISE[\widehat{m}_h]&\sim &\overbrace{\frac{h^4}{4}\left(\int x^2k(x)dx\right)^2\int\big[m''(x)+2m'(x)\frac{f'(x)}{f(x)}\big]^2dx}^{\text{biais}^2} +\overbrace{\frac{\sigma^2}{nh}\int k^2(x)dx \cdot\int\frac{dx}{f(x)}}^{\text{variance}},
\end{eqnarray*}
lorsque $n\rightarrow \infty$ et $nh\rightarrow\infty$. On retrouve ici une relation asymptotique qui rappelle l'ordre de grandeur de \citeNP{Silverman},
$$
h^\star =n^{-\frac{1}{5}}\left(\frac{C_1\int \frac{dx}{f(x)}}{C_2\int \big[m''(x)+2m'(x)\frac{f'(x)}{f(x)}\big]dx}\right)^{\frac{1}{5}},
$$
sauf que beaucoup de termes ici sont inconnus. On verra, le machine-learning propose des techniques computationnelles, lorsque l'\'econom\`etre avait pris l'habitude de chercher des propri\'et\'es asymptotiques.}

\subsection{Famille exponentielle et mod\`eles lin\'eaires}\label{sec:famille:exponentielle}

Le mod\`ele lin\'eaire Gaussien est un cas particulier d'une vaste famille de mod\`eles lin\'eaires, obtenu lorsque la loi conditionnelle de $Y$ appartient \`a la famille exponentielle
$$
f(y_i|\theta_i,\phi)=\exp\left(\frac{y_i\theta_i-b(\theta_i)}{a(\phi)}+c(y_i,\phi)\right)\quad\text{ avec }\quad \theta_i=\psi({\boldsymbol{x}_i^{\text{\sffamily{T}}}\boldsymbol{\beta}}).
$$
Les fonctions $a$, $b$ et $c$ sont sp\'ecifi\'ees en fonction du type de
loi exponentielle (\'etudi\'ee abondamment en statistique depuis les \citeNP{Darmois}, comme le rappelle \citeNP{Brown}), et $\psi$ est une fonction bijective que se donne l'utilisateur. %{\color{red} (references)}.
La log-vraisemblance a alors une expression relative simple
$$
\log\mathcal{L}(\boldsymbol{\theta},\phi|\boldsymbol{y})=\prod_{i=1}^n \log f(y_i|\theta_i,\phi)
=\frac{\sum_{i=1}^ny_i\theta_i-\sum_{i=1}^nb(\theta_i)}{a(\phi)}+
\sum_{i=1}^n c(y_i,\phi) 
$$
et la condition du premier ordre s'\'ecrit alors
$$
\frac{\partial \log \mathcal{L}(\boldsymbol{\theta},\phi|\boldsymbol{y})}{\partial \boldsymbol{\beta}} = \boldsymbol{X}^{\text{\sffamily{T}}}\boldsymbol{W}^{-1}[\boldsymbol{y}-\widehat{\by}]=\boldsymbol{0}
$$
pour reprendre les notations de \citeNP{Muller}, o\`u $\boldsymbol{W}$ est une matrice de poids (qui d\'epend de $\bbeta$). Compte tenu du lien entre $\theta$ et l'esp\'erance de $Y$, au lieu de sp\'ecifier la fonction $\psi(\cdot)$, on aura plut\^ot tendance \`a specifier la fonction de lien $g(\cdot)$ d\'efinie par 
$$
\widehat{y}=m(\bx)=\esp[Y\vert \bX=\bx]=g\inverse(\bx\transpose\bbeta).
$$
Pour la r\'egression lin\'eaire Gaussienne on prendra un lien Identit\'e, alors que pour la r\'egression de Poisson, le lien naturel (dit canonique) est le lien logarithmique. Ici, comme $\boldsymbol{W}$ d\'epend de $\bbeta$ (avec  $\boldsymbol{W}=\text{diag}(\nabla g(\widehat{\by})\text{Var}[\by])$) il n'existe en g\'en\'eral par de formule explicite pour l'estimateur du maximum de vraisemblance. Mais un algorithme it\'eratif permet d'obtenir une approximation num\'erique. En posant
$$
\bz=g(\widehat{\by})+(\by-\widehat{\by})\cdot\nabla g(\widehat{\by})
$$
correspondant au terme d'erreur d'un d\'eveloppement de Taylor \`a l'ordre 1 de $g$, on obtient un algorithme de la forme
$$
\widehat{\boldsymbol{\beta}}_{k+1} = [\boldsymbol{X}^{\text{\sffamily T}}\boldsymbol{W}_k^{-1}\boldsymbol{X}]^{-1} \boldsymbol{X}^{\text{\sffamily T}}\boldsymbol{W}_k^{-1}\boldsymbol{z}_k
$$
En it\'erant, on notera $\widehat{\boldsymbol{\beta}}=\widehat{\boldsymbol{\beta}}_\infty$, et on peut montrer que - moyennant quelques hypoth\`eses techniques (cf \citeNP{Muller}) - cet estimateur est asymptotiquement Gaussien, avec
$$
\sqrt{n}(\widehat{\boldsymbol{\beta}}-\boldsymbol{\beta})\overset{\mathcal{L}}{\rightarrow}
\mathcal{N}(\boldsymbol{0},I(\boldsymbol{\beta})^{-1}),
$$
o\`u num\'eriquement $I(\boldsymbol{\beta})=\phi\cdot  [\boldsymbol{X}^{\text{\sffamily T}}\boldsymbol{W}_\infty^{-1}\boldsymbol{X}]$.

D'un point de vue num\'erique toujours, on r\'esout la condition du premier ordre, et la loi de $Y$ n'intervient pas r\'eellement. Par exemple, on peut estimer une « r\'egression de Poisson » m\^eme lorsque $y\in\mathbb{R}_+$, pas n\'ecessairement $y\in\mathbb{N}$. Autrement dit, la loi de $Y$ n'est qu'une interpr\'etation donn\'ee ici, et l'algorithme pourrait \^etre introduit de mani\`ere diff\'erente (comme nous le verrons dans la section suivante), sans forc\'ement avoir de mod\`ele probabiliste sous-jacent.

\subsection{R\'egression logistique}\label{sec:regression:logistique}

La r\'egression logistique est le mod\`ele lin\'eaire g\'en\'eralis\'e obtenu avec une loi de Bernoulli, et une fonction de lien qui est la fonction quantile d'une loi logistique (ce qui correspond au lien canonique au sens de la famille exponentielle). Compte tenu de la forme de la loi de Bernoulli, l'\'econom\'etrie propose un mod\`ele pour $y_i\in\lbrace0,1\rbrace$, dans lequel le logarithme de la cote suit un mod\`ele lin\'eaire :
$$
\log\left(\frac{\mathbb{P}[Y=1\vert \bX=\bx]}{\mathbb{P}[Y\neq 1\vert \bX=\bx]}\right)=\beta_0+\bx\transpose\bbeta,
$$
ou encore :
$$
\esp[Y\vert \bX=\bx]=\mathbb{P}[Y=1\vert \bX=\bx]=\frac{e^{\beta_0+\bx\transpose\bbeta}}{1+e^{\beta_0+\bx\transpose\bbeta}}=H(\beta_0+\bx\transpose\bbeta), \text{\quad o\`u \quad}H(\cdot)=\frac{\exp(\cdot)}{1+\exp(\cdot)},
$$
correspondant \`a la fonction de r\'epartition de la loi logistique.
L'estimation de $(\beta_0,\bbeta)$ se fait par maximisation de la vraisemblance :
$$
\mathcal{L}=\prod_{i=1}^n \left(\frac{e^{\boldsymbol{x}_i^{\text{\sffamily T}}\boldsymbol{\beta}}}{1+e^{\boldsymbol{x}_i^{\text{\sffamily T}}\boldsymbol{\beta}}}\right)^{y_i}
\left(\frac{1}{1+e^{\boldsymbol{x}_i^{\text{\sffamily T}}\boldsymbol{\beta}}}\right)^{1-y_i}
$$
On continuera \`a parler des mod\`eles lin\'eaires car les courbes d'isoprobabilit\'es sont ici les hyperplans parall\`eles $b_0+\bx\transpose\bbeta$. \`A ce mod\`ele, popularis\'e par \citeNP{Berkson1}, certains pr\'ef\`eront le mod\`ele probit (comme le raconte \citeNP{Berkson2}), introduit par \citeNP{Bliss}. Dans ce mod\`ele :
$$
\esp[Y\vert \bX=\bx]=\mathbb{P}[Y=1\vert \bX=\bx]=\Phi(\beta_0+\bx\transpose\bbeta),
$$
o\`u $\Phi$ d\'esigne la fonction de r\'epartition de la loi normale centr\'ee r\'eduite. Ce mod\`ele pr\'esente l'avantage d'avoir un lien direct avec le mod\`ele lin\'eaire Gaussien, puisque 
$$
y_i=\boldsymbol{1}(y_i^\star>0)\text{\quad avec \quad}
y_i^\star=\beta_0+\bx\transpose\bbeta+\varepsilon_i$$
o\`u les r\'esidus sont Gaussiens, de loi $\mathcal{N}(0,\sigma^2)$. Un alternative est d'avoir des r\'esidus centr\'es de variance unitaire, et de consid\'erer une mod\'elisation latente de la forme $y_i=\boldsymbol{1}(y_i^\star>\xi)$ (o\`u $\xi$ sera \`a fixer). On le voit, ces techniques sont fondamentalement li\'ees \`a un mod\`ele stochastique sous-jacent. Mais dans la section \ref{sec:3:classif}, nous pr\'esenterons plusieurs techniques alternatives - tir\'ees de la litt\'erature en apprentissage - pour ce probl\`eme de classification (avec deux classes, ici $0$ et $1$).

\revision{\subsection{R\'egression en grande dimension}\label{Section:reg-grande-dim}}

\revision{Comme nous l'avions mentionn\'e auparavant, la condition du premier ordre $\boldsymbol{X}\transpose(\boldsymbol{X}\widehat{\boldsymbol{\beta}}-\boldsymbol{y})=\boldsymbol{0}$ se r\'esout num\'eriquement en effectuant une d\'ecomposition QR, pour un co\^ut en $O(np^2)$ op\'erations (o\`u $p$ est le rang de $\bX\transpose\bX$). Num\'eriquement, ce calcul peut \^etre long (soit parce que $p$ est grand, soit parce que $n$ est grand), et une strat\'egie plus simple peut \^etre de faire du sous-\'echantillonnage. Soit $n_s\ll n$, et consid\'erons un sous-\'echantillon de taille $n_s$ de $\{1,\cdots,n\}$. Alors  $\widehat{\boldsymbol{\beta}}_s=(\bX_s\transpose\bX_s)^{-1}\bX_s\transpose\boldsymbol{y}_s$ est une bonne approximation de $\widehat{\boldsymbol{\beta}}$ comme le montre \citeNP{Dhilonetal2013}. Cet algorithme est toutefois dangereux si certains points ont un pouvoir de levier important (i.e. $L_i=\bx_i(\bX\transpose\bX)^{-1}\bx_i\transpose$). \citeNP{Tropp2011} propose de transformer les donn\'ees (de mani\`ere lin\'eaire), mais une approche plus populaire est de faire du sous-\'echantillonnage non uniforme, avec une probabilit\'e li\'ee \`a l'influence des observations (d\'efinie par $I_i=\widehat{\varepsilon}_iL_i/(1-L_i)^2$, et qui malheureusement ne peut \^etre calcul\'ee qu'une fois le mod\`ele estim\'e).\\}

\revision{De manière générale, on parlera de données massives lorsque  la table de données de taille $n \times p$ ne tient pas en mémoire RAM de l'ordinateur. Cette situation est souvent rencontr\'ee en apprentissage statistique de nos jours avec très souvent $p \ll n$. C'est la raison pour laquelle, en pratique de nombreuses \revision{biblioth\`eques} d'algorithmes assimilées à de l'apprentissage machine\footnote{comme, par exemple, celles du langage Python.} utilisent des méthodes itératives pour résoudre la condition du premier ordre. Lorsque le modèle paramétrique à calibrer est effectivement convexe et semi-différentiable, il est possible d'utiliser par exemple la méthode de descente de gradient stochastique comme le suggère \citeNP{bottou}. Ce dernier permet de s'affranchir à chaque itération du calcul du gradient sur chaque observation de notre base d'apprentissage. 
Plutôt que d'effectuer une descente moyenne à chaque itération, on commence par tirer (sans remise) une observation $X_i$ parmi les $n$ disponibles. On corrige ensuite les paramètres du modèle de sorte à ce que la prédiction faite à partir de $X_i$ soit la plus proche possible de la vraie valeur $y_i$. On réitère ensuite la méthode jusqu'à avoir parcourue l'ensemble des données. Dans cet algorithme il y a donc autant d'itération que d'observations. Contrairement à l'algorithme de descente de gradient (ou méthode de Newton) à chaque itération un seul vecteur de gradient est calculé (et non plus $n$). Il est néanmoins parfois nécessaire d'exécuter cette algorithme plusieurs fois pour augmenter la convergence des paramètres du modèle.\\}
    
\revision{Si l'objectif est par exemple de minimiser l'erreur quadratique $\ell$ entre l'estimateur $f_{\beta}(X)$ et $y$ l'algorithme peut se résumer ainsi : }

\noindent Etape 0: Mélange des données\\
Etape d'itérations: Pour $t=1,...,n$, on tire $i \in\{ 1,\cdots, n \}$ sans remise, on pose
$$\beta^{t+1} = \beta^{t} - \gamma_t\frac{ \partial{l(y_i,f_{\beta^t}(X_i)) } }{ \partial{ \beta}} $$
%\noindent$\forall k \in [[1,p]]$ (p : dimension de $\beta$)} \\
  
\revision{Cet algorithme peut être réitérée plusieurs fois dans son ensemble selon le besoin de l'utilisateur.
L'avantage de cette méthode est qu'à chaque itération, il n'est pas nécessaire de calculer le gradient sur toutes les observations (plus de somme). Elle est donc adaptée aux bases de données volumineuses. Cet algorithme s'appuie sur une convergence en probabilité vers un voisinage de l'optimum (et non pas l'optimum lui même)}.

\

% En machine learning le probl\`eme des donn\'ees massives, \'evoqu\'e dans la section \ref{Section:reg-grande-dim} dans le contexte de la r\'egression continue \`a se poser. 

%\begin{figure}
% \begin{center}
%\includegraphics[width=.8\textwidth]{regularization_l012}
%\end{center}
%\caption{issage .}\label{Fig:massif}
%\end{figure}

\subsection{Qualit\'e d'un ajustement et choix de mod\`ele}\label{sec:goodness}

Dans le mod\`ele lin\'eaire Gaussien, le coefficient de d\'etermination - not\'e $R^2$ - est souvent utilis\'e comme mesure de la qualit\'e d'ajustement. Compte tenu de la formule de d\'ecomposition de la variance
$$
\underbrace{\frac{1}{n}\sum_{i=1}^n (y_i-\bar{y})^2}_{\text{variance totale}}=\underbrace{\frac{1}{n}\sum_{i=1}^n (y_i-\widehat{y}_i)^2}_{\text{variance r\'esiduelle}}+\underbrace{\frac{1}{n}\sum_{i=1}^n (\widehat{y}_i-\bar{y})^2}_{\text{variance expliqu\'ee}}
$$
on d\'efinit le $R^2$ comme le ratio de variance expliqu\'ee et de la variance totale, autre interp\'etation du coefficient que nous avions introduit \`a partir de la g\'eom\'etrie des moindres carr\'es.
$$
R^2 = \frac{\sum_{i=1}^n (y_i-\bar{y})^2-\sum_{i=1}^n (y_i-\widehat{y}_i)^2}{\sum_{i=1}^n (y_i-\bar{y})^2}
$$
Les sommes des carr\'es d'erreurs dans cette \'ecriture peut se r\'e\'ecrire comme une log-vraisemblance. Or rappelons qu'\`a une constante pr\`es, dans les mod\`eles lin\'eaires g\'en\'eralis\'es, la d\'eviance est d\'efinie par
$$
\text{Deviance}(\bbeta) = -2\log[\mathcal{L}]
$$
que l'on peut aussi noter  $\text{Deviance}(\widehat{\by})$. On peut d\'efinir une d\'eviance nulle comme celle obtenue sans utiliser les variables explicatives $\bx$, de telle sorte que $\widehat{y}_i=\overline{y}$. On peut alors d\'efinir, dans un contexte plus g\'en\'eral
$$
R^2=
\frac{\text{Deviance}(\overline{y})-\text{Deviance}(\widehat{\boldsymbol{y}})}{\text{Deviance}(\overline{y})}=1-
\frac{\text{Deviance}(\widehat{\boldsymbol{y}})}{\text{Deviance}(\overline{y})}.
$$

Toutefois, cette mesure ne peut \^etre utilis\'ee pour choisir un mod\`ele, si on souhaite avoir au final un mod\`ele relativement simple, car elle augmente artificiellement avec l'ajout de variables explicatives sans effet significatif. On aura alors tendance \`a pr\'ef\'erer le $R^2$ ajust\'e
$$
\bar R^2 = {1-(1-R^{2}){n-1 \over n-p}} = R^{2}-\underbrace{(1-R^{2}){p-1 \over n-p}}_{\text{p\'enalisation}},
$$
o\`u $p$ est le nombre de param\`etres du mod\`ele (not\'e plus g\'en\'eralement $\nu$ dans la section \ref{sec:nonlineaire}). \`A la mesure de la qualit\'e de l'ajustement, on va p\'enaliser les mod\`eles trop complexes.

Cette id\'ee va se retrouver dans le crit\`ere d'Akaike, o\`u $
AIC = \text{Deviance} + 2\cdot p
$ ou dans le crit\`ere de Schwarz, $
BIC = \text{Deviance} + \log(n)\cdot p
$. En grande dimension (typiquement $p>\sqrt{n}$), on aura tendance \`a utiliser un AIC corrig\'e, d\'efini par
$$
AICc = \text{Deviance} + 2\cdot p\cdot\frac{n}{n-p-1}
$$

Ces crit\`eres sont utilis\'es dans les m\'ethodes dites « {\em stepwise} », introduisant les m\'ethodes ensemblistes. Dans la m\'ethode dite « {\em forward} », on commence par r\'egresser sur la constante, puis on ajoute une variable \`a la fois, en retenant celle qui fait le plus baisser le crit\`ere AIC, jusqu'\`a ce que rajouter une variable augmente le crit\`ere AIC du mod\`ele. Dans la m\'ethode dite « {\em backward} », on commence par r\'egresser sur toutes les variables, puis on enl\`eve une variable \`a la fois, en retirant celle qui fait le plus baisser le crit\`ere AIC, jusqu'\`a ce que retirer une variable augmente le crit\`ere AIC du mod\`ele. 

%Pour cela, pour $\mathcal{J}\subset\lbrace1,\cdots,p\rbrace$, posons $AIC(\mathcal{J})$ le crit\`ere AIC obtenu comme r\'egression de $Y$ sur les variables $X_j$ pour $j\in\mathcal{J}$. La m\'ethode dite « {\em forward} » consiste \`a it\'erer les calculs suivants
%\begin{enumerate}
%\item trouver $j_1^\star=\displaystyle{\underset{j\in\lbrace\emptyset,1,\cdots,n\rbrace}{\text{argmin}}\left\lbrace AIC(\lbrace j \rbrace) \right\rbrace}$
%\item trouver $j_2^\star=\displaystyle{\underset{j\in\lbrace\emptyset,1,\cdots,n\rbrace\backslash\lbrace j_1^\star\rbrace}{\text{argmin}}\left\lbrace AIC(\lbrace j_1^\star,j \rbrace )\right\rbrace}$
%\item ... jusqu'\`a $j^\star=\emptyset$
%\end{enumerate}
%alors que la m\'ethode dite « {\em backward} » consiste \`a it\'erer les calculs suivants
%\begin{enumerate}
%\item trouver $j_1^\star=\displaystyle{\underset{j\in\lbrace\emptyset,1,\cdots,n\rbrace}{\text{argmin}}\left\lbrace AIC(\lbrace1,\cdots,n\rbrace\backslash\lbrace j \rbrace) \right\rbrace}$
%\item trouver $j_2^\star=\displaystyle{\underset{j\in\lbrace\emptyset,1,\cdots,n\rbrace\backslash\lbrace j_1^\star\rbrace}{\text{argmin}}\left\lbrace AIC(\lbrace1,\cdots,n\rbrace\backslash\lbrace j_1^\star,j \rbrace) \right\rbrace}$
%\item ... jusqu\`a $j^\star=\emptyset$
%\end{enumerate}
%Cette proc\'edure s'av\`ere toutefois tr\`es lente en grande dimension, et souvent inefficace.
%
%PARLER DE MALLOWS $C_p$

Une autre justification de cette notion de p\'enalisation (nous reviendrons sur cette id\'ee en apprentissage) peut \^etre la suivante. Consid\'erons un estimateur dans la classe des pr\'edicteurs lin\'eaires,
$$
\mathcal{M}=\big\lbrace m:~m(\bx)=s_h(\bx)\transpose\by \text{ o\`u }S=(s(\bx_1),\cdots,s(\bx_n)\transpose\text{ est la matrice de lissage}\big\rbrace
$$
et supposons que $\boldsymbol{y}=m_0(\boldsymbol{x})+\boldsymbol{\varepsilon}$, avec $\mathbb{E}[\boldsymbol{\varepsilon}]=\boldsymbol{0}$ et $\text{Var}[\boldsymbol{\varepsilon}]=\sigma^2\mathbb{I}$, de telle sorte que 
$m_0(\boldsymbol{x})=\mathbb{E}[Y|\boldsymbol{X}=\boldsymbol{x}]$.
D'un point de vue th\'eorique, le risque quadratique, associ\'e \`a un mod\`ele estim\'e $\widehat{m}$, s'\'ecrit 
$$
\mathcal{R}(\widehat{m})=\mathbb{E}\big[(Y-\widehat{m}(\boldsymbol{X}))^2\big]
=
\underbrace{\mathbb{E}\big[(Y-m_0(\boldsymbol{X}))^2\big]}_{\text{erreur}}+
\underbrace{\mathbb{E}\big[(m_0(\boldsymbol{X})-\mathbb{E}[\widehat{m}(\bX)])^2\big]}_{\text{biais}}+
\underbrace{\mathbb{E}\big[(\mathbb{E}[\widehat{m}(\bX)]-\widehat{m}(\bX))^2\big]}_{\text{variance}}
$$
si $m_0$ d\'esigne le vrai mod\`ele. Le premier terme est parfois appel\'e « erreur de Bayes », et ne d\'epend pas de l'estimateur retenu, $\widehat{m}$.

Le risque empirique \revision{quadratique}, associ\'e \`a un mod\`ele $m$, est ici :
$$\widehat{\mathcal{R}}_n(m)=\frac{1}{n}\sum_{i=1}^n (y_i-m(\boldsymbol{x}_i))^2 =\frac{1}{n}\|\boldsymbol{y}-m(\boldsymbol{x})\|^2$$
(par convention). On reconna\^it ici l'erreur quadratique moyenne, {\sc mse}, qui donnera plus g\'en\'eralement le « risque » du mod\`ele $m$ quand on utilise une autre fonction de perte (comme nous le discuterons dans la partie suivante). Notons que:
$$\displaystyle{\mathbb{E}[\widehat{\mathcal{R}}_n(m)]=\frac{1}{n}\|m_0(\boldsymbol{x})-m(\boldsymbol{x})\|^2+\frac{1}{n}
\mathbb{E}\big(\|\boldsymbol{y}-m_0(\boldsymbol{x})\|^2\big)}$$
On peut montrer que :
$$
n\mathbb{E}\big[\widehat{\mathcal{R}}_n(\widehat{m})\big]
=\mathbb{E}\big(
\|\boldsymbol{y}-\widehat{m}(\boldsymbol{x})\|^2\big)
=
\|(\mathbb{I}-\boldsymbol{S})m_0\|^2+\sigma^2
\|\mathbb{I}-\boldsymbol{S}\|^2$$
de telle sorte que le (vrai) risque de $\widehat{m}$ est :
$$
{\mathcal{R}}_n(\widehat{m})=
\mathbb{E}\big[\widehat{\mathcal{R}}_n(\widehat{m})\big]+2\frac{\sigma^2}{n}\text{trace}(\boldsymbol{S}).
$$
Aussi, si $\text{trace}(\boldsymbol{S})\geq 0$, le risque empirique sous-estime le vrai risque de l'estimateur. On reconna\^it ici le nombre de degr\'es de libert\'e du mod\`ele, le terme de droite correspondant au $C_p$ de Mallow, introduit dans \citeNP{Mallows} utilisant non pas la d\'eviance mais le $R^2$.

\subsection{\'Econom\'etrie et tests statistiques}\label{sec:tests}

Le test le plus classique en \'econom\'etrie est probablement le test de significativit\'e, correspondant \`a la nullit\'e d'un coefficient dans un mod\`ele de r\'egression lin\'eaire. Formellement, il s'agit du test de $H_0:\beta_k=0$ contre $H_1:\beta_k\neq 0$.
Le test de Student, bas\'e  sur la statistique $t_k=\widehat{\beta}_k/\text{se}_{\widehat{\beta}_k}$, permet a priori de trancher entre les deux alternatives, \`a l'aide de la $p$-value du test, d\'efinie par $\mathbb{P}[|T|>|t_k|]$ avec $T\overset{\mathcal{L}}{\sim} Std_\nu$, o\`u $\nu$ est le nombre de degr\'es de libert\'e du mod\`ele ($\nu=p+1$ pour le mod\`ele lin\'eaire standard).
En grande dimension, cette statistique est n\'eanmoins d'un int\'er\^et tr\`es limit\'e, compte tenu d'un FDR (False Discovery Ratio) important. Classiquement, avec un niveau de significativit\'e $\alpha=0.05$, $5\%$ des variables sont faussement significatives. Supposons que nous disposions de $p=100$ variables explicatives, mais que $5$ (seulement) sont r\'eellement significatives. On peut esp\'erer que ces 5 variables passeront le test de Student, mais on peut aussi s'attendre \`a
ce que 5 variables suppl\'ementaires (test faussement positif) ressortent. On aura alors 10 variables per\c{c}ues comme significatives, alors que seulement la moiti\'e le sont, soit un ratio FDR de 50\%. Afin d'\'eviter cet \'ecueil r\'ecurent dans les tests multiples, il est naturel d'utiliser la proc\'edure de \citeNP{Benjamini}.

%{\color{red} a compl\'eter ??}

%\subsection{Aspects Num\'eriques}\label{sec:mapreduce}
%
%L'estimateur usuel dans un mod\`ele lin\'eaire est $\boldsymbol{\beta}=(\boldsymbol{X}^{\text{\sffamily{T}}}\boldsymbol{X})^{-1} \boldsymbol{X}^{\text{\sffamily{T}}}\boldsymbol{y}$. IL est alors n\'ecessaire de pouvoir calculer, num\'eriquement, $\boldsymbol{X}^{\text{\sffamily{T}}}\boldsymbol{X}$ et $ \boldsymbol{X}^{\text{\sffamily{T}}}\boldsymbol{y}$.

\subsection{Quitter la corr\'elation pour quantifier un effet causal}\label{sec:causal}

\revision{Les mod\`eles \'econom\'etriques sont utilis\'es pour mettre en oeuvre des politiques publiques. Il est alors fondamental de bien comprendre les m\'ecanismes sous-jacents pour savoir quelles variables permettent effectivement d'agir sur une variable d'int\'er\^et. Mais on passe alors dans une autre dimension importante de l'\'econom\'etrie.} C'est \`a Jerry Neyman que l'on doit les premiers travaux sur l'identification de m\'ecanismes causaux, c'est \citeNP{Rubin} qui a formalis\'e le test, appel\'e « mod\`ele causal de Rubin » dans \citeNP{Holland}. Les premi\`eres approches autour de la notion de causalit\'e en \'econom\'etrie se sont faites avec l'utilisation des variables instrumentales, des mod\`eles avec discontinuit\'e de r\'egression, l'analyse de diff\'erences dans les diff\'erences, ainsi que des exp\'eriences naturelles ou pas. La causalit\'e est g\'en\'eralement d\'eduite en comparant l'effet d'une politique - ou plus g\'en\'eralement d'un traitement - avec son contrefactuel, id\'ealement donn\'e par un groupe t\'emoin, al\'eatoire.  L'effet causal du traitement est alors d\'efini comme $\Delta=y_1-y_0$, c'est \`a dire la diff\'erence entre ce que serait la situation avec traitement (not\'e $t=1$) et sans traitement (not\'e $t=0$). Le souci est que seul $y=t\cdot y_1+(1-t)y_0$ et $t$ sont observ\'es. Autrement dit l'effet causal de la variable $t$ sur $y$ n'est pas observ\'e (puisque seule une des deux variables potentielles - $y_0$ ou $y_1$ est observ\'ee pour chaque individu), mais il est aussi individuel, et donc fonction de covariables $\bx$. G\'en\'eralement, en faisant des hypoth\`eses sur la distribution du triplet $(Y_0,Y_1,T)$, certains param\`etres de la distribution de l'effet causal deviennent identifiables, \`a partir de la densit\'e 
des variables observables $(Y, T)$. Classiquement, on sera int\'eress\'e par les moments de cette distribution, en particulier l’effet moyen du traitement dans la population, $\mathbb{E}[\Delta]$, voire juste l'effet moyen du traitement en cas de traitement $\mathbb{E}[\Delta|T=1]$. Si le r\'esultat $(Y_0,Y_1)$ est ind\'ependant de la variable d’acc\`es au traitement $T$, on peut montrer que
$\mathbb{E}[\Delta]=\mathbb{E}[Y|T=1]-\mathbb{E}[Y|T=0]$. Mais si cette hypoth\`ese d'ind\'ependance n'est pas v\'erifi\'ee, on a un biais de s\'election, souvent associ\'e \`a $\mathbb{E}[Y_0|T=1]-\mathbb{E}[Y_0|T=0]$. \citeNP{RosenbaumRubin} proposent d'utiliser un score de propension \`a être trait\'e, $p(\bx)=\mathbb{P}[T=1|\boldsymbol{X}=x]$, en notant que si la variable $Y_0$ est ind\'ependante de l'acc\`es au traitement $T$ conditionnellement aux variables explicatives $\bX$, alors elle est ind\'ependante de $T$ conditionnellement au score $p(\bX)$ : il suffit de les apparier \`a l'aide de leur score de propension.
\citeNP{Heckman2} propose ainsi un estimateur \`a noyau sur le score de propension, ce qui permet d'avoir simplement un estimateur de l'effet du traitement, conditionnellement au fait d’être trait\'e.

\section{Philosophie des m\'ethodes de {\em machine learning}}\label{sec:2:ML}

\revision{Parall\`element \`a ces outils d\'evelopp\'es par et pour des \'economistes, toute une litt\'erature a \'et\'e d\'evelopp\'ee sur des questions similaires, centr\'ees autour de la pr\'evision. Pour \citeNP{Breiman}, une premi\`ere diff\'erence vient du fait que la statistique s'est d\'evelopp\'ee autour du principe d'inf\'erence (ou d'expliciter la relation liant $y$ aux variables $\bx$) alors qu'une autre culture s'int\'eresse avant tout \`a la pr\'ediction. Dans une discussion qui suit l'article, David Cox l'affirme tr\`es clairement « {\em predictive success} (...) {\em is not the primary basis for model choice} ». }% Une diff\'erence profonde est qu'en} apprentissage, on n'a pas besoin de mod\`ele probabiliste. Les probabilit\'es sont un fondement de l'\'econom\'etrie, alors que les probabilit\'es servent plut\^ot d'outil pour interpr\'eter un mod\`ele d'apprentissage machine. Cela n'emp\^eche pas de supposer qu'il y a une composante al\'eatoire dans le mod\`ele, comme dans le « perceptron » de \citeNP{Rosenblatt}, o\`u les cellules sont stimul\'ees avec une certaine probabilit\'e, et r\'epondent ensuite \`a ce stimuli avec une autre probabilit\'e. 
Nous allons pr\'esenter les fondements des techniques du {\em machine learning} (les exemples d'algorithmes \'etant pr\'esent\'es dans les sections suivantes). Le point important, comme nous allons le voir, est que la principale pr\'eoccupation de l'apprentissage machine est li\'ee aux propri\'et\'es de g\'en\'eralisation d'un mod\`ele, c'est-\`a-dire sa performance - selon un crit\`ere choisi a priori - sur des donn\'ees nouvelles, et donc des tests hors \'echantillon.

\revision{\subsection{Apprentissage par une machine}}

\revision{Aujourd'hui, on parle de « {\em machine learning} » pour d\'ecrire tout un ensemble de techniques, souvent computationnelles, alternatives \`a l'approche d\'ecrite auparavant (correspondant \`a l'\'econom\'etrie classique). Avant de les caract\'eriser autant que possible, notons juste qu'historiquement d'autres noms ont pu \^etre donn\'es. Par exemple, \citeNP{Friedman97} propose de faire le lien entre la statistique (qui ressemble beaucoup aux techniques \'econom\'etriques - test d'hypoth\`eses, ANOVA, r\'egression lin\'eaire, logistique, GLM, etc) et ce qu'il appelait alors « {\em data mining} » (qui englobait alors les arbres de d\'ecisions, les m\'ethodes des plus proches voisins, les r\'eseaux de neurones, etc.). Le pont qu'il contribuera \`a construire correspond aux techniques d'apprentissages statistiques, d\'ecrites dans \citeNP{HastieEtal}, mais l'apprentissage machine est un tr\`es vaste champ de recherche.}

\revision{L'apprentissage dit « naturel » (par opposition \`a celui d'une machine) est celui des enfants, qui apprennent \`a parler, \`a lire, \`a jouer. Apprendre \`a parler signifie segmenter et cat\'egoriser des sons, et les associer \`a des significations. Un enfant apprend aussi simultan\'ement la structure de sa langue maternelle et acquiert un ensemble de mots d\'ecrivant le monde qui l'entoure. Plusieurs techniques sont possible, allant d'un apprentissage par coeur, par g\'en\'eralisation, par d\'ecouverte, apprentissage plus ou moins supervis\'e ou autonome, etc. L'id\'ee en intelligence artificielle est de s'inspirer du fonctionnement du cerveau pour apprendre, pour permettre un apprentissage « artificiel » ou « automatique », par une machine. Une premi\`ere application a \'et\'e d'apprendre \`a une machine à jouer \`a un jeux ({\em tic-tac-toe}, \'echecs, go, etc). Une \'etape indispensable est d'expliquer l'objectif qu'il doit atteindre pour gagner. Une approche historique a \'et\'e de lui apprendre les r\`egles du jeu. Si cela permet de jouer, cela ne permettra pas \`a la machine de {\em bien } jouer. En supposant que la machine connaisse les r\`egles du jeu, et qu'elle a le choix entre plusieurs dizaines de coups possible, lequel doit-elle choisir ? L'approche classique en intelligence artificielle utilise l'algorithme dit {\em min-max} utilisant une fonction d'\'evaluation : dans cet algorithme, la machine effectue une recherche en avant dans l'arbre des coups possibles, aussi loin que les ressources de
calcul le lui permettent (une dizaine de coups aux échecs, par exemple). Ensuite, elle calcule diff\'erents crit\`eres (qui lui ont \'et\'e indiqu\'e au pr\'ealable) pour toutes les positions (nombre de pi\`eces prises, ou perdues, occupation du centre, etc. dans notre exemple du jeu d'\'echec), et finalement, la machine joue le coup qui lui permet de maximiser son gain. Un autre exemple peut \^etre celui de la classification et de la reconnaissance d'images ou de formes. Par exemple, la machine doit identifier un chiffre dans une \'ecriture manuscrite (ch\`eque, code postal). Il s'agit de pr\'edire la valeur d'une variable $y$, en sachant qu'a priori $y\in\{0,1,2,\cdots,8,9\}$. Un strat\'egie classique est de fournir \`a la machine des bases d'apprentissage, autrement dit ici des millions d'images lab\'elis\'ees (identifi\'ees) de chiffres manuscrits. Une strat\'egie simple et naturelle est d'utiliser un critère de d\'ecision bas\'e sur les plus proches voisins dont on conna\^it l'\'etiquette (\`a l'aide d'une m\'etrique pr\'ed\'efinie).}

\revision{La m\'ethode des plus proches voisins (« $k$-{\em nearest neighbors} ) peut \^etre d\'ecrit de la mani\`ere suivante : on consid\`ere (comme dans la partie pr\'ec\'edante) un ensemble de $n$ observations, c'est \`a dire des paires $(y_i,\boldsymbol{x}_i)$ avec $\boldsymbol{x}_i\in\mathbb{R}^p$. Consid\'erons une distance $\Delta$ sur $\mathbb{R}^p$ (la distance Euclienne ou la distance de Mahalanobis, par exemple). \'Etant donn\'ee une nouvelle observation $\boldsymbol{x}\in\mathbb{R}^p$, supposons les observations ordonn\'ees en fonction de la distance entre les $\boldsymbol{x}_i$ et $\boldsymbol{x}$, au sens o\`u
$$
\Delta(\boldsymbol{x}_1,\boldsymbol{x})\leq
\Delta(\boldsymbol{x}_2,\boldsymbol{x})\leq
\cdots\leq 
\Delta(\boldsymbol{x}_n,\boldsymbol{x})
$$
alors on peut consid\'erer comme pr\'ediction pour $y$ la moyenne des $k$ plus proches voisins,
$$
m_k(\boldsymbol{x})=\frac{1}{k}\sum_{i=1}^k y_i.
$$
L'apprentissage fonctionne ici par induction, \`a partir d'un \'echantillon (appel\'e base d'apprentissage).}

\revision{Le Machine Learning englobe ces algorithmes qui donnent aux ordinateurs la capacit\'e d'apprendre sans \^etre explicitement programm\'e (comme l'avait d\'efini Arthur Samuel en 1959). La machine va alors explorer les donn\'ees avec un objectif pr\'ecis (comme chercher les plus proches voisins dans l'exemple que nous venons de d\'ecrire). Tom Mitchell a proposé une définition plus précise en 1998 : on dit qu'un programme d'ordinateur apprend de l'exp\'erience $E$ par rapport \`a une t\^ache $T$ et une mesure de performance $P$, si sa performance sur $T$, mesur\'ee par $P$, s'am\'eliore avec l'exp\'erience $E$. La t\^ache $T$ peut \^etre un score de d\'efaut par exemple, et la performance $P$ peut \^etre le pourcentage d'erreurs commise. Le syst\`eme apprend si le pourcentage de d\'efauts pr\'edit augmente avec l'exp\'erience}.

\revision{On le voit, l'apprentissage machine est fondamentalement un probl\`eme d'optimisation d'un crit\`ere \`a partir de donn\'ees (dites d'apprentissage). Nombreux sont les ouvrages de programmation qui proposent des algorithmes, sans jamais faire mention d'un quelconque mod\`ele probabiliste. Dans \citeNP{Watt} par exemple, il n'est fait mention du mot « probabilité » qu'une seule fois, avec cette note de bas de page qui surprendra et fera sourire les \'econom\`etres, « {\em logistic regression can also be interpreted from a probabilistic perspective} » (page 86). Mais beaucoup d'ouvrages r\'ecents proposent une relecture des approches d'apprentissage machine \`a l'aide de th\'eories probabilistes, suite aux travaux de Vaillant et Vapnik. En proposant le paradigme de l'apprentissage « {\em probablement à peu près correct} » (PAC), une saveur probabiliste a \'et\'e rajout\'e \`a l'approche jusqu'alors tr\`es computationnelle, en quantifiant l'erreur de l'algorithme d'apprentissage (dans un probl\`eme de classification).}

\revision{\subsection{Le tournant des ann\'ees 80/90 et le formalisme probabiliste}}

\revision{On dispose d'un \'echantillon d'apprentissage, avec des observations $(\boldsymbol{x}_i,y_i)$ o\`u les variables $y$ sont dans un ensemble $\mathcal{Y}$. Dans le cas de la classification, $\mathcal{Y}=\{-1,+1\}$, mais on peut imaginer un ensemble relativement g\'en\'eral. Un pr\'edicteur est une fonction $m$ \`a valeurs dans $\mathcal{Y}$, permettant d'\'etiqueter (ou de classer) les nouvelles observations \`a venir. On suppose que les \'etiquettes sont produites par un classifieur $f$ appel\'e cible. Pour un statisticien, cette fonction serait le vrai mod\`ele. Naturellement, on veut construire $m$ le plus proche possible de $f$. Soit $\mathbb{P}$ une distribution (inconnue) sur $\mathcal{X}$. L'erreur de $m$ relativement à la cible $f$ est définie par
$$
\mathcal{R}_{\mathbb{P},f}(m)=\mathbb{P}[m(\boldsymbol{X})\neq f(\boldsymbol{X})]\text{ o\`u }\boldsymbol{X}\sim\mathbb{P},
$$
ou \'ecrit de mani\`ere \'equivalente,
$$
\mathcal{R}_{\mathbb{P},f}(m)=\mathbb{P}\big[\{\boldsymbol{x}\in\mathcal{X}:m(\boldsymbol{x})\neq f(\boldsymbol{x})\}\big].
$$
Pour trouver notre classifieur, il devient n\'ecessaire de supposer qu'il existe un lien entre les donn\'ees de notre \'echantillon et le couple $(\mathbb{P},f)$, c'est \`a dire un mod\`ele de g\'en\'eration des donn\'ees. On va alors supposer que les $\boldsymbol{x}_i$ sont obtenus par des tirages ind\'ependants suivant $\mathbb{P}$, et qu'ensuite $y_i=f(\boldsymbol{x}_i)$ .}

\revision{On peut ici d\'efinir le risque empirique d'un mod\`ele $m$,
$$
\widehat{\mathcal{R}}(m)=\frac{1}{n}\sum_{i=1}^n \boldsymbol{1}(m(\boldsymbol{x}_i)\neq y_i).
$$}

\revision{Il est alors important d'admettre qu'on ne peut pas trouver un mod\`ele parfait, au sens o\`u $\mathcal{R}_{\mathbb{P},f}(m)=0$. En effet, si on consid\`ere le cas le plus simple qui soit, avec $\mathcal{X}=\{\bx_1,\bx_2\}$ et que $\mathbb{P}$ soit telle que $\mathbb{P}(\{\bx_1\})=p$ et $\mathbb{P}(\{\bx_2\})=1-p$. Il est aussi possible d'observer $x_1$ et $x_2$, et malgr\'e tout, de se tromper sur les \'etiquettes. Aussi, au lieu de chercher un mod\`ele parfait, on peut tenter d'avoir un mod\`ele approximativement correct. On va alors chercher \`a trouver $m$ tel que $\mathcal{R}_{\mathbb{P},f}(m)\leq \epsilon$, o\`u $\epsilon$ est un seuil sp\'ecifi\'e a priori.}

\revision{Une fois admis ce premier \'ecueil, qui fait penser \`a l'errreur de mod\`ele, notons aussi un second probl\`eme. Sur notre exemple \`a deux valeurs, la probabilit\'e de ne jamais observer $x_2$ parmi $n$ tirages suivant $\mathbb{P}$ est $p^n$. Il sera alors impossible de trouver $m(x_2)$ car cette valeur n'aura jamais \'et\'e observ\'ee. Autrement dit, aucun algorithme ne peut nous assurer d'avoir avec certitude, avec $n$ observations, $\mathcal{R}_{\mathbb{P},f}(m)\leq \epsilon$. On va alors chercher \`a \^etre probablement approximativement correct (PAC). Pour se faire, on autorise l'algorithme \`a se tromper avec une probabilit\'e $\delta$, l\`a aussi fix\'ee a priori.}

\revision{Aussi, quand on construit un classifieur, on ne conna\^it ni $\mathbb{P}$, ni $f$, mais on se donne un crit\`ere de pr\'ecision $\epsilon$, et un param\`etre de confiance $\delta$, et on dispose de $n$ observations. Notons que $n$, $\epsilon$ et $\delta$ peuvent \^etre li\'es. On cherche alors un mod\`ele $m$ tel que $\mathcal{R}_{\mathbb{P},f}(m)\leq\epsilon$ avec probabilit\'e (au moins) $1-\delta$, de mani\`ere \`a \^etre probablement approximativement correct.}

\revision{\citeNP{Wolpert96} a montr\'e (d\'etaill\'e dans \citeNP{Wolpert97}) qu'il n'existe pas d'algorithme d'apprentissage universel. En particulier, on peut montrer qu'il existe $\mathbb{P}$ telle que $\mathcal{R}_{\mathbb{P},f}(m)$ soit relativement grande, avec une probabilit\'e (elle aussi) relativement grande.}

\revision{L'interpr\'etation qui en est faite est qu'il est n\'ecessaire d'avoir un biais pour apprendre. Comme on ne peut pas apprendre (au sens PAC) sur l'ensemble des fonctions $m$, on va alors contraindre $m$ \`a appartenir une classe particuli\`ere, not\'ee $\mathcal{M}$. Supposons pour commencer que $\mathcal{M}$ contienne un nombre fini de mod\`eles possibles. On peut alors montrer que pour tout $\epsilon$ et $\delta$, que pour tout $\mathbb{P}$ et $f$, si on dispose d'assez d'observations (plus pr\'ecis\'ement $n\geq \epsilon^{-1}\log[\delta^{-1}|\mathcal{M}|]$, alors avec une probabilit\'e plus grande que $1-\delta$, $\mathcal{R}_{\mathbb{P},f}(m^\star)\leq \epsilon$ o\`u
$$
m^\star \in \underset{m\in\mathcal{M}}{\text{argmin}}\Big\lbrace
\frac{1}{n}\sum_{i=1}^n \boldsymbol{1}(m(\boldsymbol{x}_i)\neq y_i)\Big\rbrace
$$
autrement dit $m^\star$ est un mod\`ele dans $\mathcal{M}$ qui minimise le risque empirique.}

\revision{Un peut aller un peu plus loin, en restant dans le cas o\`u $\mathcal{Y}=\{-1,+1\}$. Une classe $\mathcal{M}$ de classifieurs sera dite PAC-apprenable s'il existe $n_{\mathcal{M}}:[0,1]^2\rightarrow \mathbb{N}$ tel que, pour tout $\epsilon$, $\delta$, $\mathbb{P}$ et si on suppose que la cible $f$ appartient \`a $\mathcal{M}$, alors en utilisant $n>n_{\mathcal{M}}(\epsilon,\delta)$ tirages d'observations $\boldsymbol{x}_i$ suivant $\mathbb{P}$, \'etiquet\'es $y_i$ par $f$, alors il existe $m\in\mathcal{M}$ tel que, avec probabilit\'e $1-\delta$, $\mathcal{R}_{\mathbb{P},f}(m)\leq \epsilon$. La fonction $n_{\mathcal{M}}$ est alors appel\'ee complexit\'e d'\'echantillon pour apprendre. En particulier, nous avons vu que si $\mathcal{M}$ contient un nombre fini de classifieurs, alors $\mathcal{M}$ est PAC-apprenable avec la complexit\'e $n_{\mathcal{M}}(\epsilon,\delta)=\epsilon^{-1}\log[\delta^{-1}|\mathcal{M}|]$.}

\revision{Naturellement, on souhaiterait avoir un r\'esultat plus g\'en\'eral, en particulier si $\mathcal{M}$ n'est pas fini. Pour cela, il faut utiliser la dimension VC de Vapnik-Chervonenkis, qui repose sur l'id\'ee de pulv\'erisation de nuages de points (pour une classification binaire). Consid\'erons $k$ points $\{\boldsymbol{x}_1,\cdots\boldsymbol{x}_k\}$, et consid\'erons l'ensemble 
$$
\mathcal{E}_k=\big\lbrace(m(\boldsymbol{x}_1),\cdots,m(\boldsymbol{x}_k))\text{ pour }m\in\mathcal{M})\big\rbrace.
$$
Notons que les \'el\'ements de $\mathcal{E}_k$ appartiennent \`a $\{-1,+1\}^k$. Autrement dit $|\mathcal{E}_k|\leq 2^k$. On dira que $\mathcal{M}$ pulv\'erise l'ensemble des points si toutes les combinaisons sont possibles, c'est \`a dire $|\mathcal{E}_k|= 2^k$. Intuitivement, les \'etiquettes de l'ensemble de points ne procurent pas assez d'information sur la cible $f$, car tout est possible. La dimension VC de $\mathcal{M}$ est alors 
$$
VC(\mathcal{M})=\sup\big\lbrace
k\text{ tel que }\mathcal{M}\text{ pulv\'erise }\{\boldsymbol{x}_1,\cdots\boldsymbol{x}_k\}
\big\rbrace.
$$
}

\revision{Par exemple si $\mathcal{X}=\mathbb{R}$ et que l'on consid\`ere l'ensemble des mod\`eles (simples) de la forme $m_{a,b}=\boldsymbol{1}_{\pm}(x\in[a,b])$. Aucun ensemble de points $\{x_1,x_2,x_3\}$ ordonn\'es ne peut \^etre pulv\'eris\'e car il suffit d'assigner respectivement $+1$, $-1$ et $+1$ \`a $x_1$, $x_2$ et $x_3$ respectivement, donc $VC<3$. En revanche $\{0,1\}$ est pulv\'eris\'e donc $VC\geq 2$. La dimension de cet ensemble de pr\'edicteur est $2$. 
Si on augmente d'une dimension, $\mathcal{X}=\mathbb{R}^2$ et que l'on consid\`ere l'ensemble des mod\`eles (simples) de la forme $m_{\boldsymbol{a},\boldsymbol{b}}=\boldsymbol{1}_{{\tiny \pm}}(x\in[\boldsymbol{a},\boldsymbol{b}])$ (o\`u $[\boldsymbol{a},\boldsymbol{b}]$ d\'esigne le rectangle), alors la dimension de $\mathcal{M}$ est ici $4$.}

\revision{Pour introduire les SVM, pla\c{c}ons nous dans le cas o\`u $\mathcal{X}=\mathbb{R}^k$, et consid\'erons des s\'eparations par des hyperplans passant par l'origine (on dira homog\`enes), au sens o\`u $m_{\boldsymbol{w}}(\boldsymbol{x})=\boldsymbol{1}_{\pm}(\boldsymbol{w}^{\text{\sffamily T}}\boldsymbol{x}\geq 0)$. On peut montrer qu'aucun ensemble de $k+1$ points ne peut \^etre pulv\'eris\'e par ces deux espaces homog\`enes dans $\mathbb{R}^k$, et donc $VC(\mathcal{M})=k$. Si on rajoute une constante, au sens o\`u $m_{\boldsymbol{w},b}(\boldsymbol{x})=\boldsymbol{1}_{\pm}(\boldsymbol{w}^{\text{\sffamily T}}\boldsymbol{x}+b\geq 0)$, on peut montrer qu'aucun ensemble de $k+2$ points ne peut \^etre pulv\'eris\'e par ces deux espaces (non homog\`enes) dans $\mathbb{R}^k$, et donc $VC(\mathcal{M})=k+1$.}

\revision{De cette dimension VC, on d\'eduit le th\'eor\`eme dit fondamental de l'apprentisssage : si $\mathcal{M}$ est une classe de dimension $d=VC(\mathcal{M})$, alors il existe des constante positives $\underline{C}$ et $\overline{C}$ telles que la complexit\'e d'\'echantillon pour que $\mathcal{M}$ soit PAC-apprenable v\'erifie
$$
\underline{C}\epsilon^{-1}\big(d+\log[\delta^{-1}]\big)
\leq n_{\mathcal{M}}(\epsilon,\delta) \leq 
\overline{C}\epsilon^{-1}\big(d\log[\epsilon^{-1}]+\log[\delta^{-1}]\big).
$$
}

\revision{\subsection{Le choix de l'objectif et la fonction de perte}}\label{sec:fonctions:pertes}%\label{sec:proba:ML}

\revision{Ces choix sont essentiels, et dépendent du problème considéré.}
Commen\c{c}ons par d\'ecrire un mod\`ele historiquement important, le « perceptron » de \citeNP{Rosenblatt}, introduit dans des probl\`emes de classification, o\`u $y\in\{-1,+1\}$, inspir\'e par \citeNP{McCulloch}. On dispose de donn\'ees $\lbrace(y_i,\bx_i)\rbrace$, et on va construire de mani\`ere it\'erative un ensemble de mod\`eles $m_k(\cdot)$, o\`u \`a chaque \'etape, on va apprendre des erreurs du mod\`ele pr\'ec\'edent. Dans le perceptron, on consid\`ere un mod\`ele lin\'eaire de telle sorte que :
$$
m(\bx)=\boldsymbol{1}_{\pm}(\beta_0+\bx\transpose \boldsymbol{\beta}\geq 0)=\left\lbrace
\begin{array}{l}
+1\text{ si }\beta_0+\bx\transpose \boldsymbol{\beta}\geq 0\\
-1\text{ si }\beta_0+\bx\transpose \boldsymbol{\beta}< 0
\end{array}
\right.,
$$
o\`u les coefficients $\bbeta$ sont souvent interpr\'et\'es comme des « poids » attribu\'es \`a chacune des variables explicatives. On se donne des poids initiaux $(\beta_0^{(0)},\boldsymbol{\beta}^{(0)})$, que l'on va mettre \`a jour en tenant compte de l'erreur de pr\'ediction commise, entre $y_i$ et la pr\'ediction $\widehat{y}_i^{(k)}$ :
$$
\widehat{y}_i^{(k)}=m^{(k)}(\bx_i)=\boldsymbol{1}_{\pm}(\beta_0^{(k)}+\bx\transpose \boldsymbol{\beta}^{(k)}\geq 0),
$$ 
avec, dans le cas du perceptron :
$$
\beta_j^{(k+1)}={\beta}_j^{(k)}+\eta\underbrace{(\boldsymbol{y}-\widehat{\boldsymbol{y}}^{(k)})^{\text{\sffamily T}}}_{=\ell({\boldsymbol{y}},\widehat{\boldsymbol{y}}^{(k)})}\bx_j
$$
o\`u ici $\ell(y,y')=\boldsymbol{1}(y \neq y')$ est une fonction de perte, qui permettra de donner un prix \`a une erreur commise, en pr\'edisant $y'=m(\bx)$ et en observant $y$. Pour un probl\`eme de r\'egression, on peut consid\'erer une erreur quadratique $\ell_2$, telle que $\ell(y,m(\bx))=(y - m(\bx))^2$ ou en valeur absolue $\ell_1$, avec $\ell(y,m(\bx))=\vert y - m(\bx)\vert $. Ici, pour notre probl\`eme de classification, nous utilisions une indicatrice de mauvaise qualification (on pourrait discuter le caract\`ere sym\'etrique de cette fonction de perte, laissant croire qu'un faux positif co\^ute autant qu'un faux n\'egatif). Une fois sp\'ecifi\'ee cette fonction de perte, on reconna\^it dans le probl\`eme d\'ecrit aurapavant une descente de gradient, et on voit que l'on cherche \`a r\'esoudre :
\begin{equation}\label{eq:ML-eq}
m^\star(\bx)=\underset{m\in\mathcal{M}}{\text{argmin}}\left\lbrace
\sum_{i=1}^n \ell(y_i,m(\bx_i))
\right\rbrace
\end{equation}
pour un ensemble de pr\'edicteurs $\mathcal{M}$ pr\'ed\'efini. 
Tout probl\`eme d'apprentissage machine est math\'ematiquement formul\'e comme un probl\`eme d'optimisation, dont la solution d\'etermine un ensemble de param\`etres de mod\`ele (si la famille $\mathcal{M}$ est d\'ecrite par un ensemble de param\`etres - qui peuvent \^etre des coordonn\'ees dans une base fonctionnelle). On pourra noter $\mathcal{M}_0$ l'espace des hyperplans de $\mathbb{R}^p$ au sens o\`u
$$
m\in\mathcal{M}_0 \text{\quad signifie \quad}m(\bx)=\beta_0+\bbeta\transpose\bx\text{\quad avec \quad }\bbeta\in\mathbb{R}^p,
$$
engendrant la classe des pr\'edicteurs lin\'eaires. \revision{On aura alors l'estimateur qui minimise le risque empirique. Une partie des travaux r\'ecents en apprentissage statistique vise \`a \'etudier les propri\'et\'es de l'estimateur $\widehat{m}^{\star}$, dit « oracle », dans une famille d'estimateurs $\mathcal{M}$,
$$
\widehat{m}^{\star} = \underset{\widehat{m}\in\mathcal{M}}{\text{argmin}}\big\lbrace
\mathcal{R}(\widehat{m},m)
\big\rbrace.
$$
Cet estimateur est, bien entendu, impossible \`a d\'efinir car il d\'epend de $m$, le vrai mod\`ele, inconnu.}

\revision{Mais revenons un peu davantage sur ces fonctions de perte. Une fonction de perte $\ell$ est une fonction $\mathbb{R}^d\times\mathbb{R}^d\rightarrow\mathbb{R}_+$, symm\'etrique, qui v\'erifie l'in\'egalit\'e triangulaire, et telle que $\ell(\bx,\by)=0$ si et seulement si $\bx=\by$. La norme associ\'ee est $\|\cdot\|$, telle que $\ell(\bx,\by)=\|\bx-\by\|=\ell(\bx-\by,\boldsymbol{0})$ (en utilisant le fait que $\ell(\bx,\by+\boldsymbol{z})=\ell(\bx-\by,\boldsymbol{z})$ - nous reverrons cette propri\'et\'e fondamentale par la suite).}

Pour une fonction de perte quadratique, on notera que l'on peut avoir une interpr\'etation particuli\`ere de ce probl\`eme, puisque :
$$
\displaystyle{\overline{y} = \underset{m\in\mathbb{R}}{\text{argmin}} \left\lbrace  \sum_{i=1}^n \frac{1}{n} [y_i-m]^2 \right\rbrace = \underset{m\in\mathbb{R}}{\text{argmin}} \left\lbrace  \sum_{i=1}^n  \ell_2(y_i,m) \right\rbrace }, 
$$
o\`u $\ell_2$ est la distance quadratique usuelle
Si l'on suppose - comme on le faisait en \'econom\'etrie - qu'il existe un mod\`ele probabiliste sous-jacent, et en notant que :
$$
\displaystyle{\mathbb{E}(Y) = \underset{m\in\mathbb{R}}{\text{argmin}} \left\lbrace \Vert Y-m\Vert^2_{\ell_2}\right\rbrace = 
\underset{m\in\mathbb{R}}{\text{argmin}} \left\lbrace \mathbb{E}\left( [Y-m]^2 \right)  \right\rbrace=
\underset{m\in\mathbb{R}}{\text{argmin}} \left\lbrace  \mathbb{E}\big[\ell_2(Y,m)\big] \right\rbrace } 
$$
on notera que ce que l'on essaye d'obtenir ici, en r\'esolvant le probl\`eme (\ref{eq:ML-eq}) en prenant pour $\ell$ la norme $\ell_2$, est une approximation (dans un espace fonctionnel donn\'e, $\mathcal{M}$) de l'esp\'erance conditionnelle $\bx\mapsto \esp[Y\vert \bX=\bx]$.
Une autre fonction de perte particuli\`erement int\'eressante est la perte $\ell_1$, $\ell_1(y,m)=\vert y-m\vert $. Rappelons que
$$
\displaystyle{\text{m\'ediane}(\boldsymbol{y}) =  \underset{m\in\mathbb{R}}{\text{argmin}} \left\lbrace  \sum_{i=1}^n  \ell_1(y_i,m) \right\rbrace }. 
$$
Le probl\`eme d'optimisation :
$$
\widehat{m}^{\star}=\underset{m\in\mathcal{M}_0}{\text{argmin}}\left\lbrace
\sum_{i=1}^n \vert  y_i-m(\bx_i)\vert 
\right\rbrace 
$$
est obtenu en \'econom\'etrie si on suppose que la loi conditionnelle de $Y$ suit une loi de Laplace centr\'ee sur $m(\bx)$, et en maximisant la (log) vraisemblance \revision{(la somme des valeurs absolues de erreurs correspond \`a la log-vraimenblance d'une loi de Laplace). On pourra noter d'ailleurs que si la loi conditionnelle de $Y$ est summ\'etrique par rapport à 0, la m\'ediane et la moyenne con\"incident}
Si on r\'e\'ecrit cette fonction de perte $\ell_1(y,m)=\vert (y-m)(1/2-\boldsymbol{1}_{y\leq m})\vert $, on peut obtenir une g\'en\'eralisation pour $\tau\in(0,1)$ :
$$
\widehat{m}^\star_\tau=\underset{m\in\mathcal{M}_0}{\text{argmin}}\left\lbrace
\sum_{i=1}^n \ell_\tau^{\text{\sffamily q}} (y_i,m(\bx_i)) 
\right\rbrace\text{\quad avec \quad}\ell_{\tau}^{\text{\sffamily q}}(x,y)= (x-y)(\tau-\boldsymbol{1}_{x\leq y})
$$
est alors la r\'egression quantile de niveau $\tau$ (voir \citeNP{Koenker} et \citeNP{dHaultefoeuille}). Une autre fonction de perte, introduite par \citeNP{Aigneretal} et analys\'ee dans \citeNP{Walltrup}, est la fonction associ\'ee \`a la notion d'expectiles :
$$
\displaystyle{\ell}^{\text{\sffamily e}}_{\tau}(x,y)= (x-y)^2\cdot\big\vert\tau-\boldsymbol{1}_{x\leq y}\big\vert
$$
avec $\tau\in[0,1]$. On voit le parall\`ele avec la fonction quantile :
$$
\displaystyle{\ell}^{\text{\sffamily q}}_{\tau}(x,y)= \vert x-y\vert \cdot\big\vert\tau-\boldsymbol{1}_{x\leq y}\big\vert.
$$
\revision{\citeNP{KoenkerMachado} et \citeNP{YuMoyeed} ont d'ailleurs not\'e un lien entre cette condition et la recherche du maximum de vraisemblance lorsque la loi conditionnelle de $Y$ suit une loi de Laplace assymétrique.}

En lien avec cette approche, \citeNP{Gneiting} a introduit la notion de « {\em statistique ellicitable} » - ou de « {\em mesure  ellicitable} » dans sa version probabiliste (ou distributionnelle) : $T$ sera dite « {\em ellicitable} » s'il existe une fonction de perte $\ell: \mathbb{R}\times \mathbb{R}\rightarrow \mathbb{R}_+$ telle que :
$$
T(Y)=\underset{x\in\mathbb{R}}{\text{argmin}}\left\lbrace
\int_{\mathbb{R}} \ell(x,y)dF(y)\right\rbrace
=\underset{x\in\mathbb{R}}{\text{argmin}}\left\lbrace
\mathbb{E}\big[ \ell(x,Y)\big]\text{ o\`u }Y\overset{\mathcal{L}}{\sim} F
\right\rbrace.
$$
La moyenne (esp\'erance math\'ematique) est ainsi ellicitable par la distance quadratique, $\ell_2$, alors que la m\'ediane est ellicitable par la distance $\ell_1$. Selon \citeNP{Gneiting}, cette propri\'et\'e est essentielle pour construire des pr\'edictions. Il peut alors exister un lien fort entre des mesures associ\'ees \`a des mod\`eles probabilistes et les fonctions de perte. Enfin, la statistique Bay\'esienne propose un lien direct entre la forme de la loi a priori et la fonction de perte, comme l'ont \'etudi\'e \citeNP{Berger} et \citeNP{Bernardo}. \revision{Nous reviendrons sur l'utilisation de ces diff\'erentes normes dans la section sur la p\'enalisation.}

\

\subsection{Boosting et apprentissage s\'equentiel}\label{sec:boosting}

\revision{Nous l'avons vu auparavant: la modélisation repose ici sur la résolution d'un problème d'optimisation, et} r\'esoudre le probl\`eme d\'ecrit par l'\'equation (\ref{eq:ML-eq}) est d'autant plus complexe que l'espace fonctionnel $\mathcal{M}$ est volumineux. L'id\'ee du Boosting,  tel qu'introduit par \citeNP{ShapireFreund}, est d'apprendre, lentement, \`a partir des erreurs du mod\`ele, de mani\`ere it\'erative. \`A la premi\`ere \'etape, on estime un mod\`ele $m_1$ pour $\by$, \`a partir de $\bX$, qui donnera une erreur $\bepsilon_1$. \`A la seconde \'etape, on estime un mod\`ele $m_2$ pour $\bepsilon_1$, \`a partir de $\bX$, qui donnera une erreur $\bepsilon_2$, etc. On va alors retenir comme mod\`ele, au bout de $k$ it\'eration :
\begin{equation}\label{eq:boosting}
m^{(k)}(\cdot)=\underbrace{m_1(\cdot)}_{\sim \by}+\underbrace{m_2(\cdot)}_{\sim \bepsilon_1}+\underbrace{m_3(\cdot)}_{\sim \bepsilon_2}+\cdots+
\underbrace{m_k(\cdot)}_{\sim \bepsilon_{k-1}}=m^{(k-1)}(\cdot)+m_k(\cdot).
\end{equation}
Ici, l'erreur $\varepsilon$ est vue comme la diff\'erence entre $y$ et le mod\`ele $m(\bx)$, mais elle peut aussi \^etre vue comme le gradient associ\'e \`a la fonction de perte quadratique. Formellement, $\bepsilon$ peut \^etre vu comme un $\nabla\ell$ dans un contexte plus g\'eneral (on retrouve ici une interpr\'etation qui fait penser aux r\'esidus dans les mod\`eles lin\'eaires g\'en\'eralis\'es).

L'\'equation (\ref{eq:boosting}) peut se voir comme une descente du gradient, mais \'ecrit de mani\`ere duale. En effet, la descente de gradient permettant d'obtenir le minimum d'une fonction $f$ repose sur une \'equation de la forme
$$
\underbrace{f(\boldsymbol{x}_k)}_{\langle f,\boldsymbol{x}_k \rangle} \sim \underbrace{f(\boldsymbol{x}_{k-1})}_{\langle f,\boldsymbol{x}_{k-1} \rangle}  + \underbrace{(\boldsymbol{x}_k-\boldsymbol{x}_{k-1}) }_{\alpha_k}\underbrace{\nabla{f}(\boldsymbol{x}_{k-1})}_{\langle \nabla f,\boldsymbol{x}_{k-1} \rangle}
$$
Le probl\`eme (\ref{eq:ML-eq}) est dual dans le sens o\`u c'est la fonction $f$ qui doit \^etre optimis\'ee. On pourrait alors \'ecrire une descente de gradient de la forme :
$$
\underbrace{f_k(\boldsymbol{x})}_{\langle f_k,\boldsymbol{x} \rangle} \sim \underbrace{f_{k-1}(\boldsymbol{x})}_{\langle f_{k-1},\boldsymbol{x} \rangle}  + \underbrace{(f_k-f_{k-1})}_{\beta_k} \underbrace{\star}_{\langle  f_{k-1},\nabla\boldsymbol{x} \rangle}
$$
o\`u le terme $\star$ peut \^etre interpr\'et\'e comme un gradient, mais dans un espace fonctionnel, et non plus dans $\mathbb{R}^{p}$. Le probl\`eme (\ref{eq:boosting}) va alors se r\'e\'ecrire comme un probl\`eme d'optimisation :
\begin{equation}\label{eqboot}
m^{(k)}=m^{(k-1)}+\underset{h\in\mathcal{H}}{\text{argmin}}\left\lbrace \sum_{i=1}^n \ell(\underbrace{y_i-m^{(k-1)}(\boldsymbol{x}_i)}_{\varepsilon_{k,i}},h(\boldsymbol{x}_i))\right\rbrace
\end{equation}
o\`u l'astuce consiste \`a consid\'erer un espace $\mathcal{H}$ relativement simple (on parlera de « {\em weak learner} »~). Classiquement, les fonctions $\mathcal{H}$ sont des fonctions en escalier (que l'on retrouvera dans les arbres de classification et de r\'egression) appel\'es {\em stumps}. Afin de s'assurer que l'apprentissage est effectivement lent, il n'est pas rare d'utiliser un param\`etre de « shrinkage », et au lieu de poser, par exemple, $\varepsilon_1=y-m_1(\bx)$, on posera $\varepsilon_1=y-\alpha \cdot m_1(\bx)$ avec $\alpha\in[0,1]$. On notera que c'est parce qu'on utilise pour $\mathcal{H}$ un espace non-lin\'eaire, et que l'apprentissage est lent, que cet algorithme fonctionne bien. Dans le cas du mod\`ele lin\'eaire Gaussien, rappelons en effet que les r\'esidus $\widehat{\bepsilon}=\by-\bX\widehat{\bbeta}$ sont orthogonaux aux variables explicatives, $\bX$, et il est alors impossible d'apprendre de nos erreurs. La principale difficult\'e est de s'arr\^eter \`a temps, car apr\`es trop d'it\'erations, ce n'est plus la fonction $m$ que l'on approxime, mais le bruit. Ce probl\`eme est appel\'e sur-apprentissage.

\revision{Cette pr\'esentation a l'avantage d'avoir une heuristique faisant penser \`a un mod\`ele \'econom\'etrique, en mod\'elisant de mani\`ere int\'erative les r\'esidus par un mod\`ele (tr\`es) simple. Mais ce n'est souvent pas la pr\'esentation retenue dans la litt\'erature en apprentissage, qui insiste davantage sur une heuristique d'algorithme d'optimisation (et d'approximation du gradient). La fonction est apprise de mani\`ere it\'erative, en partant d'une valeur constante,
\begin{equation*}
m^{(0)}=\underset{m\in\mathbb{R}}{\text{argmin}}\left\lbrace
\sum_{i=1}^n \ell(y_i,m)
\right\rbrace.
\end{equation*}
puis on consid\`ere l'apprentissage suivant
\begin{equation}\label{eqboot2}
{\displaystyle m^{(k)}=m^{(k-1)}+{\underset {h\in {\mathcal {H}}}{\text{argmin} }}\sum _{i=1}^{n}\ell(y_{i},m^{(k-1)}(\bx_{i})+h(\bx_{i}))},
\end{equation}
qui peut s'\'ecrire, si $\mathcal {H}$ est un ensemble de fonctions diff\'erentiables,
\begin{equation}
{\displaystyle m^{(k)}=m^{(k-1)}-\gamma _{k}\sum _{i=1}^{n}\nabla _{m^{(k-1)}}\ell(y_{i},m^{(k-1)}(\bx_{i})),}
\end{equation}
o\`u
$$
{\displaystyle \gamma _{k}=\underset{\gamma }{\text{argmin }}\sum _{i=1}^{n}\ell\left(y_{i},m^{(k-1)}(\bx_{i})-\gamma \nabla _{m^{(k-1)}}\ell(y_{i},m^{(k-1)}(\bx_{i}))\right).}
$$
Pour mieux comprendre le lien avec l'approche d\'ecrite auparavant, \`a l'\'etape $k$, on d\'efinit des pseudo-r\'esidus en posant
$$
r_{i,k}=-\left.\frac{\partial \ell(y_i,m(\bx_i))}{\partial m(\bx_i)}\right\vert_{m(\bx)=m^{(k-1)}(\bx)}
\text{ pour }i=1,\cdots,n.
$$
On cherche alors un mod\`ele simple pour expliquer ces pseudo-r\'esidus en fonction des variables explicatives $\bx_i$, i.e. $r_{i,k}=h^\star(\bx_i)$, o\`u $h^\star\in\mathcal{H}$. Dans un second temps, on cherche un multiplicateur optimal en r\'esolvant
$$
\gamma_k = \underset{\gamma\in\mathbb{R}}{\text{argmin}}\left\lbrace
\sum_{i=1}^n \ell(y_i,m^{(k-1)}(\bx_i)+\gamma h^\star(\bx_i))
\right\rbrace
$$
puis on met \`a jour le mod\`ele en posant
$
m_k(\bx)=m_{k-1}(\bx)+\gamma_k h^\star(\bx)
$. Plus formellement, on passe de l'\'equation (\ref{eqboot}) - qui montre clairement qu'on construit un mod\`ele sur les r\'esidus - \`a l'\'equation (\ref{eqboot2}) - qui sera ensuite retraduit comme une probl\`eme de calcul de gradient - en notant que $\ell(y,m+h)=\ell(y-m,h)$. Classiquement, les fonctions $\mathcal{H}$ sont construites avec des arbres de r\'egression. Il est aussi possible d'utiliser une forme de p\'enalisation en posant $m_k(\bx)=m_{k-1}(\bx)+\nu\gamma_k h^\star(\bx)$, avec $\nu\in(0,1)$. Mais revenons un peu plus longuement sur l'importance de la p\'enalisation avant de discuter les aspects num\'eriques de l'optimisation.}

\subsection{P\'enalisation et choix de variables}\label{sec:penalization}

Dans la section \ref{sec:goodness}, nous avions \'evoqu\'e le principe de parcimonie, populaire en \'econom\'etrie. Le crit\`ere d'Akaike \'etait bas\'e sur une p\'enalisation de la vraisemblance en tenant compte de la complexit\'e du mod\`ele (le nombre de variables explicatives retenues). Si en \'econom\'etrie, il est d'usage de maximiser la vraisemblance (pour construire un estimateur asymptotiquement sans biais), et de juger de la qualit\'e du mod\`ele ex-post en p\'enalisant la vraisemblance, la strat\'egie ici sera de p\'enaliser ex-ante dans la fonction objectif, quitte \`a construire un estimateur biais\'e. Typiquement, on va construire :
\begin{equation}\label{eq:penalization}
(\widehat{\beta}_{0,\lambda},\widehat{\bbeta}_{\lambda})=\text{argmin}\left\lbrace
\sum_{i=1}^n \ell(y_i,\beta_0+\bx\transpose\bbeta)+\lambda \text{ p\'enalisation}( \boldsymbol{\beta})
\right\rbrace,
\end{equation}
o\`u la fonction de p\'enalisation sera souvent une norme $\Vert\cdot\Vert$ choisie a priori, et un param\`etre de p\'enalisation $\lambda$ (on retrouve en quelque sorte la distinction entre AIC et BIC si la fonction de p\'enalisation est la complexit\'e du mod\`ele - le nombre de variables explicatives retenues). Dans le cas de la norme $\ell_2$, on retrouve l'estimateur Ridge, et pour la norme $\ell_1$, on retrouve l'estimateur {\sc lasso} (« {\em Least Absolute Shrinkage and Selection Operator }\nolinebreak »). La p\'enalisation utilis\'ee auparavant faisait intervenir le nombre de degr\'es de libert\'e du mod\`ele, il peut alors para\^itre surprenant de faire intervenir $\Vert\boldsymbol{\beta}\Vert_{\ell_2}$ comme dans la r\'egression Ridge. On peut toutefois envisager une vision Bay\'esienne de cette p\'enalisation. Rappelons que dans un mod\`ele Bay\'esien :
$$
\underbrace{\mathbb{P}[\boldsymbol{\theta}\vert\boldsymbol{y}]}_{\text{a posteriori}} \propto \underbrace{\mathbb{P}[\boldsymbol{y}\vert\boldsymbol{\theta}]}_{\text{vraisemblance}} \cdot \underbrace{\mathbb{P}[\boldsymbol{\theta}]}_{\text{a priori}}
\text{ ~~ soit ~~}
\log\mathbb{P}[\boldsymbol{\theta}\vert\boldsymbol{y}]= \underbrace{\log \mathbb{P}[\boldsymbol{y}\vert\boldsymbol{\theta}]}_{\text{log vraisemblance}} + \underbrace{\log\mathbb{P}[\boldsymbol{\theta}]}_{\text{{p\'enalisation}}}.
$$
Dans un mod\`ele lin\'eaire Gaussien, si on suppose que la loi \textit{a priori} de $\boldsymbol{\theta}$ suit une loi normale centr\'ee, on retrouve une p\'enalisation bas\'ee sur une forme quadratique des composantes de $\boldsymbol{\theta}$.

%Le probl\`eme d'optimisation d\'ecrit par l'\'equation (\ref{eq:penalization}) peut se voir comme la minimisation d'un Lagrangien, obtenu \`a l'aide d'un programme d'optimisation sous contrainte, de la forme :$$(\widehat{\beta}_0,\widehat{\bbeta})=\underset{\boldsymbol{\beta};\Vert\boldsymbol{\beta}\Vert\leq s}{\text{argmin}} \left\lbrace \sum_{i=1}^n \ell(y_i,\beta_0+\bx\transpose\bbeta) \right\rbrace.$$

Avant de revenir en d\'etails sur ces deux estimateurs, obtenus en utilisant la norme $\ell_1$ ou la norme $\ell_2$, revenons un instant sur un probl\`eme tr\`es proche : celui du meilleur choix de variables explicatives. Classiquement (et \c{c}a sera encore plus vrai en  grande dimension), on peut disposer d'un grand nombre de variables explicatives, $p$, mais beaucoup sont juste du bruit, au sens o\`u $\beta_j=0$ pour un grand nombre de $j$. Soit $s$ le nombre de covariables (r\'eellement) pertinentes, $s=\# \mathcal{S}$
avec $\mathcal{S}(\bbeta)=\lbrace j=1,\cdots,p; \beta_j\neq 0\rbrace$. Si on note $\bX_{\mathcal{S}}$ la matrice constitu\'ee des variables pertinentes (en colonnes), alors on suppose que le vrai mod\`ele
est de la forme $y=\bx_{\mathcal{S}}\transpose \bbeta_{\mathcal{S}}+\varepsilon$. Intuitivement, un estimateur int\'eressant serait alors $\widehat{\boldsymbol{\beta}}_{\mathcal{S}}=[\bX_{\mathcal{S}}\transpose\bX_{\mathcal{S}}]^{-1}\bX_{\mathcal{S}}\by$, mais cet estimateur n'est que th\'eorique car ${\mathcal{S}}$ est ici inconnue. Cet estimateur est l'estimateur oracle \'evoqu\'e auparavant.
On peut alors \^etre tent\'e de r\'esoudre 
$$
(\widehat{\beta}_{0,s},\widehat{\bbeta}_{s})=\text{argmin}\left\lbrace
\sum_{i=1}^n \ell(y_i,\beta_0+\bx\transpose\bbeta)\right\rbrace,\text{ sous la contrainte } \# \mathcal{S}(\bbeta)=s.
$$

\revision{Ce probl\`eme a \'et\'e introduit par \citeNP{FosterGeorge94} en introduisant la norme $\ell_0$. Plus pr\'ecis\'ement, d\'efinissons ici les trois normes suivantes
$$
\Vert\boldsymbol{a} \Vert_{\ell_0}=
\sum_{i=1}^d \boldsymbol{1}(a_i\neq 0), ~~ 
\Vert\boldsymbol{a} \Vert_{\ell_1}=
\sum_{i=1}^d |a_i|~~\text{ et }~~
\Vert\boldsymbol{a} \Vert_{\ell_2}=\left(
\sum_{i=1}^d a_i^2\right)^{1/2},\text{ pour }\boldsymbol{a}\in\mathbb{R}^d.
$$}

\revision{
%\begin{center}
\begin{table}
\begin{tabular}{|lccr|}\hline
&optimisation contrainte & p\'enalisation &\\\hline
meilleur groupe &
$\displaystyle{\underset{\boldsymbol{\beta};\Vert\boldsymbol{\beta}\Vert_{\ell_0}\leq s}{\text{argmin}} \left\lbrace \sum_{i=1}^n \ell(y_i,\beta_0+\bx\transpose\bbeta) \right\rbrace}$ & 
$\displaystyle{\underset{\boldsymbol{\beta},\lambda}{\text{argmin}} \left\lbrace \sum_{i=1}^n \ell(y_i,\beta_0+\bx\transpose\bbeta) +\lambda \Vert\boldsymbol{\beta}\Vert_{\ell_0} \right\rbrace}$&($\ell$0) \\
{\sc Lasso} &
$\displaystyle{\underset{\boldsymbol{\beta};\Vert\boldsymbol{\beta}\Vert_{\ell_1}\leq s}{\text{argmin}} \left\lbrace \sum_{i=1}^n \ell(y_i,\beta_0+\bx\transpose\bbeta) \right\rbrace}$ & 
$\displaystyle{\underset{\boldsymbol{\beta},\lambda}{\text{argmin}} \left\lbrace \sum_{i=1}^n \ell(y_i,\beta_0+\bx\transpose\bbeta) +\lambda \Vert\boldsymbol{\beta}\Vert_{\ell_1} \right\rbrace}$&($\ell$1) \\
Ridge &
$\displaystyle{\underset{\boldsymbol{\beta};\Vert\boldsymbol{\beta}\Vert_{\ell_2}\leq s}{\text{argmin}} \left\lbrace \sum_{i=1}^n \ell(y_i,\beta_0+\bx\transpose\bbeta) \right\rbrace}$ & 
$\displaystyle{\underset{\boldsymbol{\beta},\lambda}{\text{argmin}} \left\lbrace \sum_{i=1}^n \ell(y_i,\beta_0+\bx\transpose\bbeta) +\lambda \Vert\boldsymbol{\beta}\Vert_{\ell_2} \right\rbrace}$&($\ell$2) \\\hline
\end{tabular}
\caption{Optimisation contrainte et r\'egularisation.}\label{tab:012}
\end{table}
%\end{center}
}

\revision{Consid\'erons les probl\`emes d'optimisation de la Table \ref{tab:012}. Si on consid\`ere le probl\`eme classique o\`u $\ell$ est la norme quadratique, les deux probl\`emes de l'\'equation ($\ell$1) de la Table \ref{tab:012} sont \'equivalents, au sens o\`u, pour toute solution $(\boldsymbol{\beta}^\star,s^\star)$ au probl\`eme de gauche, il existe $\lambda^\star$ tel que $(\boldsymbol{\beta}^\star,\lambda^\star)$ soit solution du probl\`eme de droite; et inversement. Le r\'esultat est \'egalement vrai pour les probl\`emes ($\ell$2)\footnote{ Pour ($\ell$1), s'il y a \'equivalence au niveau th\'eorique, il peut exister des soucis num\'eriques car il n'y a pas forc\'ement unicit\'e de la solution.}. Il s'agit en effet de probl\`emes convexes. En revanche, les deux probl\`emes ($\ell0$) ne sont pas \'equivalents : si pour $(\boldsymbol{\beta}^\star,\lambda^\star)$ solution du probl\`eme de droite, il existe $s^\star$ tel que $\boldsymbol{\beta}^\star$ soit solution du probl\`eme de gauche, la r\'ecriproque n'est pas vraie. Plus g\'en\'eralement, si on veut utiliser une norme $\ell_p$, la sparsit\'ee est obtenue si $p\leq 1$ alors qu'il faut avoir $p\geq 1$ pour avoir la convexit\'e du programme d'optimisation.}

\revision{On peut \^etre tent\'e de r\'esoudre le programme p\'enalis\'e ($\ell0$) directement, comme le sugg\`ere \citeNP{FosterGeorge94}. Num\'eriquement, c'est un probl\`eme combinatoire complexe en grande dimension (\citeNP{Natarajan} note que c'est un probl\`eme NP-difficile), mais il est possible de montrer que si $\lambda \sim \sigma^2 \log(p)$, alors
$$
\mathbb{E}\big([\bx\transpose \widehat{\bbeta}-\bx\transpose \bbeta_0]^2\big) \leq \underbrace{\mathbb{E}\big(\bx_{\mathcal{S}}\transpose\widehat{\bbeta}_{\mathcal{S}}-\bx\transpose \bbeta_0]^2\big)}_{=\sigma^2 \#\mathcal{S}}\cdot \big(4\log p+2+o(1)\big).
$$
Notons quand dans ce cas
$$
\widehat{\bbeta}_{\lambda,j}^{\text{\sffamily sub}} = \left\lbrace
\begin{array}{l}
0 \text{ si } j\notin \mathcal{S}_\lambda(\bbeta)\\
\widehat{\bbeta}_{j}^{\text{\sffamily ols}} \text{ si } j\in \mathcal{S}_\lambda(\bbeta),
\end{array}
\right.
$$
o\`u $\mathcal{S}_\lambda(\bbeta)$ d\'esigne l'ensemble des coordonn\'ees non nulle lors de la r\'esolution de ($\ell0$).}

\revision{Le probl\`eme  ($\ell$2) est strictement convexe si $\ell$ est la norme quadratique, autrement dit, l'estimateur Ridge est toujours bien d\'efini, avec en plus une forme explicite pour l'estimateur, 
$$
\widehat{\boldsymbol{\beta}}_\lambda^{\text{\sffamily ridge}}=(\bX\transpose\bX+\lambda\mathbb{I})^{-1}\bX\transpose\by
=(\bX\transpose\bX+\lambda\mathbb{I})^{-1}(\bX\transpose\bX)\widehat{\boldsymbol{\beta}}^{\text{\sffamily ols}}.
$$
Aussi, on peut en d\'eduire que
$$
\text{biais}[\widehat{\boldsymbol{\beta}}_\lambda^{\text{\sffamily ridge}}]=-\lambda[\bX\transpose\bX+\lambda\mathbb{I}]^{-1}~\widehat{\boldsymbol{\beta}}^{\text{\sffamily ols}}\text{ et }
\text{Var}[\widehat{\boldsymbol{\beta}}_\lambda^{\text{\sffamily ridge}}]=
\sigma^2[\bX\transpose\bX+\lambda\mathbb{I}]^{-1}\bX\transpose\bX[\bX\transpose\bX+\lambda\mathbb{I}]^{-1}.
$$
Avec une matrice de variables explicatives orthonorm\'ees (i.e. $\bX\transpose\bX=\mathbb{I}$), les expressions se simplifient
$$
\text{biais}[\widehat{\boldsymbol{\beta}}_\lambda^{\text{\sffamily ridge}}]=\frac{\lambda}{1+\lambda}~\widehat{\boldsymbol{\beta}}^{\text{\sffamily ols}}\text{ et }
\text{Var}[\widehat{\boldsymbol{\beta}}_\lambda^{\text{\sffamily ridge}}]=\frac{\sigma^2}{(1+\lambda)^2}\mathbb{I}.
$$
Notons que $\text{Var}[\widehat{\boldsymbol{\beta}}_\lambda^{\text{\sffamily ridge}}]<\text{Var}[\widehat{\boldsymbol{\beta}}^{\text{\sffamily ols}}]$. En notant que
$$
\text{mse}[\widehat{\boldsymbol{\beta}}_\lambda^{\text{\sffamily ridge}}]=\frac{k\sigma^2}{(1+\lambda)^2}+\frac{\lambda^2}{(1+\lambda)^2}\bbeta\transpose\bbeta,
$$
on obtient une valeur optimale pour $\lambda$: $\lambda^\star=k\sigma^2 /\bbeta\transpose\bbeta $.
}

\begin{figure}
 \begin{center}
\includegraphics[width=.8\textwidth]{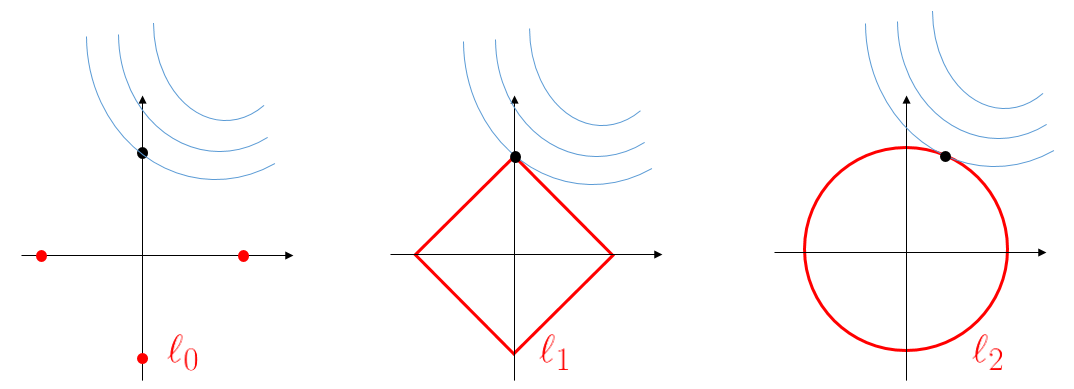}
\end{center}
\caption{P\'enalisation bas\'ee sur la norme $\ell_0$, $\ell_1$ et $\ell_2$ de $\bbeta$, respectivement (inspir\'e de Hastie {\em et al.} (2016)).}\label{Fig:lasso}
\end{figure}

\revision{En revanche, si $\ell$ n'est plus la norme quadratique mais la norme $\ell_1$, le probl\`eme  ($\ell$1) n'est pas toujours strictement convexe, et en particulier, l'optimum n'est pas toujours unique (par exemple si $\bX\transpose\bX$ est singuli\`ere). Mais le fait que $\ell$ soit strictement convexe $\bX\widehat{\boldsymbol{\beta}}$ sera unique. Notons de plus que deux solutions sont forc\'ement coh\'erentes en terme de signe des coefficients : il n'est pas possible d'avoir $\widehat{\beta}_j<0$ pour une solution et $\widehat{\beta}_j>0$ pour une autre. D'un point de vue heuristique, le programme ($\ell1$) est int\'eressant car il permet d'obtenir dans bon nombre de cas une solution en coin, qui correspond \`a une r\'esolution de probl\`eme de type $(\ell0)$ - comme le montre de mani\`ere visuelle la Figure \ref{Fig:lasso}.} 

\revision{Considérons un modèle très simple: $y_i=x_i\beta+\varepsilon$, avec une pénalité $\ell_1$ et une fonction de perte $\ell_2$. Le probl\`eme $(\ell2)$ s\'ecrit alors
$$
\min\big\{\by\transpose\by-2\by\transpose\bx\beta+\beta\bx\transpose\bx\beta+2\lambda|\beta|\big\} 
$$
La condition du premier ordre est alors
$$
-2\by\transpose\bx + 2\bx\transpose\bx\widehat{\beta}\pm 2\lambda=0.
$$
le signe du dernier terme d\'epend du signe de $\widehat{\beta}$.
Supposons que l'estimateur par moindre carr\'es (obtenu en posant $\lambda=0$) soit (strictement positif), autrement dit $\by\transpose\bx>0$. Si $\lambda$ n'est pas trop grand, on peut imaginer que $\widehat{\beta}$ soit du m\^eme signe que $\widehat{\beta}^{\text{\sffamily mco}}$, et donc la condition devient
$$
-2\by\transpose\bx + 2\bx\transpose\bx\widehat{\beta}+ 2\lambda=0.
$$
et la solution est 
$$
\widehat{\beta}_{\lambda}^{\text{\sffamily lasso}}=\frac{\by\transpose\bx-\lambda}{\bx\transpose\bx}.
$$
En augmentant $\lambda$, on va arriver \`a un moment o\`u $\widehat{\beta}_{\lambda}=0$. Si on augmente encore un peu $\widehat{\beta}_{\lambda}$ ne devient pas n\'egatif car dans ce cas le dernier terme de la condition du premier ordre change, et dans ce cas on cherche \`a r\'esoudre
$$
-2\by\transpose\bx + 2\bx\transpose\bx\widehat{\beta}- 2\lambda=0.
$$
dont la solution est alors
$$
\widehat{\beta}_{\lambda}^{\text{\sffamily lasso}}=\frac{\by\transpose\bx+\lambda}{\bx\transpose\bx}.
$$
Mais cette solution est positive (nous avions suppos\'e $\by\transpose\bx>0$), et donc il est possible d'avoir en m\^eme temps $\widehat{\beta}_{\lambda}<0$. Aussi, au bout d'un moment, $\widehat{\beta}_{\lambda}=0$, qui est alors une solution de coin. Les choses sont bien entendu plus compliqu\'ees en dimension plus grande (\citeNP{TibWasserman} revient longuement sur la g\'eom\'etrie des solutions) mais comme le note \citeNP{Candes}, sous des hypothèses minimales garantissant que les pr\'edicteurs ne sont pas fortement corrélées, le {\sc Lasso} obtient une erreur quadratique presque aussi bonne que si l'on dispose d'un oracle fournissant des informations parfaites sur quels $\beta_j$ sont non nulles.
Moyennant quelques hypoth\`eses techniques suppl\'ementaires, on peut montrer que cet estimateur est « sparsistent »  au sens o\`u le support de $\widehat{\beta}_{\lambda}^{\text{\sffamily lasso}}$ est celui de $\bbeta$, autement dit {\sc Lasso} a permis de faire de la s\'election de variables (plus de discussions sur ce point peuvent \^etre obtenues dans \citeNP{HTT}).}

\revision{De mani\`ere plus g\'en\'erale, on peut montrer que $
\widehat{\beta}_{\lambda}^{\text{\sffamily lasso}}$ est un estimateur biais\'e, mais qui peut \^etre de variance suffisamment faible pour que l'erreur quadratique moyenne soit plus faible qu'en utilisant des moindres carr\'es. Pour comparer les trois techniques, par rapport \`a l'estimateur par moindre carr\'ees (obtenu quand $\lambda=0)$, si on suppose que les variables explicatives sont orthonorm\'ees, alors
$$
\widehat{\beta}_{\lambda,j}^{\text{\sffamily sub}}=\widehat{\beta}_{j}^{\text{\sffamily ols}}\boldsymbol{1}_{|\widehat{\beta}_{\lambda,j}^{\text{\sffamily sub}}|>b}, ~~
\widehat{\beta}_{\lambda,j}^{\text{\sffamily ridge}}=\frac{\widehat{\beta}_{j}^{\text{\sffamily ols}}}{1+\lambda}~\text{ et }~
\widehat{\beta}_{\lambda,j}^{\text{\sffamily lasso}}=\text{signe}[\widehat{\beta}_{j}^{\text{\sffamily ols}}]\cdot(|\widehat{\beta}_{j}^{\text{\sffamily ols}}|-\lambda)_+.
$$}

\begin{figure}
 \begin{center}
\includegraphics[width=.8\textwidth]{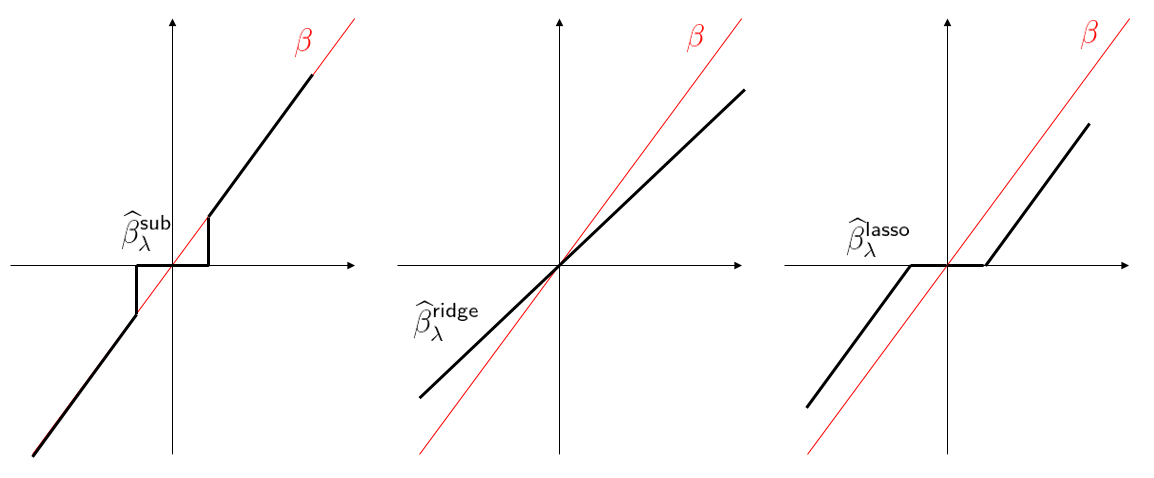}
\end{center}
\caption{P\'enalisation bas\'ee sur la norme $\ell_0$, $\ell_1$ et $\ell_2$ de $\bbeta$, respectivement (inspir\'e de Hastie {\em et al.} (2016)).}\label{Fig:lasso2}
\end{figure}

\subsection{Optimisation et aspects algorithmiques}

\revision{En \'econom\'etrie, l'optimisation (num\'erique) est devenu omnipr\'esente d\`es que l'on a quitt\'e le mod\`ele Gaussien. Nous l'avions rapidement \'evoqu\'e dans la section sur la famille exponentielle, et l'utilisation du score de Fisher (descente de gradient) pour r\'esoudre la condition du premier ordre $\boldsymbol{X}^{\text{\sffamily{T}}}\boldsymbol{W}(\boldsymbol{\beta})^{-1}[\boldsymbol{y}-\widehat{\by}]=\boldsymbol{0}$. En apprentissage, l'optimisation est l'outil central. Et il est n\'ecessaire d'avoir des algorithmes d'optimisation efficaces, pour r\'esoudre des probl\`emes de la forme :
$$
\widehat{\bbeta}\in\underset{\bbeta\in\mathbb{R}^p}{\text{argmin}}\left\lbrace
\sum_{i=1}^n \ell(y_i,\beta_0+\bx\transpose\bbeta)+\lambda \Vert \boldsymbol{\beta}\Vert
\right\rbrace.
$$}

\revision{Dans certains cas, au lieu de faire de l'optimisation globale, il suffit de consid\'erer de l'optimisation par coordonn\'ees (largement \'etudi\'ee dans \citeNP{Daubechies}). Si $f:\mathbb{R}^d\rightarrow\mathbb{R}$ est convexe et diff\'erentiable, alors 
$$
\text{si }\bx\text{ v\'erifie }f(\bx+h\boldsymbol{e}_i)\geq f(\bx)\text{ pour tout }h>0\text{ et }i\in\{1,\cdots, d\},\text{ alors }
f(\bx)=\min\{f\},
$$
où $\boldsymbol{e}=(\boldsymbol{e}_i)$ est la base canonique de  $\mathbb{R}^d$. Cette propri\'et\'e n'est toutefois pas vraie dans le cas non-diff\'erentiable. Mais si on suppose que la partie non-diff\'erentiable est s\'eparable (additivement), elle redevient vraie. Plus pr\'ecis\'ement, si 
$$
f(\bx)=g(\bx)+\sum_{i=1}^d h_i(x_i)\text{ avec }
\left\lbrace
\begin{array}{l}
g: \mathbb{R}^d\rightarrow\mathbb{R}\text{ convexe-diff\'erentiable}\\
h_i: \mathbb{R}\rightarrow\mathbb{R}\text{ convexe}.
\end{array}
\right.
$$
C'est le cas pour la r\'egression {\sc Lasso}, $f(\bbeta)=\|\by-\boldsymbol{X}\bbeta\|_{\ell_2}+\lambda\|\bbeta\|_{\ell_1}$, comme le montre \citeNP{Tsen}.}
\revision{On peut alors utiliser un algorithme de descente par coordonn\'ees: \`a partir d'une valeur initiale $\bx^{(0)}$, on consid\`ere (en it\'erant)
$$
x_j^{(k)}\in\text{argmin}\big\lbrace
f(x_1^{(k)},\cdots,x_{k-1}^{(k)},x_k,x_{k+1}^{(k-1)},
\cdots,x_n^{(k-1)})
\big\rbrace ~\text{ pour }j=1,2,\cdots,n.
$$
Ces probl\`emes algorithmiques peuvent para\^itre secondaires \`a des \'econom\`etres. Ils sont pourtant essentiels en apprentissage machine: une technique est int\'eressante s'il existe un algorithme stable et rapide, qui permet d'obtenir une solution.} 
 \revision{Ces techniques d'optimisation sont d'ailleurs transposables : par exemple, on pourra utiliser cette technique de descente par coordonn\'ees dans le cas des m\'ethodes {\sc svm} (dit « à support vecteur\nolinebreak ») lorsque l'espace n'est pas lin\'eairement s\'eparable, et qu'il convient de p\'enaliser l'erreur de classification (nous reviendrons sur cette technique dans la prochaine section).}

\subsection{\textit{In-sample}, \textit{out-of-sample} et validation crois\'ee}\label{sec:in:out:sample}%\label{sec:validation:croisee}

Ces techniques semblent intellectuellement int\'eressantes, mais nous n'avons pas encore abord\'e le choix du param\`etre de p\'enalisation $\lambda$. Mais ce probl\`eme est en fait plus g\'en\'eral, car comparer deux param\`etres $\widehat{\bbeta}_{\lambda_1}$ et $\widehat{\bbeta}_{\lambda_2}$ revient en fait \`a comparer deux mod\`eles. En particulier, si on utilise une m\'ethode de type {\sc Lasso}, avec des seuils $\lambda$ diff\'erents, on compare des mod\`eles qui n'ont pas la m\^eme dimension. Dans la section \ref{sec:goodness}, nous avions abord\'e le probl\`eme de la comparaison de mod\`eles sous l'angle \'econom\'etrique (en p\'enalisant les mod\`eles trop complexes). \revision{Dans la litt\'erature en apprentissage, juger de la qualit\'e d'un mod\`ele sur les donn\'ees qui ont servi \`a le construire ne permet en rien de savoir comment le mod\`ele se comportera sur des nouvelles donn\'ees. Il s'agit du probl\`eme dit de « g\'en\'eralisation ». L'approche classique consiste alors \`a s\'eparer l'\'echantillon (de taille $n$) en deux : une partie qui servira \`a entra\^iner le mod\`ele (la base d'apprentissage, {\em in-sample}, de taille $m$) et une partie qui servira \`a tester le mod\`ele (la base de test, {\em out-of-sample}, de taille $n-m$). Cette derni\`ere permet alors de mesure un vrai risque pr\'edictif. Supposons que les donn\'ees soient g\'en\'er\'ees par un mod\`ele lin\'eaire $y_i=\bx_i\transpose \bbeta_0 + \varepsilon_i$ o\`u les $\varepsilon_i$ sont des r\'ealisations de lois ind\'ependantes et centr\'ees. Le risque quadratique empirique {\em in-sample} est ici
$$
\frac{1}{m}\sum_{i=1}^m\mathbb{E}\big([\bx_i\transpose \widehat{\bbeta}-\bx_i\transpose \bbeta_0]^2\big)=\mathbb{E}\big([\bx_i\transpose \widehat{\bbeta}-\bx_i\transpose \bbeta_0]^2\big),
$$
pour n'importe quelle observation $i$. En supposant les r\'esidus $\varepsilon$ Gaussiens, alors on peut montrer que ce risque vaut $\sigma^2\text{trace}(\boldsymbol{\Pi}_{\mathcal{X}})/m$ soit $\sigma^2p/m$. En revanche le risque quadratique empirique {\em out-of-sample} est ici
$$
\mathbb{E}\big([\bx\transpose \widehat{\bbeta}-\bx\transpose \bbeta_0]^2\big)
$$
o\`u $\bx$ est une nouvelle observation, ind\'ependante des autres. On peut noter que 
$$
\mathbb{E}\big([\bx\transpose \widehat{\bbeta}-\bx\transpose \bbeta_0]^2\big\vert \bx\big)=
%\text{Var}\big(\bx\transpose \widehat{\bbeta}\big\vert \bx\big)=
\sigma^2\bx\transpose(\bX\transpose\bX)^{-1}\bx,
$$
et en int\'egrant par rapport \`a $\bx$,
$$
\mathbb{E}\big([\bx\transpose \widehat{\bbeta}-\bx\transpose \bbeta_0]^2\big)=
\mathbb{E}\big(\mathbb{E}\big([\bx\transpose \widehat{\bbeta}-\bx\transpose \bbeta_0]^2\big\vert \bx\big)\big)=
\sigma^2\text{trace}\big(\mathbb{E}[\bx\bx\transpose]\mathbb{E}\big[(\bX\transpose\bX)^{-1}\big]\big).
$$
L'expression est alors diff\'erente de celle obtenue {\em in-sample}, et en utilisation la majoration de \citeNP{GrovesRothenberg69}, on peut montrer que
$$
\mathbb{E}\big([\bx\transpose \widehat{\bbeta}-\bx\transpose \bbeta_0]^2\big) \geq \sigma^2\frac{p}{m},
$$
ce qui est assez intuitif, finalement. Hormis certains cas simple, il n'y a pas de formule simple. Notons toutefois que si $\bx \sim\mathcal{N}(\boldsymbol{0},\sigma^2\mathbb{I})$, alors $\bx\transpose\bx$ suit une loi de Wishart, et on peut montrer que
$$
\mathbb{E}\big([\bx\transpose \widehat{\bbeta}-\bx\transpose \bbeta_0]^2\big)=\sigma^2\frac{p}{m-p-1}.
$$
Si on regarde maintenant la version empirique: si $\widehat{\boldsymbol{\beta}}$ est estim\'e sur les $m$ premi\`eres observations,
$$
\widehat{\mathcal{R}}^{~\text{\sffamily IS}}=\sum_{i=1}^m [y_i-\boldsymbol{x}_i\transpose\widehat{\boldsymbol{\beta}}]^2
\text{\quad et \quad}
\widehat{\mathcal{R}}^{\text{\sffamily OS}}=\sum_{i=m+1}^{n} [y_i-\boldsymbol{x}_i\transpose\widehat{\boldsymbol{\beta}}]^2,
$$
et comme l'a not\'e \citeNP{Leeb}, $\widehat{\mathcal{R}}^{~\text{\sffamily IS}} -\widehat{\mathcal{R}}^{~\text{\sffamily OS}}  \approx 2 \cdot\nu$ o\`u $\nu$ repr\'esente le nombre de degr\'es de libert\'es, qui n'est pas sans rappeler la p\'enalisation utilis\'ee dans le crit\`ere d'Akaike.}

\begin{figure}
\begin{center}
\includegraphics[width=0.4\textwidth]{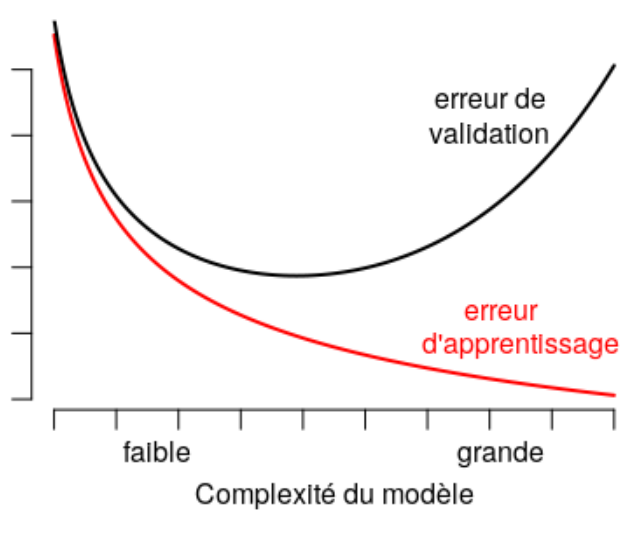}~
\includegraphics[width=0.4\textwidth]{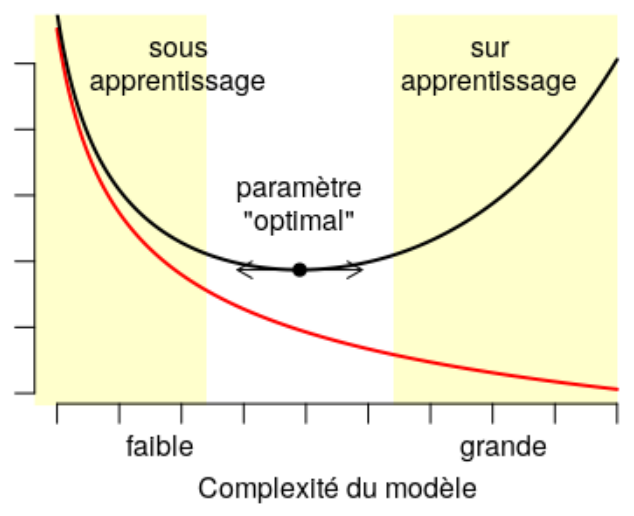}
\end{center}
\caption{G\'en\'eralisation, et sur-apprentissage.}\label{Fig:tradeoff2}
\end{figure}

La Figure \ref{Fig:tradeoff2} montre l'\'evolution respective de $\widehat{\mathcal{R}}^{\text{\sffamily IS}}$ et $\widehat{\mathcal{R}}^{\text{\sffamily OS}}$ en fonction de la complexit\'e du mod\`ele (nombre de degr\'es dans une r\'egression polynomiale, nombre de noeuds dans des splines, etc). Plus le mod\`ele est complexe, plus $\widehat{\mathcal{R}}^{\text{\sffamily IS}}$ va diminuer (c'est la courbe rouge). Mais ce n'est pas ce qui nous int\'eresse ici : on veut un mod\`ele qui pr\'edise bien sur de nouvelles donn\'ees (autrement dit {\em out-of-sample}). Comme le montre la Figure \ref{Fig:tradeoff2}, si le mod\`ele est trop simple, il pr\'edit mal (tout comme sur les donn\'ees {\em in-sample}). Mais ce que l'on peut voir, c'est que si le mod\`ele est trop complexe, on est dans une situation de « sur-apprentissage » : le mod\`ele va commencer \`a mod\'eliser le bruit.

\revision{Au lieu de s\'eparer la base en deux, avec une partie des donn\'ees qui vont servir \`a calibrer le mod\`ele et une autre \`a \'etudier sa performance, il est aussi possible d'utiliser la validation crois\'ee. Pour pr\'esenter l'id\'ee g\'en\'erale, on peut revenir au « jackknife », introduit par \citeNP{Quenouille1} (et formalis\'e par \citeNP{Quenouille2} et \citeNP{Tukey}) et utilis\'e en statistique pour r\'eduire le biais.} En effet, si on suppose que $\lbrace y_1,\cdots,y_n\rbrace$ est un \'echantillon tir\'e suivant une loi $F_{\theta}$, et que l'on dispose d'un estimateur $T_n(\by)=T_n(y_1,\cdots,y_n)$, mais que cet estimateur est biais\'e, avec
$\esp[T_n(\bY)]=\theta+O\left(n^{-1}\right)$,  
il est possible de r\'eduire le biais en consid\'erant :
$$
\widetilde{T}_n(\by)=\frac{1}{n}\sum_{i=1}^n T_{n-1}(\by_{(i)})\text{ \quad avec \quad }\by_{(i)}=(y_1,\cdots,y_{i-1},y_{i+1},\cdots,y_n).
$$
On peut alors montrer que $\esp[\widetilde{T}_n(\bY)]=\theta+O\left(n^{-2}\right)$.

L'id\'ee de la validation crois\'ee repose sur l'id\'ee de construire un estimateur en enlevant une observation. Comme on souhaite construire un mod\`ele pr\'edictif, on va comparer la pr\'evision obtenue avec le mod\`ele estim\'e, et l'observation manquante :
$$
\widehat{\mathcal{R}}^{\text{\sffamily CV}}=\frac{1}{n}\sum_{i=1}^n \ell(y_i,\widehat{m}_{(i)}(\bx_i))
$$
On parlera ici de m\'ethode « {\em leave-one-out} » ({\sc loocv}).

On utilise classiquement cette technique pour trouver le param\`etre optimal dans les m\'ethodes de lissage exponentiel, pour des s\'eries chronologiques. Dans le lissage simple, on va construire une pr\'ediction de la forme ${}_{t}\widehat{y}_{t+1}=\alpha\cdot {}_{t-1}\widehat{y}_{t}+(1-\alpha)\cdot y_t$, avec $\alpha\in[0,1]$, et on va consid\'erer :
$$
\alpha^\star = \underset{\alpha\in[0,1]}{\text{argmin}}\left\lbrace \sum_{t=2}^T \ell({}_{t-1}\widehat{y}_{t},y_{t}) \right\rbrace,
$$
comme le d\'ecrit \citeNP{Hyndman}.

Le principal probl\`eme de la m\'ethode « {\em leave-one-out} » est qu'elle n\'ecessite de calibrer $n$ mod\`eles, ce qui peut être probl\'ematique en grande dimension. Une m\'ethode alternative est la validation crois\'ee par $k$-blocs (dit « $k$-{\em fold cross validation} ») qui consiste \`a utiliser une partition de $\lbrace1,\cdots,n\rbrace$ en $k$ groupes (ou blocs) de m\^eme taille,
$\mathcal{I}_1,\cdots,\mathcal{I}_k$, et notons $\mathcal{I}_{\overline{j}}=\lbrace
1,\cdots,n\rbrace\backslash \mathcal{I}_j$. En notant $\widehat{m}_{(j)}$ construit sur l'\'echantillon $\mathcal{I}_{\overline{j}}$, on pose alors :
$$
\widehat{\mathcal{R}}^{k-\text{\sffamily CV}}=\frac{1}{k}\sum_{j=1}^k \mathcal{R}_j\text{\quad o\`u \quad }\mathcal{R}_j=\frac{k}{n}\sum_{i\in\mathcal{I}_{{j}}} \ell(y_i,\widehat{m}_{(j)}(\bx_i)).
$$
La validation croisée standard, où une seule observation est enlevée à chaque fois ({\sc loocv}), est un cas particulier, avec  $k=n$. Utiliser $k=5,10$ a un double avantage par rapport à $k=n$~: (1) le nombre d'estimations à effectuer est beaucoup plus faible, 5 ou 10 plutôt que $n$ ; (2) les échantillons utilisés pour l'estimation sont moins similaires et donc, moins corrélés les uns aux autres, ce qui tend à éviter les excès de variance, commme le rappelle \citeNP{James}.

%%generate $I_1,\cdots,I_B$ bootstrapped samples from $\lbrace 1,\cdots, n\rbrace$
%%
%%set 	$n_i=\boldsymbol{1}_{i\notin I_1}+\cdots+\boldsymbol{1}_{i\notin I_B}$
%%
%%$$
%%\widehat{R}=\frac{1}{n}\sum_{i=1}^n \frac{1}{n_i}\sum_{b:i\notin I_b} \ell(y_i,\widehat{m}_{b}(\boldsymbol{x}_i)
%%$$
%%
%%{\bf Remark} Probability that $i$th raw is not selection $(1-n^{-1})^n\rightarrow e^{-1}\sim 36.8\%$, 

Une autre alternative consiste \`a utiliser des \'echantillons boostrapp\'es. Soit $\mathcal{I}_b$ un \'echantillon de taille $n$ obtenu en tirant avec remise dans $\lbrace1,\cdots,n\rbrace$ pour savoir quelles observations $(y_i,\bx_i)$ seront gard\'ees dans la population d'apprentissage (\`a chaque tirage). Notons $\mathcal{I}_{\overline{b}}=\lbrace
1,\cdots,n\rbrace\backslash \mathcal{I}_b$. En notant $\widehat{m}_{(b)}$ construit sur l'\'echantillon $\mathcal{I}_{b}$, on pose alors :
$$
\widehat{\mathcal{R}}^{\text{\sffamily B}}=\frac{1}{B}\sum_{b=1}^B \mathcal{R}_b\text{\quad o\`u \quad }\mathcal{R}_b=\frac{n_{\overline{b}}}{n}\sum_{i\in\mathcal{I}_{\overline{b}}} \ell(y_i,\widehat{m}_{(b)}(\bx_i)),
$$
o\`u $n_{\overline{b}}$ est le nombre d'observations qui n'ont pas \'et\'e conserv\'ees dans $\mathcal{I}_b$. On notera qu'avec cette technique, en moyenne $e^{-1}\sim 36.7\%$ des observations ne figurent pas dans l'\'echantillon boostrapp\'e, et on retrouve un ordre de grandeur des proportions utilis\'ees en cr\'eant un \'echantillon de calibration, et un \'echantillon de test.
 En fait, comme l'avait montr\'e \citeNP{Stone}, la minimization du AIC est \`a rapprocher du crit\`ere de validation crois\'ee, et \citeNP{Shao97} a montr\'e que la minimisation du BIC correspond \`a de la validation crois\'ee de type $k$-fold, avec $k=n/\log n$.

\section{Quelques \revision{outils} de {\em machine learning}}\label{sec:algo}

\subsection{R\'eseaux de Neurones}\label{sec:selection:nn}

%\subsubsection{Introduction}

\revision{Les réseaux de neurones sont des modèles semi-paramétriques. Néanmoins, cette famille de modèles peut être appréhendée de la même manière que les modèles non-paramétriques: la structure des réseaux de neurones (présentée par la suite) peut être modifiée afin d'étendre la classe des fonctions utilisées pour approcher une variable d'intérêt. Plus précisément, \citeNP{Cybenko} a démontré que l'ensemble des fonctions neuronales est dense dans l'espace des fonctions continues sur un compact. En d'autres termes, on a un cadre théorique permettant de garantir une forme d'approximation universelle. Il impose en outre une définition d'un neurone et met en avant l'existence d'un nombre de neurones suffisant pour approcher toute fonction continue sur un compact. Ainsi, un phénomène continue peut être approché par une suite de neurones: on appellera cette suite « réseau de neurones à une couche ». Si ce théorème d'approximation universelle est démontré en 1989, le premier neurone artificiel fonctionnel fut introduit par Franck Rosenblatt au milieu du XXi\`eme si\`ecle, dans \citeNP{Rosenblatt}. Ce neurone, qualifi\'e de nos jours de « neurone \'el\'ementaire », porte le nom de « Perceptron ». Il a permis dans ses premi\`eres utilisations de d\'eterminer le sexe d'un individu pr\'esent\'e aux travers d'une photo. Si ce premier neurone est important, c'est qu'il introduit le premier formalisme math\'ematique d'un neurone biologique. On peut d\'ecrire un neurone artificiel par analogie avec une cellule nerveuse :} 

\begin{itemize}
\item[-]  les synapses apportant l'information \`a la cellule sont formalis\'es par un vecteur r\'eel. La dimension du vecteur d'entr\'ee du neurone (qui n'est d'autre qu'une fonction) correspond biologiquement au nombre de connections synaptiques;
\item[-]  chaque signal apport\'e par un synapse est ensuite analys\'e par la cellule. Math\'ematiquement, ce sch\'ema est transcrit par la pond\'eration des diff\'erents \'el\'ements constitutifs du vecteur d'entr\'ee;
\item[-] en fonction de l'information acquise, le neurone d\'ecide de retransmettre ou non un signal. Ce ph\'enom\`ene est r\'epliqu\'e par l'introduction d'une fonction d'activation. Le signal de sortie est mod\'elis\'e par un nombre r\'eel calcul\'e comme image par la fonction d'activation du vecteur d'entr\'ee pond\'er\'e.\\
\end{itemize}

Ainsi, un neurone artificiel est un mod\`ele semi-param\'etrique. Le choix de la fonction d'activation est en effet laiss\'e \`a l'utilisateur. Nous introduisons dans le paragraphe qui suit une formalisation rigoureuse qui nous permettra de poser le mod\`ele, \revision{et de faire le lien avec les notations \'econom\'etriques usuelles}. On peut alors d\'efinir un neurone \'el\'ementaire formellement par : 

\begin{enumerate}

\item[-]  un espace d'entr\'ee $\mathcal{X}$, g\'en\'eralement $\mathbb{R}^k$ avec $k \in \mathbb{N}^*$;
\item[-]  un espace de sortie $\mathcal{Y}$, g\'en\'eralement $\mathbb{R}$ ou un ensemble fini (classiquement $\{0,1\}$, mais on pr\'ef\'erera ici $\{-1,+1\}$);
\item[-]  un vecteur de param\`etres $\boldsymbol{w} \in \mathbb{R}^p$ %où $p$ peut être diff\'erent de $m$ (par exemple lorsque l'on souhaite introduire un offset dans le cas d'un Kernel Trick);
\item[-]  une fonction d'activation $\phi : \mathbb{R} \rightarrow \mathbb{R} $.  Cette fonction doit être dans l'id\'eal monotone, d\'erivable et born\'ee (on dira ici « saturante  ») afin de s'assurer de certaines propri\'et\'es de convergence. %d'apr\`es le th\'eor\`eme de Cybenko (ressemble aux th\'eor\`emes de Stone-Weirstrass et sera pr\'esent\'e plus loin). 
\end{enumerate}
Cette derni\`ere fonction $\phi$ fait penser aux transformations logistique ou probit, populaire en \'econom\'etrie (qui sont des fonctions de r\'epartition, \`a valeur dans $[0,1]$, id\'eal quand $\mathcal{Y}$ est l'ensemble $\{0,1\}$). Pour les r\'eseaux de neurones, on utilisera plut\^ot la tangente hyperbolique, la fonction arctangente ou les fonctions sigmo\"ides pour des probl\`emes de classification ($\mathcal{Y}=\{-1,+1\}$.
On appellera neurone toute application $f_{w}$ de $\mathcal{X}$ dans $\mathcal{Y}$ d\'efinie par : 
$$ y=f_{\boldsymbol{w}}(\bx) = \phi (\boldsymbol{w}\transpose\boldsymbol{x}), ~~ \forall \boldsymbol{x}\in \mathcal{X}.$$

%Voici une liste de fonctions d'activation $\phi$ d\'efinies sur $\mathbb{R}$ couramment utilis\'ees pour les r\'eseaux de neurones : \\
%\begin{table}
%\centering
%\caption{Exemple de fonctions d'activation $\phi$ utilis\'ees}
%\begin{tabular}{|c|c|c|c|}
%Fonction & D\'efinition & Image   \\ \hline
% Tanh & $x \rightarrow \frac{e^x - e^{-x}}{e^x + e^{-x}}$ & $[-1,1]$  \\
% Sigmoide\tablefootnote{Le terme sigmoide est ici abusif car il d\'esigne en r\'ealit\'e une classe de fonctions plus g\'en\'erale}, $\alpha \in \mathbb{R}^+$ & $x \rightarrow \frac{1}{1 + e^{-\alpha x}}$  & $[0,1]$ \\
% Arctan & $x \rightarrow arctan(x)$  & $]-\frac{\pi}{2},\frac{\pi}{2}[$
%\end{tabular}
%\label{activation}
%\end{table}

Pour le perceptron introduit par \citeNP{Rosenblatt}, on assimile un neurone \'el\'ementaire \`a la fonction :
$$ % : \in \mathbb{R}^p \rightarrow 
y=f_{\boldsymbol{w}}(\boldsymbol{x})=\text{signe}(\boldsymbol{w}\transpose \boldsymbol{x})~~ \forall \boldsymbol{x}\in \mathcal{X}$$ 

On remarque que selon cette formalisation, beaucoup de mod\`eles statistiques comme par exemple les r\'egressions logistiques pourraient être vus comme des neurones. En effet si l'on regarde d'un peu plus pr\`es, tout mod\`ele {\sc glm} (« {\em Generalized Linear Model} ») pourrait s'interpr\'eter comme un neurone formel où la fonction d'activation $\phi$ n'est d'autre que l'inverse de la fonction de lien canonique (par exemple). Si $g$ d\'esigne la fonction de lien du {\sc glm} , $\boldsymbol{w}$ le vecteur de param\`etres, $y$ la variable \`a expliquer et $\boldsymbol{x}$ le vecteur des variables explicatives de même dimension que $\boldsymbol{w}$ : 
$$ g(\mathbb{E}[Y|\boldsymbol{X}=\boldsymbol{x}])=\boldsymbol{w}\transpose \boldsymbol{x} $$ 
On retrouve la mod\'elisation neuronale en prenant $\phi = g^{-1} $. Cependant, l\`a où r\'eside la diff\'erence majeure entre les {\sc glm} et le mod\`ele neuronale est que ce dernier n'introduit aucune hypoth\`ese de distribution sur $Y|\boldsymbol{X}$ (on n'a d'ailleurs pas besoin d'introduire ici de mod\`ele probabiliste). D'autre part, lorsque le nombre de neurones par couche augmente, la convergence n'est pas n\'ecessairement garantie si la fonction d'activation ne v\'erifie pas certaines propri\'et\'es (qu'on ne retrouve pas dans la majorit\'e des fonctions de liens canoniques des {\sc glm}). 
Cependant, comme \'enonc\'e pr\'ec\'edemment, la th\'eorie des r\'eseaux de neurones introduit des contraintes math\'ematiques suppl\'ementaires sur la fonction $g$ (d\'etaill\'e dans \citeNP{Cybenko}). Ainsi par exemple, une r\'egression logistique peut être perçue comme un neurone \revision{alors que les r\'egressions lin\'eraires généralisées ne v\'erifient pas toutes les hypoth\`eses n\'ecessaires.}

%\subsubsection{Les r\'eseaux multicouches}

Toujours par analogie avec le fonctionnement du syst\`eme nerveux, il est alors possible de connecter diff\'erents neurones entre eux. On parlera de structure de r\'eseaux de neurones par couche. Chaque couche de neurones recevant \`a chaque fois le même vecteur d'observation. %Ainsi un r\'eseau de neurones \`a une couche par exemple sera constitu\'e de plusieurs neurones recevant tous le même signal d'entr\'ee. %De façon plus formelle, chaque couche peut être associ\'ee \`a une fonction de transfert qui prend en entr\'ee un vecteur et fournit en sortie un autre vecteur. Dans le cadre des r\'eseaux de neurones, chaque fonction de transfert est une fonction param\'etrique avec un nombre de param\`etres proportionnel aux nombres de neurones de chaque couche. 

\begin{figure}[h]
\centering
\caption{Exemple de notations associ\'ees aux r\'eseaux de neurones multicouche.}\label{fig:nn}
%source: intelligenceartificielle.org}
\vspace{0.5cm}
\includegraphics[scale=0.45]{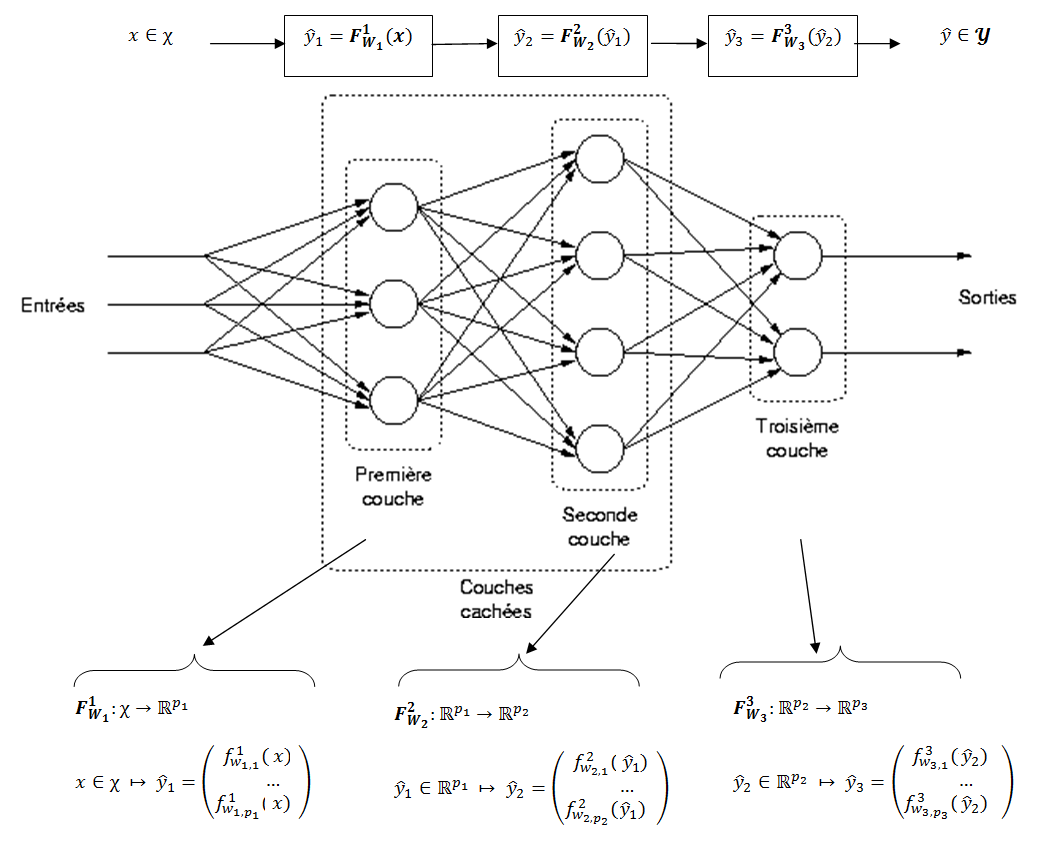}
\end{figure}

\revision{Pour revenir à une analogie plus économétrique, on peut imaginer passer par une étape intermédiaire (on reviendra sur cette construction dans la Figure \ref{fig:nn}), par exemple en ne faisant pas une régression sur les variables brutes $\bx$ mais un ensemble plus faible de variables orthogonales, obtenues par exemple suite à une analyse en composantes principales. Soit $\boldsymbol{A}$ la matrice associée à cette transformation linéaire, avec $\boldsymbol{A}$ de taille $k\times p$ si on souhaite utiliser les $p$ premières composantes. Notons $\boldsymbol{z}$ la transformation de $\boldsymbol{x}$, au sens où $\boldsymbol{z} = \boldsymbol{A}\transpose \bx$, ou encore $z_j=\boldsymbol{A}_j\transpose \bx$. Un généralisation du mod\`ele précédant peut être de poser
$$ y=f(x) = \phi (\boldsymbol{w}\transpose\boldsymbol{z})= \phi (\boldsymbol{w}\transpose\boldsymbol{A}\transpose \bx)=, ~~ \forall \boldsymbol{x}\in \mathcal{X},$$
où cette fois $\boldsymbol{w}\in\mathbb{R}^p$. On a ici une transformation linéaire (en considérant une analyse en composante principale) mais on peut imaginer une généralisation avec des transformée non-linéaire, avec une fonction de la forme
$$ y=f(x) = \phi (\boldsymbol{w}\transpose F_{\boldsymbol{A}}(\bx)=, ~~ \forall \boldsymbol{x}\in \mathcal{X},$$
o\`u $F$ est ici une fonction $\mathbb{R}^k\rightarrow \mathbb{R}^p$. C'est le réseau de neurone à deux couches.} Plus généralement, pour formaliser la construction, on introduit les notations suivantes : 

\begin{itemize}
\item[-] $K \in \mathbb{N}^*$ : nombre de couches;
\item[-] $\forall k \in \{ 1 ,\cdots K \}$ , $p_k$ repr\'esente le nombre de neurones dans la couche $k$;
\item[-] $\forall k \in \{ 1 ,\cdots K \}$ ,$W_k$ d\'esigne la matrice des param\`etres associ\'es \`a la couche $k$. Plus pr\'ecis\'ement, $W_k $ est une matrice $p_k\times p_{k-1}$ et pour tout $l \in \{ 1 ,\cdots p_k \}$ , $w_{k,l} \in \mathbb{R}^{ p_{k-1}}$ d\'esigne le vecteur de poids associ\'e au neurone \'el\'ementaire $l$ de la couche $k$;
\item[-] on appellera $W = \{W_1,..,W_K\}$, l'ensemble des param\`etres associ\'es au r\'eseau de neurones.
\item[-] $F^k_{W_k} : \mathbb{R}^{p_{k-1}} \rightarrow \mathbb{R}^{p_{k}}$ d\'esigne la fonction de transfert associ\'e \`a la couche $k$. Pour des raisons de simplification, on pourra \'egalement \'ecrire $F^k$;
\item[-] $\hat{y}_k \in \mathbb{R}^{p_k}$ repr\'esentera le vecteur image de la couche $k \in \{ 1 ,\cdots, K \}$;
\item[-] on appellera $F= F_W = F^1 \circ \cdots\circ F^K$ la fonction de transfert associ\'ee au r\'eseau global. A ce titre, si $\boldsymbol{x}\in \mathcal{X}$, on pourra noter $\widehat{ \boldsymbol{y} }= F_W(\boldsymbol{x})$.\\
\end{itemize}

La Figure \ref{fig:nn} permet d'illustrer les notations pr\'esent\'ees ici\footnote{Source: {\sffamily http://intelligenceartificielle.org}.}. Chaque cercle repr\'esente un neurone \'el\'ementaire. Chaque rectangle englobant plusieurs cercles repr\'esente une couche. On parle de couche d'entr\'ee pour la premi\`ere couche prenant en « input » les observation $\boldsymbol{x}\in \mathcal{X}$, de couche de sortie pour la couche fournissant en « output » la pr\'ediction $\hat{\boldsymbol{y}} \in \mathcal{Y}$. Les autres couches sont couramment appel\'ees couches cach\'ees. 

Un r\'eseau de neurones multicouches est donc \'egalement un mod\`ele semi-param\'etrique dont les param\`etres sont l'ensemble des composantes des matrices $W_k$ pour tout entier $k$ de $\lbrace 1,\cdots, K \rbrace$. Chaque fonction d'activation associ\'ee \`a chaque neurone (chaque cercle de la Figure \ref{fig:nn}) est \`a d\'eterminer par l'utilisateur.

%\subsubsection{Le choix de la fonction de perte pour le calibrage des coefficients du r\'eseau de neurones}\label{nnet_modele}

Une fois que les param\`etres \`a calibrer du mod\`ele sont identifi\'es (ici les r\'eels constituant les matrices $W_k$ pour chaque couche $k \in \{1,\cdots,K \}$) , il est n\'ecessaire de fixer une fonction de perte $\ell$. En effet, on rappelle que l'objectif de l'apprentissage supervis\'e sur une base d'apprentissage de $n \in \mathbb{N}^*$ couples $(y_i,\boldsymbol{x}_i) \in \mathcal{Y}\times \mathcal{X}$ est de minimiser le risque empirique : 
$$\widehat{\mathcal{R}}_n(F_W) = \frac{1}{n} \sum_{i=1}^n \ell(y_i,F_W(\boldsymbol{x}_i))$$
        
\revision{Afin d'illustrer les propos, intéressons nous à l'exemple suivant qui illustrera également la démarche opérée. Supposons que nous observons un phénomène $y$ aux travers de $n$ observations $y_i \in [-1,1]$. On souhaiterait expliquer ce phénomène à partir des variables explicatives $\bx$ que l'on suppose à valeurs réelles. La « théorie de l'approximation universelle » nous indique qu'un réseau à une couche de neurones devrait permettre de modéliser le phénomène (sous hypothèse qu'il soit continue). On note toutefois que ce théorème ne donne pas de vitesse de convergence. Il est alors laissé à l'utilisateur le choix de la structure. Ainsi par exemple une première structure pourrait être un simple neurone dont la fonction d'activation serait la fonction tangente hyperbolique.}

\revision{On aurait ainsi comme premier modèle:
$$y_1 = \tanh(w_0 + w_1 x) $$ 
où les paramètres $w_0,~w_1$  sont les paramètres à optimiser de sorte que sur les données d'apprentissage, le risque empirique soit minimal. 

 Si l'on suit toujours la philosophie du théorème d'approximation universelle, en ajoutant plusieurs neurones, l'erreur est censée diminuer. Cependant,  ne connaissant pas la fonction à estimer, on ne peut l'observer qu'aux travers de l'échantillon. Ainsi, mécaniquement, on s'attend  à ce que plus on ajoute de paramètres, plus l'erreur sur la base d'apprentissage diminue. L'analyse de l'erreur sur la base de test permet alors d'évaluer notre capacité à généraliser (cf partie précédente).
 
On peut ainsi s'intéresser à un second modèle qui cette fois utilise plusieurs neurones. Par exemple, considérons le modèle
$$y_2 = w_a \tanh(w_0 + w_1 x) + w_b \tanh(w_2 + w_3 x) + w_c \tanh(w_4 + w_5 x)  $$ 
où les paramètres $w_0,..,w_5$ ainsi que $w_a, w_b, w_c$  sont les paramètres à optimiser. %La figure \ref{illapprox} permet de comparer les deux modélisations. 
Calibrer un réseaux de neurones revient alors à réitérer ces étapes de modification de la structure jusqu'à minimisation du risque sur la base de test.} 
 
%\begin{figure}[h]
%    \centering
%    \includegraphics[scale=0.5]{nnet_exempleTanh.png}
%    \caption{Illustatrion du théorème d'approximation universelle: $y$ %(fonction cible), $y_1$ (1 neurone) et $y_2$ (3 neurones)}
%    \label{illapprox}
%\end{figure}

Pour une structure de r\'eseau de neurones fix\'ee (c'est-\`a-dire nombre de couches, nombre de neurones par couches et fonctions d'activation fix\'es), le programme revient donc \`a d\'eterminer l'ensemble de param\`etres $W^*=(W_1,...,W_K)$ de sorte que :
$$W^{*} \in \underset{ W=(W_1,...,W_K)}{\text{argmin}}\left\lbrace \frac{1}{n} \sum_{i=1}^n \ell(y_i,F_W(\bx_i))
\right\rbrace.
$$
De cette formule apparaît l'importance du choix de la fonction $\ell$. Cette fonction de perte, elle quantifie l'erreur moyenne commise par notre mod\`ele $F_W$ sur la base d'apprentissage. \textit{A priori} $\ell$ peut être choisie arbitrairement. Cependant, dans l'optique de r\'esoudre un programme d'optimisation, on pr\'ef\'era des fonctions de coût sous-diff\'erentiables et convexes afin de garantir la convergence des algorithmes d'optimisation. 
Parmi les fonctions de perte classiques, en plus de la fonction de perte quadratique $\ell_2$ on retiendra la fonction dite « {\em Hinge} » - $ \ell(y,\hat{y})=\max(0, 1-y\hat{y})$ - ou la fonction dite logistique - $ \ell(y,\hat{y})=\log(1-e^{-y\hat{y}})$.

%Le tableau ci-dessous pr\'esente quelques exemples de fonctions de perte : 
%\begin{table}[h]
%\centering
%\caption{Exemple de fonctions de perte utilis\'ees, $y$ repr\'esente la "vraie" valeur et $\hat{y}$ la valeur pr\'edite par le mod\`ele}
%\vspace{0.5cm}
%\begin{tabular}{|c|c|}
%Fonction & D\'efinition   \\ \hline
% Hinge & $\hat{y} \rightarrow max(0, 1-y\hat{y})$  \\
% SquareLoss  & $\hat{y}\rightarrow ||y-\hat{y}||_2$   \\
% Logistic & $\hat{y}\rightarrow ln(1-e^{-y\hat{y}})$ 
%\end{tabular}
%\label{loss}
%\end{table}

%Cette liste est loin d'être exhaustive et peut être compl\'et\'ee selon les contraintes de l'utilisateur. \\

En d\'efinitive les r\'eseaux de neurones sont des mod\`eles semi-param\'etriques dont le nombre de param\`etres est croissant avec le nombre de couches et de neurones par couche. Il est laiss\'e \`a l'utilisateur de choisir les fonctions d'activation et la structure du r\'eseau. Ceci explique l'analogie avec la philosophie des modèles non-paramétriques faite auparavant.

%\subsubsection{Le deep learning}

Les r\'eseaux de neurones ont \'et\'e utilis\'es tr\`es t\^ot en \'economie et en finance, en particulier sur les d\'efauts d'entreprises - \citeNP{Tam} ou \citeNP{Altman} - ou plus r\'ecemment la notation de cr\'edit - \citeNP{Blanco} ou \citeNP{Khashman}. Cependant les structures telles que pr\'esent\'ees pr\'ec\'edemment sont g\'en\'eralement limit\'ees. L'apprentissage profond (ou « {\em deep learning} ») caract\'erise plus particuli\`erement des r\'eseaux de neurones plus complexes (parfois plus d'une dizaine de couches avec parfois des centaines de neurones par couche). Si aujourd'hui ces structures sont tr\`es populaires en analyse du signal (image, texte, son) c'est qu'elles sont capables \`a partir d'une quantit\'e d'observations tr\`es importante d'extraire des informations que l'humain ne peut percevoir et de faire face \`a des probl\`emes non lin\'eaires, comme le rappelle \citeNP{LeCun}. 

L'extraction d'informations peut, par exemple, se faire grâce \`a la convolution. Proc\'ed\'e non supervis\'e, il a permis notamment d'obtenir d'excellente performance dans l'analyse d'image. Techniquement, cela peut s'apparenter \`a une transformation \`a noyaux (comme utilis\'e dans les techniques SVM). Si une image peut être perçue comme une matrice dont chaque coordonn\'ee repr\'esente un pixel, une convolution reviendrait \`a appliquer une transformation sur un point (ou une zone) de cette matrice g\'en\'erant ainsi une nouvelle donn\'ee. Le proc\'ed\'e peut ainsi être r\'ep\'et\'e en appliquant des transformations diff\'erentes (d'où la notion de couches convolutives). Le vecteur final obtenu peut alors enfin alimenter un mod\`ele neuronal comme introduit dans le paragraphe pr\'ec\'edant. En fait, plus g\'en\'eralement, une couche de convolution peut être perçue comme un filtre qui permet de transformer la donn\'ee initiale.

%\begin{figure}[h]
%\centering
%\caption{Sch\'ema illustrant une convolution appliqu\'ee \`a une image 'x'}
%\vspace{0.5cm}
%\includegraphics[scale=0.45]{convolution}
%\label{convolution}
%\end{figure}

Une explication intuitive pour laquelle l'apprentissage approfondi, en particulier les r\'eseaux nerveux profonds, est si puissant pour d\'ecrire des relations complexes dans les donn\'ees, c'est leur construction autour de l'approximation fonctionnelle simple et l'exploitation d'une forme de hi\'erarchie, comme le note \citeNP{Lin}.
N\'eanmoins les mod\`eles de type « {\em deep learning} » sont plus difficiles \`a appr\'ehender car ils n\'ecessitent beaucoup de jugement empirique. En effet, si aujourd'hui les \revision{biblioth\`eques} open sources (keras, torch, etc.) permettent de parall\'eliser plus facilement les calculs en utilisant par exemple les GPU (\textit{Graphical Processor Units}), il reste n\'eanmoins \`a l'utilisateur de d\'eterminer la structure du r\'eseau de neurones le plus appropri\'e.

\subsection{Support Vecteurs Machine}\label{sec:SVM}

Comme nous l'avions noté auparavant, dans les problèmes de classification en apprentissage machine (comme en traitement du signal) on pr\'ef\'erera avoir des observations dans l'ensemble $\{-1,+1\}$ (plutôt que $\{0,1\}$, comme en économétrie). Avec cette notation, \citeNP{Cortes} ont pos\'e les bases th\'eorique des mod\`eles dit {\sc svm}, proposant une alternative aux r\'eseaux de neurones alors tr\`es populaires comme algorithme de classification dans la communaut\'e de l'apprentissage machine. L'id\'ee initiale des m\'ethodes de « {\em Support Vectors Machine} » ({\sc svm}) consiste \`a trouver un hyperplan s\'eparateur divisant l'espace en deux ensembles de points le plus homog\`ene possible (i.e. contenant des labels identiques). En dimension deux, l'algorithme consiste \`a d\'eterminer une droite s\'eparant l'espace en deux zones les plus homog\`enes possibles. La r\'esolution de ce probl\`eme poss\'edant parfois une infinit\'e de solution (il peut en effet exister une infinit\'e de droites qui s\'eparent l'espace en deux zones distinctes et homog\`enes), on rajoute g\'en\'eralement une contrainte suppl\'ementaire. L'hyperplan s\'eparateur doit se trouver le plus \'eloign\'e possible des deux sous-ensembles homog\`enes qu'il engendre. On parlera ainsi de marge. L'algorithme ainsi d\'ecrit est alors un {\sc svm} lin\'eaire \`a marge.

\begin{figure}[h]
\centering
\caption{Sch\'ema d'illustration d'un SVM \`a marge, Vert (2017).}
\vspace{0.5cm}
\includegraphics[scale=0.45]{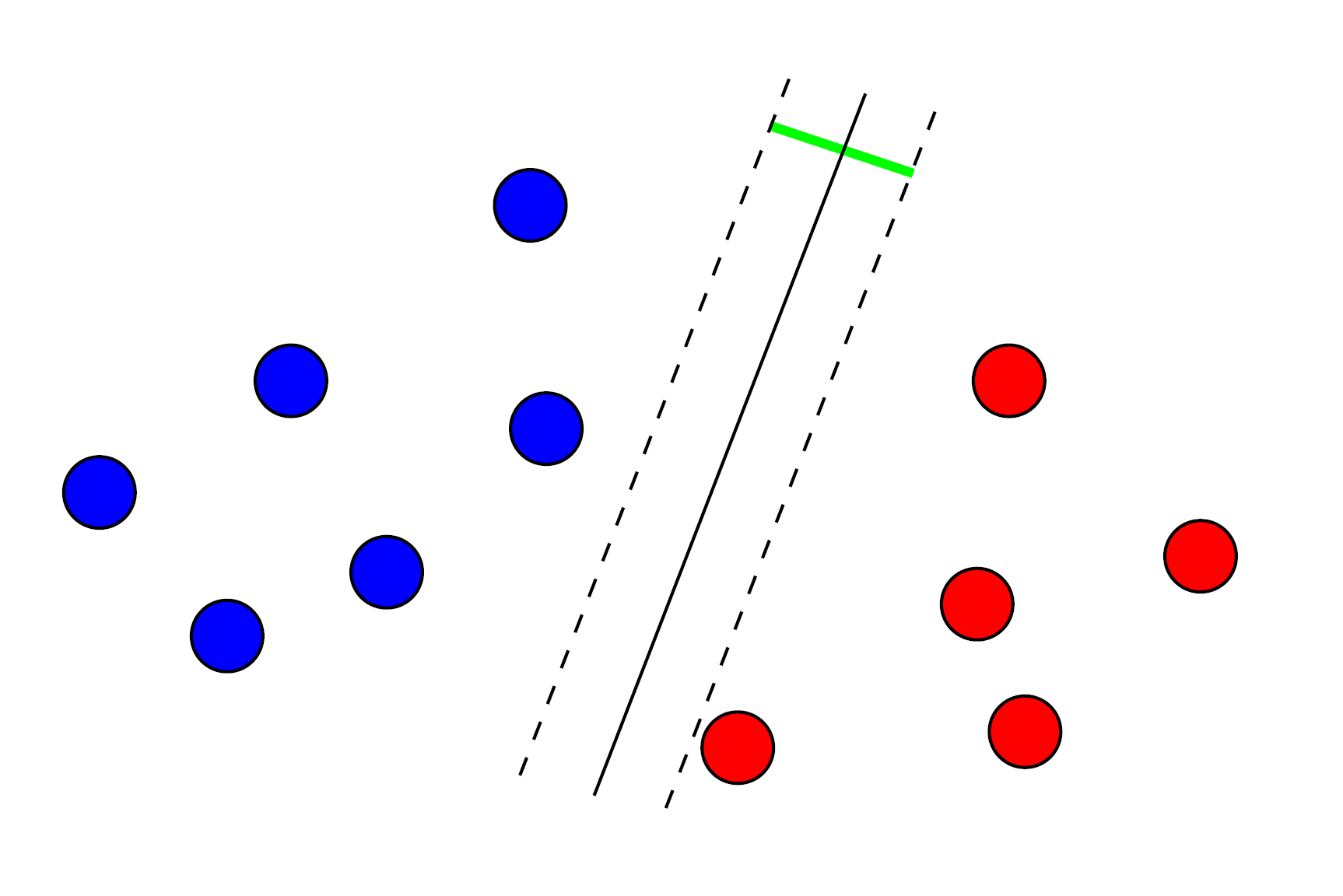}
\label{SVM}
\end{figure}

Si un plan peut être caract\'eris\'e enti\`erement par un vecteur directeur $\boldsymbol{w}$ orthogonal \`a ce dernier et une constante $b$, appliquer un algorithme SVM \`a un ensemble de $n\in \mathbb{N}^*$ points $\bx_i$ de $\mathbb{R}^p$ labellis\'es par $y_i \in \{-1,1\}$ revient alors \`a r\'esoudre un programme d'optimisation sous contrainte similaire \`a celui d'un {\sc lasso} (distance quadratique sous contrainte linéaire). Plus particuli\`erement, on sera amen\'e \`a r\'esoudre :
$$ (\boldsymbol{w}^\star,b^\star) = \underset{\boldsymbol{w},b}{\text{argmin}}\left\lbrace\|\boldsymbol{w}\|^2\right\rbrace=\underset{\boldsymbol{w},b}{\text{argmin}}\left\lbrace \boldsymbol{w}\transpose\boldsymbol{w}\right\rbrace, $$
$$\text{sous contrainte } ~\forall i \in \{ 1, \cdots, n \},~\left\lbrace 
\begin{array}{l}
\boldsymbol{\omega}^{\text{\sffamily T}}\boldsymbol{x}_i+b \geq +1\text{\quad lorsque \quad }y_i=+1\\
\boldsymbol{\omega}^{\text{\sffamily T}}\boldsymbol{x}_i+b \leq -1\text{\quad lorsque \quad}y_i=-1\\
\end{array}
\right. $$ 

La contrainte peut être relâch\'ee en autorisant que dans un sous-ensemble, un point puisse ne pas être du même label que la majeure partie des points de ce sous-ensemble \`a condition de ne pas être trop loin de la fronti\`ere. C'est ce qu'on appelle les SVM lin\'eaire \`a marge l\'eg\`ere (\textit{soft margin}). \revision{De mani\`ere heuristique, comme en pratique, bien souvent, on ne peut pas avoir $y_i(\boldsymbol{w}\transpose\boldsymbol{x}_i +b)-1  \geq 0$ pour tout $i\in\{ 1, \cdots, n \}$, on rel\^ache en introduisant des variables positives $\boldsymbol{\xi}$ telle que 
\begin{equation}\label{eq:slack}
\left\lbrace 
\begin{array}{l}
\boldsymbol{\omega}^{\text{\sffamily T}}\boldsymbol{x}_i+b \geq +1-\xi_i\text{\quad lorsque \quad}y_i=+1\\
\boldsymbol{\omega}^{\text{\sffamily T}}\boldsymbol{x}_i+b \leq -1+\xi_i\text{\quad lorsque \quad}y_i=-1\\
\end{array}
\right.
\end{equation}
avec $\xi_i\geq 0$. On a une erreur de classification si $\xi_i>1$, et on va alors introduire une p\'enalit\'e, un co\^ut \`a payer pour chaque erreur commise. On cherche alors \`a r\'esoudre un probl\`eme quadratique
$$
\min\left\lbrace \frac{1}{2}\boldsymbol{\omega}^{\text{\sffamily T}}\boldsymbol{\omega}+{C} \boldsymbol{1}^{\text{\sffamily{T}}}\boldsymbol{1}_{\boldsymbol{\xi}>1} \right\rbrace
$$
sous la contrainte (\ref{eq:slack}), qui pourra \^etre r\'esolu de mani\`ere num\'erique tr\`es efficacement par descente de coordonn\'ees (décrit auparavant).}

S'il n'est pas possible de s\'eparer les points, une autre astuce possible consiste \`a les transformer dans une dimension sup\'erieure, de sorte que les donn\'ees deviennent alors lin\'eairement s\'eparables. Trouver la bonne transformation qui s\'epare les donn\'ees est toutefois tr\`es difficile. Cependant, il existe une astuce math\'ematique pour r\'esoudre ce probl\`eme avec \'el\'egance, en d\'efinisant les transformations $T (\cdot)$ et les produits scalaires via un noyau
$ K (\boldsymbol {x} _1, \boldsymbol {x} _2) = \langle T (\boldsymbol {x} _1),T (\boldsymbol {x} _2)\rangle $.
L'un des choix les plus courants pour une fonction de noyau est la fonction de base radiale (noyau gaussien)
$ K (\boldsymbol {x} _1, \boldsymbol {x} _2) = \exp \big (-\Vert \boldsymbol{x} _1- \boldsymbol {x} _2) \Vert ^ 2 \big) $.
Il n'existe n\'eanmoins pas de r\`egles \`a ce jour permettant de choisir le « meilleur » noyau. Comme mentionné au début de la section précédante, cette technique est basé sur de la minimisation de distance, et il n'a aucune prévision de la probabilité d'être positif ou négatif (mais une interprétation probabiliste est néanmoins possible, comme le montre \citeNP{Grandvalet}, par exemple).

%\citeNP{Grandvalet} propose une interpr\'etation probabiliste en voyant la fonction objectif comme l'oppos\'e d'une log-vraisemblance , alors que \citeNP{Platt} construit une probabilit\'e \`a l'aide d'une fonction logistique.

\subsection{Arbres, Bagging et For\^ets Al\'eatoires}\label{sec:arbres:forets}

Les arbres de classification ont \'et\'e introduits dans \citeNP{Quinlan} mais c'est surtout \citeNP{Breiman} qui a assur\'e la popularit\'e de l'algorithme. On parle de mod\`ele CART pour « {\em Classification And Regression Tree} ». 
L'id\'ee est de diviser cons\'ecutivement (par une notion de branchement) les donn\'ees d'entr\'ee jusqu'\`a ce qu'un crit\`ere d'affectation (par rapport \`a la variable cible) soit atteint, selon une r\`egle pr\'ed\'efinie. 

L'intuition de la construction des arbres de classification est la suivante. L'entropie $H (\boldsymbol{x})$ est associ\'ee \`a la quantit\'e de d\'esordre dans les donn\'ees $\boldsymbol{x}$ par rapport aux modalit\'es prises par la variable de classification $y$, et chaque partition vise \`a r\'eduire ce d\'esordre. L'interpr\'etation probabiliste est de cr\'eer les groupes les plus homog\`enes possible, en r\'eduisant la variance par groupe (variance intra), ou de mani\`ere \'equivalente en cr\'eant deux groupes aussi diff\'erents que possible, en augmentant la variance entre les groupe (variance inter).
\`A chaque \'etape, nous choisissons la partition qui donne la plus forte r\'eduction de d\'esordre (ou de variance). L'arbre de d\'ecision complet se d\'eveloppe en r\'ep\'etant cette proc\'edure sur tous les sous-groupes, o\`u chaque \'etape $k$ aboutit \`a une nouvelle partition en 2 branches, qui subdivise notre ensemble de donn\'ees en 2. Enfin, on d\'ecide quand mettre fin \`a cette constitution de nouvelles branches, en proc\'edant \`a des affectations finales (n\oe{}uds dits foliaires). Il existe plusieurs options pour mettre fin \`a cette croissance. L'une est de construire un arbre jusqu'\`a ce que toutes les feuilles soient pures, c'est \`a dire compos\'ees d'une seule observation. Une autre option est de d\'efinir une r\`egle d'arrêt li\'ee \`a la taille, ou \`a la d\'ecomposition, des feuilles. Les exemples de r\`egles d'arrêt peuvent être d'une taille minimale (au moins 5 \'el\'ements par feuille), ou une entropie minimale. On parlera alors d'\'elagage de l'arbre: \revision{on laisse l'arbre grossir, puis on coupe certaines branches {\em a posteriori} (ce qui est diff\'erent de l'introduction d'un critère d'arrêt {\em a priori} au processus de croissance de l'arbre - par exemple en imposant une taille minimale aux feuilles, ou d'autres critères discutés dans \citeNP{Breiman}).}

\`A un n\oe{}ud donn\'e, constitu\'e de $n_0$ observations $(\boldsymbol{x}_i,y_i)$ avec $i\in\mathcal{I}_0$, on va couper en deux branches (une \`a gauche et une \`a droite), partitionnant ainsi $\mathcal{I}_0$ en $\mathcal{I}_{\text{g}}$ et $\mathcal{I}_{\text{d}}$. Soit $I$ le crit\`ere d'int\'er\^et, comme l'entropie du n\oe{}ud (ou plut\^ot du n\oe{}ud vu en tant que feuille):
$$
I(\boldsymbol{y}_0)=-n_0 p_0\log p_0 \text{\quad o\`u \quad}p_0=\frac{1}{n_0}\sum_{i\in\mathcal{I}_0}y_i, 
$$
ou la variance du n\oe{}ud:
$$
I(\boldsymbol{y}_0)=n_0 p_0(1- p_0) \text{\quad o\`u \quad}p_0=\frac{1}{n_0}\sum_{i\in\mathcal{I}_0}y_i, 
$$
ce dernier \'etant \'egalement l'indice \revision{d'impureté} de Gini. 

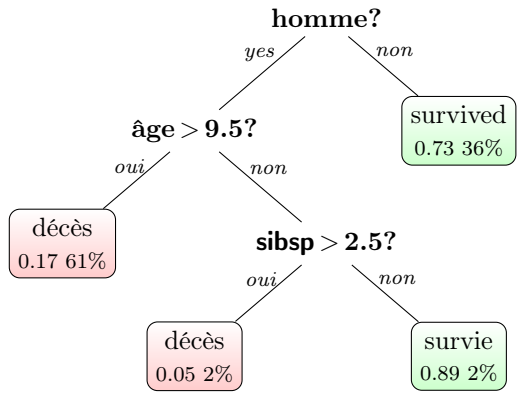
\begin{figure}[h]
\centering
\caption{Sch\'ema d'illustration d'un arbre de décision permettant de prédire le taux de survie d'un individu du Titanic.}
\vspace{0.5cm}
\begin{tikzpicture}[sibling distance=10em,
  good/.style = {shape=rectangle, rounded corners,
    draw, align=center, top color=white, bottom color=green!20},
  bad/.style = {shape=rectangle, rounded corners,
    draw, align=center, top color=white, bottom color=red!20}
]]
  \node(A) {\bf homme?}
    child { node(B1) {\bf \^age$\,>\,$9.5? } 
	child { node[bad](C1) {d\'ec\`es \\ {\footnotesize 0.17 61\%}} }
	child { node(C2) {\bf {\sffamily sibsp}$\,>\,$2.5?} 
		child { node[bad](D1) {d\'ec\`es \\ {\footnotesize 0.05 2\%}}} 
		child { node[good](D2) {survie \\ {\footnotesize 0.89 2\%}}} 
}
}
    child { node[good](B2) {survived\\ {\footnotesize 0.73 36\%}} };
  % Labels
  \begin{scope}[nodes = {draw = none}]
    \path (A)     -- (B1) node [near start, left]  {\footnotesize \em yes};
    \path (A)     -- (B2) node [near start, right] {\footnotesize \em non};
    \path (B1) -- (C1) node [near start, left] {\footnotesize \em oui};
    \path (B1) -- (C2) node [near start, right] {\footnotesize \em non};
    \path (C2) -- (D1) node [near start, left] {\footnotesize \em oui};
    \path (C2) -- (D2) node [near start, right] {\footnotesize \em non};
 \end{scope}
\end{tikzpicture}
\label{cart_titanic}
\end{figure}

On partitionnera entre la branche gauche et la branche droite si le gain $I(\boldsymbol{y}_0)-\big[I(\boldsymbol{y}_{\text{g}})+I(\boldsymbol{y}_{\text{d}})\big]$ 
est suffisamment important. Lors de la construction des arbres, on va chercher la partition qui donne le gain le plus important possible. Ce probl\`eme combinatoire \'etant complexe, le crit\`ere sugg\'er\'e par \citeNP{Breiman} est de consid\'erer un d\'ecoupage suivant une des variables, avec $\mathcal{I}_{\text{g}}=\lbrace i\in\mathcal{I}_0:x_{k,i}<s\rbrace$ et $\mathcal{I}_{\text{d}}=\lbrace i\in\mathcal{I}_0:x_{k,i}>s\rbrace$, pour une variable $k$ et un seuil $s$ (si la variable est continue, sinon on consid\`ere des regroupements de modalit\'es pour des variables qualitatives). 

Les arbres de d\'ecision ainsi d\'ecrits sont simples \`a obtenir et faciles \`a interpr\'eter (comme le montre la Figure \ref{cart_titanic} sur les donn\'ees du Titanic\footnote{Ce jeu de donn\'ees, contenant des informations sur tous les passagers (et membres d'\'equipage) du Titanic, dont la variable $y$ indiquant si la personne a surv\'ecu a \'et\'e abondamment utilis\'e pour illustrer les techniques de classification, voir {\sffamily https://www.kaggle.com/c/titanic/data}.}), mais ils sont peu robustes, et leur pouvoir pr\'edictif est souvent tr\`es faible, en particulier si l'arbre est tr\`es profond.
Une id\'ee naturelle est de d\'evelopper un ensemble de mod\`eles d'arbres \`a peu pr\`es ind\'ependants, qui pr\'edisent conjointement mieux qu'un mod\`ele d'arbre unique. On va utiliser le bootstrap, en tirant (avec remise) $n$ observations parmi $\{(\bx_i,y_i)\}$. Ces \revision{ensembles d'arbres - naturellement appel\'es} « for\^ets » - une fois agr\'eg\'es donnent souvent de bien meilleurs r\'esultats que les arbres isol\'es, mais elles sont difficiles \`a interpr\'eter. Ces techniques ressemblent toutefois beaucoup \`a ce qui est fait lorsque l'on utilise les techniques de bootstrap en r\'egression (par exemple pour construire des tubes de confiance dans une r\'egression fonctionnelle).

Le principe du « {\em bagging} », pour « {\em bootstrap aggregating} »,  consiste \`a g\'en\'erer des \'echantillons al\'eatoires, en tirant avec remise dans l'\'echantillon d'origine, comme pour le bootstrap. Chaque \'echantillon ainsi g\'en\'er\'e  permet d'estimer un nouvel arbre de classification, formant ainsi une for\^et \revision{d'arbres}. C'est l'aggr\'egation de tous ces arbres qui conduit \`a la pr\'evision. Le résultat global est moins sensible à l'échantillon initial et donne souvent de meilleurs résultats de pr\'evision.

Les for\^ets al\'eatoires, ou « {\em random forests} » \revision{reposent sur} le m\^eme principe que le « {\em bagging} », mais en plus, lors de la construction d'un arbre de classification, \`a chaque branche, un  \revision{sous-}ensemble de $m$ covariables est tiré al\'eatoirement. Autrement dit, chaque branche d'un arbre ne s'appuie pas sur le m\^eme ensemble de covariables. \revision{Cela permet d'amplifier la variabilit\'e entre les diff\'erents arbres et d'obtenir, au final, une forêt compos\'ee d'arbres moins corrélés les uns aux autres.} 

\subsection{S\'election de mod\`ele de classification}\label{sec:selection:modele}

\'Etant donn\'e un mod\`ele $m(\cdot)$ approchant $\esp[Y\vert \bX=\bx]$, et un seuil $s\in[0,1]$, posons
$$
\widehat{y}^{(s)}=\boldsymbol{1}[m(\bx)>s]
=
\left\lbrace
\begin{array}{l}
1 \text{ si }m(\bx)>s \\
0 \text{ si }m(\bx)\leq s
\end{array}
\right.
$$
La matrice de confusion est alors le tableau de contingence associ\'e aux comptages $\boldsymbol{N}=[N_{u,v}]$ avec
$$
N_{u,v}^{(s)}=\sum_{i=1}^n \boldsymbol{1}(\widehat{y}^{(s)}_i=u,y_j=v)
$$
pour $(u,v)\in\lbrace0,1\rbrace$. La Table \ref{Tab:confusion} pr\'esente un tel tableau, avec le nom de chacun des \'el\'ements : TP ({\em true positive}) sont les vrais positifs, correspondant aux $1$ pr\'edit en $1$, 
TN ({\em true negative}) sont les vrais n\'egatifs, correspondant aux $0$ pr\'edit en $0$, FP ({\em false positive}) sont les faux positifs, correspondant aux $0$ pr\'edit en $1$, et enfin FN ({\em false negative}) sont les faux n\'egatifs, correspondant aux $1$ pr\'edit en $0$).

\begin{table}
\begin{center}
\begin{tabular}{|c|cc|c|}\hline
                & $y=0$ & $y=1$ & \\ \hline
$\widehat{y}_s=0$ &  {TN}${}_s$   & {FN}${}_s$     & TN${}_s$+FN${}_s$ \\
$\widehat{y}_s=1$ &  {FP}${}_s$   & {TP}${}_s$     & FP${}_s$+TP${}_s$ \\ \hline
                &  TN${}_s$+FP${}_s$   & FN${}_s$+TP${}_s$    & $n$ \\ \hline
\end{tabular}
\end{center}
\caption{Matrice de confusion, ou tableau de contingence pour un seuil $s$ donn\'e.}\label{Tab:confusion}
\end{table}

Plusieurs quantit\'es sont d\'eriv\'ees de ce tableau. La sensibilit\'e correspond \`a la probabilit\'e de pr\'edire $1$ dans la population des $1$, ou taux de vrais positifs. La sp\'ecificit\'e est  la probabilit\'e de pr\'edire $0$ dans la population des $0$ ou taux de vrais n\'egatifs. On s'int\'eressera toutefois davantage au taux de faux n\'egatifs,  c'est \`a dire la probabilit\'e de pr\'edire $1$ dans la population des $0$. La repr\'esentation de ces deux valeurs lorsque $s$ varie donne la courbe ROC (« {\em receiver operating characteristic} ») :
$$
\text{ROC}_s=\left(\frac{\text{FP}_s}{\text{FP}_s+\text{TN}_s},\frac{\text{TP}_s}{\text{TP}_s+\text{FN}_s}\right)=(\text{\sf sensibility}_s,1-\text{\sf specificity}_s) \text{ pour }s\in[0,1].
$$
Une telle courbe est pr\'esent\'ee dans la partie suivante, sur des donn\'ees r\'eelles.

Les deux grandeurs intensivement utilis\'ees en {\em machine learning} sont l'indice $\kappa$, qui compare la précision observ\'ee avec celle esp\'er\'ee, avec un mod\`ele al\'eatoire (tel que d\'ecrit dans \citeNP{LandisKoch}) et l'AUC correspondant \`a l'aire sous la courbe ROC. Pour le premier indice, une fois choisi $s$, notons $\boldsymbol{N}^\perp$ le tableau de contingence correspond aux cas ind\'ependants (d\'efini \`a partir de $\boldsymbol{N}$ dans le test d'ind\'ependance du chi-deux).
On pose alors
$$
\text{pr\'ecision totale}=\frac{\text{TP+TN}}{n}%
$$
alors que
$$
\text{pr\'ecision al\'eatoire}=\frac{[\text{TN+FP}]\cdot[\text{TP+FN}] 
+[\text{TP+FP}]\cdot[\text{TN+FN}]}{n^2}
$$
On peut alors d\'efinir
$$
\kappa=\frac{\text{pr\'ecision totale}-\text{pr\'ecision al\'eatoire}}{1-\text{pr\'ecision al\'eatoire}}
$$
Classiquement $s$ sera fix\'e \'egal \`a 0.5, comme dans une classification bay\'esienne na\"ive, mais d'autres valeurs peuvent \^etre retenues, en particulier si les deux erreurs ne sont pas symm\'etriques (nous reviendrons sur ce point dans un exemple par la suite).

Il existe des compromis entre des mod\`eles simples et complexes mesur\'es par leur nombre de param\`etres (ou plus g\'en\'eralement les degr\'es de libert\'e) en mati\`ere de performance et de coût. Les mod\`eles simples sont g\'en\'eralement plus faciles \`a calculer, mais peuvent conduire \`a des ajustements plus mauvais (avec un biais \'elev\'e par exemple). Au contraire, les mod\`eles complexes peuvent fournir des ajustements plus pr\'ecis, mais risquent d'être coûteux en termes de calcul. En outre, ils peuvent surpasser les donn\'ees ou avoir une grande variance et, tout autant que des mod\`eles trop simples, ont de grandes erreurs de test. Comme nous l'avons rappel\'e auparavant, dans l'apprentissage machine, la complexit\'e optimale du mod\`ele est d\'etermin\'ee en utilisant le compromis de biais-variance.

\subsection{De la classification \`a la r\'egression}

Comme nous l'avons rappel\'e en introduction, historiquement, les m\'ethodes de {\em machine learning} se sont orient\'ees autour des probl\`emes de classification (avec \'eventuellement plus de 2 modalit\'es\footnote{Par exemple dans le cas de  reconnaissance de lettres ou de chiffres}), et assez peu dans le cas o\`u la variable d'int\'er\^et $y$ est continue. N\'eanmoins, il est possible d'adapter quelques techniques, comme les arbres et les for\^ets al\'eatoires, le boosting, ou les r\'eseaux de neurones.

Pour les arbres de r\'egression, \citeNP{Morganetal} ont propos\'e la m\'ethode AID, bas\'ee sur la formule de d\'ecomposition de la variance de l'\'equation (\ref{eq:pyt}), avec un algorithme proche de celui de la m\'ethode CART d\'ecrite auparavant. Dans le contexte de la classification, on calculait, \`a chaque n\oe{}ud (dans le cas de l'indice \revision{d'impuret\'e} de Gini) en sommant sur la feuille de gauche $\{x_{k,i}<s\}$ et celle de droite $\{x_{k,i}>s\}$
$$
I=\sum_{i:x_{k,i}<s} \overline{y}_{\text{g}} \big(1-\overline{y}_{\text{g}}\big)
+\sum_{i:x_{k,i}>s} \overline{y}_{\text{d}} \big(1-\overline{y}_{\text{d}}\big)
$$
o\`u $\overline{y}_{\text{g}}$ et $\overline{y}_{\text{d}}$ d\'esignent les fr\'equences de $1$ dans la feuille de gauche et de droite, respectivement. Dans le cas d'un arbre de r\'egression, on utilisera
$$
I=\sum_{i:x_{k,i}<s} \big(y_i-\overline{y}_{\text{g}}\big)^2
+\sum_{i:x_{k,i}>s}  \big(y_i-\overline{y}_{\text{d}}\big)^2
$$
qui va correspondre \`a la somme (pond\'er\'ee) des variances intra. Le partage optimal sera celui qui aura le plus de variance intra (on veut les feuilles les plus homog\`enes possibles) ou de mani\`ere \'equivalente, on veut maximiser la variance intra.

Dans le contexte des for\^ets al\'eatoires, on utilise souvent un crit\`ere majoritaire en classification (la classe pr\'edite sera la classe majoritaire dans une feuille), alors que pour la r\'egression, on utilise la moyenne des pr\'edictions, sur tous les arbres.

%\subsection{La notion d'apprentissage}

Dans la partie pr\'ec\'edente, nous avons pr\'esent\'e la dimension « apprentissage » du {\em machine learning} en pr\'esentant le {\em boosting}. Dans un contexte de r\'egression (variable $y$ continue), l'id\'ee est de cr\'eer une succession de mod\`eles en \'ecrivant l'\'equation (\ref{eqboot}) sous la forme :
$$
m^{(k)}(\boldsymbol{x})=m^{(k-1)}(\boldsymbol{x})+\alpha_k \underset{h\in\mathcal{H}}{\text{argmin}}\left\lbrace \sum_{i=1}^n (y_i,m^{(k-1)}(\boldsymbol{x})+h(\boldsymbol{x}))^2\right\rbrace
$$
o\`u $\alpha_k$ est un param\`etre de « {\em shrinkage} », o\`u le second terme correspond \`a un arbre de r\'egression, sur les r\'esidus, $y_i-m^{(k-1)}(\boldsymbol{x}_i)$.

Mais il existe d'autres techniques permettant d'apprendre de mani\`ere s\'equentielle. Dans un mod\`ele additif ({\sc gam}) on va chercher une \'ecriture de la forme
$$
m(\boldsymbol{x})=\sum_{j=1}^p m_j(x_j)=m_1(x_1)+\cdots+m_p(x_p)
$$
L'id\'ee de la poursuite de projection repose sur une d\'ecomposition non pas sur les variables explicatives, mais sur des combinaisons lin\'eaires. On va ainsi consid\'erer un mod\`ele
$$
m(\boldsymbol{x})=\sum_{j=1}^k g_j(\boldsymbol{\omega}_j\transpose\boldsymbol{x})=g_1(\boldsymbol{\omega}_1\transpose\boldsymbol{x})+\cdots+g_k(\boldsymbol{\omega}_k\transpose\boldsymbol{x}).
$$
Tout comme les mod\`eles additifs, les fonctions $g_1,\cdots,g_k$ sont \`a estimer, tout comme les directions $\boldsymbol{\omega}_1,\cdots,\boldsymbol{\omega}_k$. Cette \'ecriture est relativement g\'en\'erale, et permet de tenir compte d'int\'eractions et d'effets crois\'es (ce que nous ne pouvions pas faire avec les mod\`eles additifs qui ne tiennent compte que de non-lin\'earit\'es). Par exemple en dimension 2, un effet multiplicatif $m(x_1,x_2)=x_1\cdot x_2$ s'\'ecrit
$$
m(x_1,x_2)=x_1\cdot x_2=\frac{(x_1+x_2)^2}{4}-\frac{(x_1-x_2)^2}{4}
$$
autrement dit $g_1(x)=x^2/4$, $g_1(x)=-x^2/4$, $\boldsymbol{\omega}_1=(1,1)\transpose$ et $\boldsymbol{\omega}_1=(1,-1)\transpose$. Dans la version simple, avec $k=1$, avec une fonction de perte quadratique, on peut utiliser un d\'eveloppement de Taylor pour approcher $[y_i-g(\boldsymbol{\omega}\transpose\boldsymbol{x}_i)]^2$, et construire classiquement un algorithme it\'eratif. Si on dispose d'une valeur initiale $\boldsymbol{\omega}_0$, notons que
$$
\sum_{i=1}^n[y_i-g(\boldsymbol{\omega}\transpose\boldsymbol{x}_i)]^2
\approx 
\sum_{i=1}^n g'(\boldsymbol{\omega}_0\transpose\boldsymbol{x}_i)^2\left[\boldsymbol{\omega}\transpose\boldsymbol{x}_i+\frac{y_i-g(\boldsymbol{\omega}_0\transpose\boldsymbol{x}_i)}{g'(\boldsymbol{\omega}_0\transpose\boldsymbol{x}_i)}-
\boldsymbol{\omega}\transpose\boldsymbol{x}_i\right]^2
$$
qui correspondrait \`a l'approximation dans les mod\`eles lin\'eaires g\'en\'eralis\'es sur la fonction $g(\cdot)$ qui \'etait la fonction de lien (suppos\'ee connue). On reconnait  un probl\`eme de moindres carr\'es pond\'er\'es. La difficult\'e ici est que les fonctions $g_j(\cdot)$ sont inconnues.

\section{Applications}\label{sec:3:classif}

Les donn\'ees massives ont rendu n\'ecessaire le d\'eveloppement de techniques d'estimation permettant de pallier les limites des mod\`eles param\'etriques, jug\'es trop restrictifs, et  des mod\`eles non-param\'etriques classiques, dont l'estimation peut être difficile en pr\'esence d'un nombre \'elev\'e de variables. L'{\em apprentissage statistique}, ou {\em apprentissage machine}, propose de nouvelles m\'ethodes d'estimation non-param\'etriques, performantes dans un cadre g\'en\'eral et en pr\'esence d'un grand nombre de variables.\footnote{Entre autres, voir \citeNP{HastieEtal} et \citeNP{James}.} Toutefois, l'obtention d'une plus grande flexibilité s'obtient au prix d'un manque d'interpr\' etation qui peut \^etre important.

En pratique, une question importante est de savoir quel est le meilleur mod\`ele ? La r\'eponse \`a cette question d\'epend du probl\`eme sous-jacent. Si la relation entre les variables est correctement approxim\'ee par un mod\`ele lin\'eaire, un mod\`ele param\'etrique correctement sp\'ecifi\'e devrait \^etre performant. Par contre, si le modèle paramétrique n'est pas correctement spécifié, car la relation est fortement non-linéaire et/ou fait intervenir des effets croisés non-n\'egligeables, alors les méthodes statistiques issues du {\em machine learning} devraient \^etre plus performantes.

La bonne sp\'ecification d'un mod\`ele de r\'egression est une hypoth\`ese souvent pos\'ee, elle est rarement v\'erifi\'ee et justifi\'ee. Dans les applications qui suivent, nous montrons comment les m\'ethodes statistiques issues du {\em machine learning} peuvent \^etre utilis\'ees pour justifier la bonne sp\'ecification d'un mod\`ele de r\'egression param\'etrique, ou pour d\'etecter une mauvaise sp\'ecification. Des applications en classification sont pr\'esent\'ees dans un premier temps, sections \ref{appli1}, \ref{appli2} et \ref{appli3}. D'autres applications sont ensuite pr\'esent\'ees dans le contexte de r\'egression classique, sections \ref{appli4} et \ref{appli5}.

\subsection{Les ventes de si\`eges auto pour enfants (classification)}
\label{appli1}

Nous reprenons ici un exemple utilis\'e dans \citeNP{James}. Le \revision{jeu de donn\'ees} contient les ventes de si\`eges auto pour enfants dans 400 magasins ({\sf Sales}), ainsi que plusieurs variables, dont la qualit\'e de pr\'esentation en rayonnage ({\sf Shelveloc}, \'egal \`a « mauvais », « moyen », « bon ») et le prix ({\sf Price}).\footnote{C'est le \revision{jeu de donn\'ees} {\sffamily Carseats} de la \revision{biblioth\`eque} ISLR.} Une variable d\'ependante binaire est artificiellement cr\'ee, pour qualifier une forte vente ou non ({\sf High}=« oui » si $\textsf{ Sales}>8$ et \`a « non » sinon). Dans cette application, on cherche \`a \'evaluer les d\'eterminants d'un bon niveau de vente.

Dans un premier temps, on consid\`ere un mod\`ele de r\'egression lin\'eaire latent:
\begin{equation}
y^\star = \gamma + \bx\transpose\bbeta + \varepsilon, \qquad \varepsilon\sim G(0,1),
\label{eq:latent}
\end{equation}
où $\bx$ est compos\'e de $k$  variables explicatives, $\bbeta$ est un vecteur de $k$ param\`etres inconnus et $\varepsilon$ est un terme d'erreur {\em i.i.d.} avec une fonction de r\'epartition $G$ d'esp\'erance nulle et de variance \'egale \`a un. La variable d\'ependante $y^\star$ n'est pas observ\'e, mais seulement $y$, avec:
\begin{equation}
y=
\begin{cases}
1 & \text{si } y^\star > \xi, \\
0 & \text{si } y^\star \leq \xi. \\
\end{cases}
\end{equation}
On peut alors exprimer la probabilit\'e d'avoir $y$ \'egal \`a 1, comme suit~:
\begin{equation}
\mathbb{P}(Y=1)=G(\beta_0+\bx\transpose\bbeta)
\label{eq:logit}
\end{equation}
où $\beta_0=\gamma-\xi$.\footnote{$\mathbb{P}[Y=1]=\mathbb{P}[Y^\star>\xi]=\mathbb{P}[\gamma + \bx\transpose\bbeta + \varepsilon>\xi]=\mathbb{P}[ \varepsilon>\xi-\gamma - \bx\transpose\bbeta]=\mathbb{P}[\varepsilon<\gamma-\xi + \bx\transpose\bbeta]$. En posant $\gamma-\xi=\beta_0$, on obtient $\mathbb{P}[Y=1]=G(\beta_0+\bx\transpose\bbeta)$. En g\'en\'eral, on suppose que le terme d'erreur est de variance $\sigma^2$, auquel cas les param\`etres du mod\`ele (\ref{eq:logit}) deviennent $\beta_0/\sigma$ et $\bbeta/\sigma$, ce qui veut dire que les param\`etres du mod\`ele latent (\ref{eq:latent}) ne sont pas identifiables, ils sont estim\'es \`a un param\`etre d'\'echelle pr\`es.} L'estimation de ce mod\`ele se fait par maximum de vraisemblance en s\'electionnant {\em a priori} une loi param\'etrique $G$. Si on suppose que $G$ est la loi Normale, c'est un mod\`ele probit, si on suppose que $G$ est la loi logistique, c'est un mod\`ele logit. Dans un  mod\`ele  logit/probit, il y a deux sources potentielles de mauvaise sp\'ecification~:
\begin{itemize}
\item[$(i)$] la relation lin\'eaire $\beta_0+\bx\transpose\bbeta$ est mal sp\'ecifi\'ee
\item[$(ii)$] la loi param\'etrique utilis\'ee $G$ n'est pas la bonne
\end{itemize}
En cas de mauvaise sp\'ecification, de l'une ou l'autre sorte, l'estimation n'est plus valide. Le mod\`ele le plus flexible est le suivant~:
\begin{equation}
\mathbb{P}[Y=1|\boldsymbol{X}=\boldsymbol{x}]=G(h(\boldsymbol{x}))
\label{eq:logit:np}
\end{equation}
où $h$ est une fonction inconnue et $G$ une fonction de r\'epartition inconnue. Les mod\`eles de {\em bagging}, de for\^et al\'eatoire et de {\em boosting} permettent d'estimer ce mod\`ele  g\'en\'eral sans faire de choix {\em \`a priori} sur la fonction $h$  et sur la distribution $G$. L'estimation du mod\`ele logit/probit est n\'eanmoins plus performante si $h$ et $G$ sont correctement sp\'ecifi\'es.

\begin{figure}[t]
\begin{center}
\begin{minipage}{.6\linewidth}
\includegraphics[width=\textwidth,height=.35\textheight]{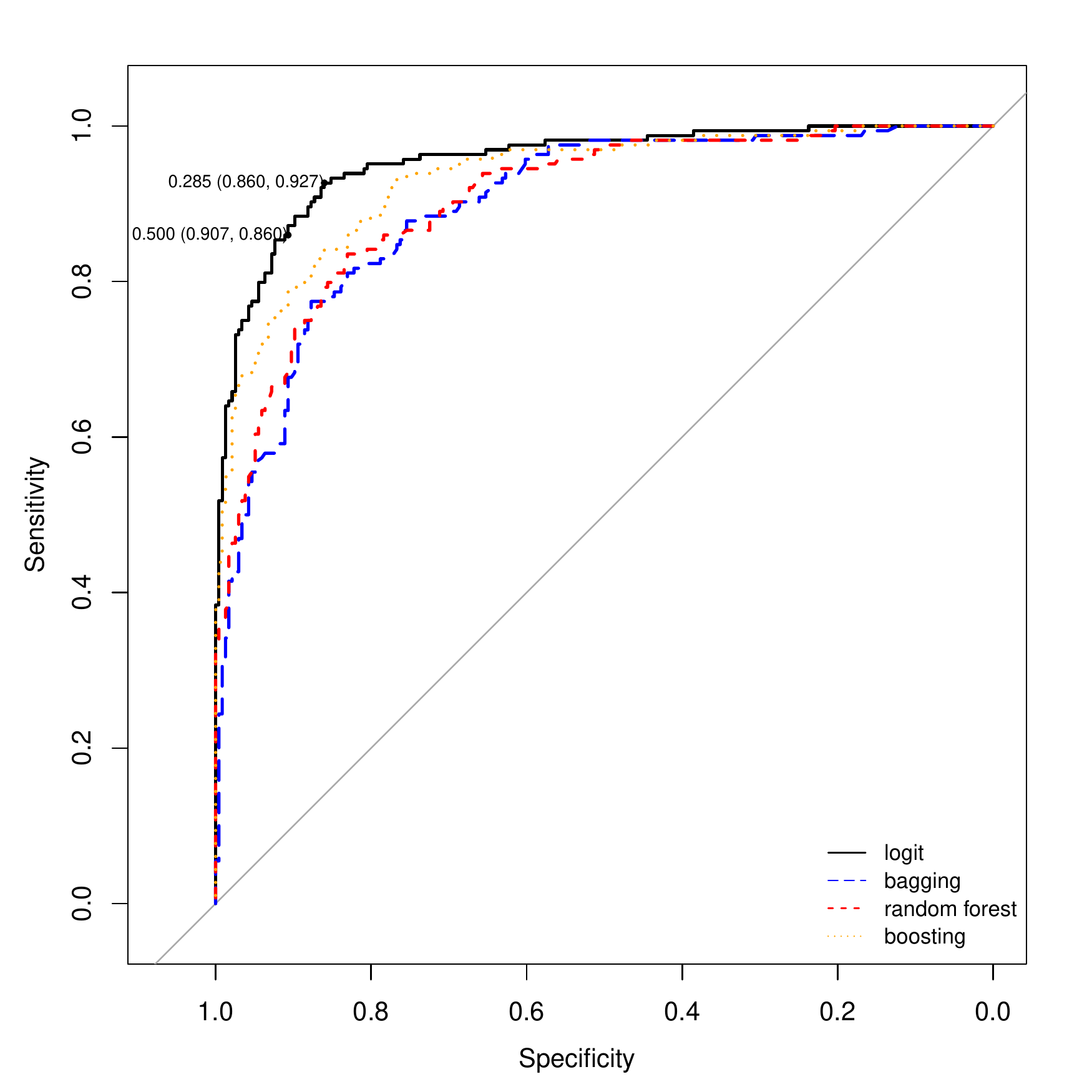}
\end{minipage}
\quad
\begin{tabular}{lr}
& AUC \\
\hline
logit &  0.9544\\
bagging & 0.8973\\
random forest & 0.9050 \\
boosting & 0.9313\\
\hline
\end{tabular}
\end{center}
\caption{Ventes de si\`eges auto: courbes ROC et aires sous la courbe (AUC).}
\label{fig:appli:roc1}
\end{figure}

Nous estimons le mod\`ele (\ref{eq:logit}) avec la loi logistique pour $G$, et le mod\`ele (\ref{eq:logit:np}) avec les m\'ethodes de {\em bagging}, de for\^et al\'eatoire et de {\em boosting}. Nous faisons une analyse de validation crois\'ee par 10 blocs. Les probabilit\'es individuelles des donn\'ees {\em out-of-sample}, c'est \`a-dire de chacun des blocs non-utilis\'ee pour l'estimation, sont utilis\'ees pour \'evaluer la qualit\'e de la classification. 

La figure \ref{fig:appli:roc1} pr\'esente la courbe ROC, ainsi que l'aire sous la courbe (AUC), pour les estimations logit, bagging, random forest et boosting. La {courbe ROC} est un graphique qui repr\'esente simultan\'ement la qualit\'e de la pr\'evision dans les deux classes, pour des valeurs diff\'erentes du seuil utilis\'e pour classer les individus (on parle de « {\em cutoff} »). Une mani\`ere naturelle de classer les individus consiste \`a les attribuer dans la classe pour laquelle ils ont la plus grande probabilit\'e estim\'ee. Dans le cas d'une variable binaire, cela revient \`a pr\'edire la classe d'appartenance pour laquelle la probabilit\'e estim\'ee est sup\'erieure \`a $0.5$. Mais un autre seuil pourrait être utilis\'e. Par exemple, dans la figure \ref{fig:appli:roc1}, un point de la courbe ROC du mod\`ele logit indique qu'en  prenant un seuil \'egal \`a 0.5, la r\'eponse "non" est correctement pr\'edite \`a 90.7\% ({\sf specificity}), et la r\'eponse « oui »  \`a  86\% ({\sf sensitivity}). Un autre  point indique qu'en prenant un seuil \'egal \`a 0.285, la r\'eponse « non » est correctement pr\'edite \`a 86\% ({\sf specificity}), et la r\'eponse « oui »  \`a  92.7\% ({\sf sensitivity}). Comme d\'ecrit auparavant, un mod\`ele de classification id\'eal aurait une courbe ROC de la forme $\Gamma$. Autrement dit, le meilleur mod\`ele est celui dont la courbe est au-dessus des autres. Un crit\`ere souvent utilis\'e pour s\'electionner le meilleur mod\`ele est celui dont l'aire sous la courbe ROC est la plus grande ($\text{AUC}$). L'avantage d'un tel crit\`ere est qu'il est simple \`a comparer et qu'il ne d\'epend pas du choix du seuil de classification.

Dans notre exemple, la courbe ROC du mod\`ele logit domine les autres courbes, et son aire sous la courbe est la plus grande (AUC=0.9544). Ces r\'esultats indiquent que ce mod\`ele  fournit les meilleures pr\'evisions de classification. N'\'etant domin\'e par aucun autre mod\`ele, ce constat sugg\`ere que le mod\`ele lin\'eaire logit est correctement sp\'ecifi\'e et qu'il n'est pas utile d'utiliser un mod\`ele plus g\'en\'eral et plus complexe.

\subsection{L'achat d'une assurance caravane (classification)}
\label{appli2}

Nous reprenons \`a nouveau un exemple utilis\'e dans \citeNP{James}. Le \revision{jeu de donn\'ees} contient 85 variables sur les charact\'eristiques d\'emographiques de 5822 individus.\footnote{C'est le \revision{jeu de donn\'ees} {\sffamily Caravan} de la \revision{biblioth\`eque} ISLR sous R.}  La variable d\'ependante ({\sf Purchase}) indique si l'individu a achet\'e une assurance caravane, c'est une variable binaire, \'egale \`a « oui » ou « non ». Dans le \revision{jeu de donn\'ees}, seulement 6\% des individus ont pris une telle assurance. Les classes sont donc fortement d\'esequilibr\'ees.

\begin{figure}[t]
\begin{center}
\begin{minipage}{.6\linewidth}
\includegraphics[width=\textwidth,height=.35\textheight]{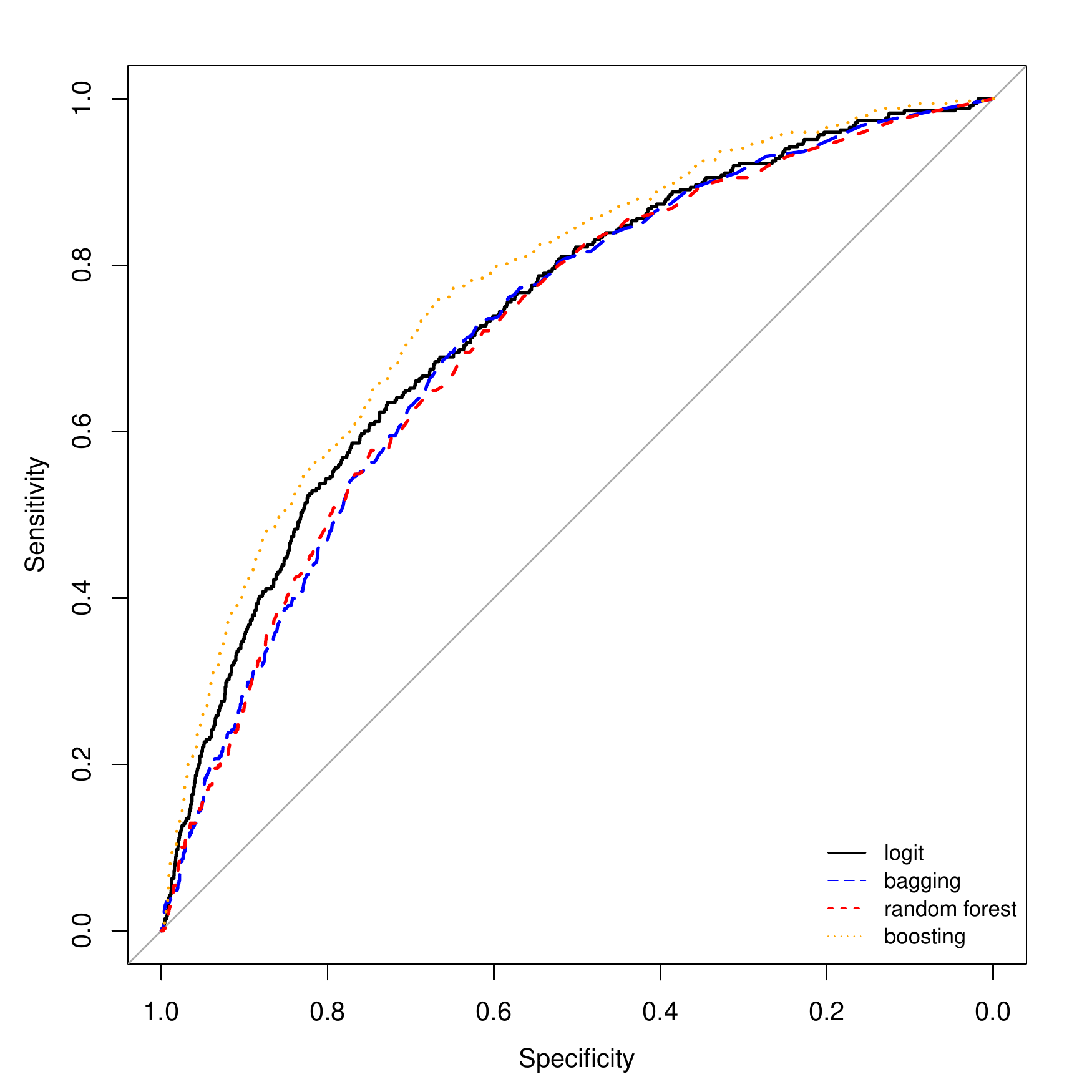}
\end{minipage}
\quad
\begin{tabular}{lr}
& AUC \\
\hline
logit &  0.7372\\
bagging & 0.7198\\
random forest & 0.7154 \\
boosting & 0.7691\\
\hline
\end{tabular}
\end{center}
\caption{Achat d'assurance: courbes ROC et aires sous la courbe  (AUC).}
\label{fig:appli:roc2}
\end{figure}

Nous estimons le mod\`ele (\ref{eq:logit}) avec la loi logistique et le mod\`ele (\ref{eq:logit:np}) avec les m\'ethodes {\em bagging}, for\^et al\'eatoire et {\em boosting} (les param\`etres de « {\em tuning} » sont ceux de \citeNP{James}, {\sffamily n.trees=1000} et {\sffamily shrinkage=0.01}).  Nous faisons une analyse de validation crois\'ee par 10 blocs. Les probabilit\'es individuelles des donn\'ees {\em out-of-sample}, c'est \`a-dire de chacun des blocs non-utilis\'ee pour l'estimation, sont utilis\'ees pour \'evaluer la qualit\'e de la classification.

La figure \ref{fig:appli:roc2} pr\'esente la courbe ROC, ainsi que l'aire sous la courbe (AUC), pour les estimations logit, bagging, random forest et boosting. La courbe du mod\`ele boosting domine les autres courbes, son aire sous la courbe est la plus grande (AUC=0.7691). Ces r\'esultats indiquent que le boosting  fournit les meilleures pr\'evisions de classification. Notons que, compar\'ees \`a l'exemple pr\'ec\'edent,  les courbes sont assez \'eloign\'ees de la forme en coude,  ce qui sugg\`ere que la classification ne sera pas aussi bonne.

Il faut faire attention aux r\'esultats d'une classification standard, c'est-\`a-dire avec un seuil de classification \'egal \`a 0.5, qui est souvent pris par d\'efaut dans les logiciels (la pr\'ediction de la r\'eponse de l'individu $i$ est « non » 1 si la probabilit\'e estim\'ee qu'il r\'eponde « non » est sup\'erieure \`a 0.5, sinon c'est « oui »). La partie gauche du tableau \ref{tab:appli:cutoff2} pr\'esente les taux de classifications correctes avec ce seuil ({\sf 0.5 cutoff}), pour les diff\'erentes m\'ethodes. Avec le meilleur mod\`ele et le seuil standard (boosting et seuil \`a 0.5), les r\'eponses « non » sont correctes \`a 99.87\% (sp\'ecificit\'e, {\sf specificity}) et les r\'eponses « oui » sont toutes fausses (sensitivit\'e, {\sf sensitivity}). Autrement dit, cela \'equivaut \`a utiliser un mod\`ele qui pr\'edit que personne n'ach\`ete d'assurance caravane. S\'electionner un tel mod\`ele est absurde pour l'analyste, qui est surtout int\'eress\'e par les 6\% des individus qui en ont pris une. Ce r\'esultat s'explique par la pr\'esence de classes fortement d\'es\'equilibr\'ees. En effet, dans notre exemple, en pr\'evoyant que personne n'ach\`ete d'assurance, on  fait « seulement » 6\% d'erreur. Mais ce sont des erreurs qui conduisent \`a ne rien expliquer.

\begin{table}[t]
\begin{center}
\begin{tabular}{lrrrrr}
\hline
& \multicolumn{2}{c}{0.5 cutoff} && \multicolumn{2}{c}{cutoff optimal}  \\
\cline{2-3}
\cline{5-6}
  & sp\'ecificit\'e & sensitivit\'e &&         sp\'ecificit\'e & sensitivit\'e  \\
\hline
logit             &  0.9967   &   0.0057    &&      0.7278  &    0.6351  \\
bagging      & 0.9779    &  0.0661      &&      0.6443 &     0.7069  \\
{\em random forest}      &  0.9892    &  0.0316      &&     0.6345   &   0.6954      \\
boosting    &   0.9987   &   0.0000    &&      0.6860   &  0.7385   \\
\hline
\end{tabular}
\end{center}
\caption{Achat d'assurance: sensibilit\'e au choix du seuil de classification.}
\label{tab:appli:cutoff2}
\end{table}	

Plusieurs m\'ethodes peuvent être utiles pour pallier \`a ce probl\`eme, li\'e aux classes fortement d\'esequilibr\'ees (pour plus d'informations, voir \citeNP{KuhnJohnson}, chapitre 16). Une solution simple consiste \`a utiliser un seuil de classification diff\'erent. La courbe ROC pr\'esente les r\'esultats en fonction de plusieurs seuils de classification, où la classification parfaite est illustr\'ee par le couple ({\sf specificity}, {\sf sensitivity})=(1,1), c'est \`a-dire par le coin sup\'erieur gauche dans le graphique. Aussi, on choisit comme seuil de classification optimal celui qui correspond au point de la courbe ROC qui est le plus proche du point (1,1), ou du coin sup\'erieur gauche. La partie droite du tableau \ref{tab:appli:cutoff2} pr\'esente les taux de classifications correctes avec les seuils optimaux ({\em optimal cutoff}), pour les diff\'erentes m\'ethodes (les seuils optimaux des m\'ethodes logit, {\em bagging}, for\^et al\'eatoire et {\em boosting} sont, respectivement, \'egaux \`a $0.0655 , 0.0365 , 0.0395, 0.0596$). Avec le boosting et un seuil optimal, les r\'eponses "non" sont correctes \`a 68.6\% ({\sf specificity}) et les r\'eponses «  oui » \`a 73.85\%  ({\sf sensitivity}). L'objet de l'analyse \'etant de pr\'evoir correctement les individus susceptibles d'acheter une assurance caravane (classe "oui"), et les distinguer suffisamment des autres (classe «  non »), le choix du seuil optimal est beaucoup plus performant que le seuil standard 0.5. Notons qu'avec un mod\`ele logit et un seuil optimal, le taux de classifications correctes de la classe "non" est de 72.78\%, celui de la classe "oui" est de 63.51\%. Par rapport au boosting, le logit pr\'edit un peu mieux la classe "non", mais nettement moins bien la classe «  oui ».

\subsection{Les d\'efauts de remboursement de cr\'edits particuliers (classification)}
\label{appli3}

Consid\'erons la base allemande de cr\'edits particuliers, utilis\'ee dans \citeNP{Nisbet} et \citeNP{Tuffery}, avec 1000 observations, et 19 variables explicatives, dont 12 qualitatives c'est \`a dire, en les disjonctant (en cr\'eant une variable indicatrice pour chaque modalit\'e), 48 variables explicatives potentielles.

Une question r\'ecurrente en mod\'elisation est de savoir quelles sont les variables qui m\'eriteraient d'\^etre utilis\'ees. La r\'eponse la plus naturelle pour un \'econom\`etre pourrait \^etre une m\'ethode de type {\em stepwise} (parcourir toutes les combinaisons possibles de variables \'etant a priori un probl\`eme trop complexe en grande dimension). La suite des variables dans une approche {\em forward} est pr\'esent\'ee dans la premi\`ere colonne du tableau \ref{tab:compare}. Une approche mentionn\'ee avant qui peut \^etre utile est le {\sc Lasso}, en p\'enalisant convenablement la norme $\ell_1$ du vecteur de param\`etres $\bbeta$. On peut ainsi, s\'equentiellement, trouver les valeurs du param\`etre de p\'enalisation $\lambda$, qui permet d'avoir une variable explicative suppl\'ementaire, non nulle. Ces variables sont pr\'esent\'ees dans la derni\`ere colonne. On note que les deux premi\`eres variables consid\'er\'ees comme non nulles (pour un $\lambda$ assez grand) sont les deux premi\`eres \`a ressortir lors d'une proc\'edure {\em forward}. Enfin, une derni\`ere m\'ethode a \'et\'e propos\'ee par  \citeNP{Breiman2001}, en utilisant tous les arbres cr\'e\'e lors de la construction d'une for\^et al\'eatoire : l'importance de la variable $x_k$ dans une for\^et de $T$ arbres est donn\'ee par:
$$
\text{Importance}(x_k)=\frac{1}{T}\sum_{t=1}^n
\sum_{j\in N_{t,k}} p_t(j)\Delta \mathcal{I}(j)
$$
o\`u $N_{t,k}$ d\'esigne l'ensemble des n\oe{}uds de l'arbre $t$ utilisant la variable $x_k$ comme variable de s\'eparation, $p_t(j)$ d\'esigne la proportion des observations au n\oe{}ud $j$, et $\Delta \mathcal(j)$ est la variation d'indice au n\oe{}ud $j$ (entre le n\oe{}ud pr\'ec\'edant, la feuille de gauche et celle de droite). Dans la colonne centrale du tableau \ref{tab:compare} sont pr\'esent\'ees les variables par ordre d'importance d\'ecroissante, lorsque l'indice utilis\'e est l'indice \revision{d'impuret\'e} de Gini.

\begin{table}
\small
\begin{tabular}{lrlrl}\hline
                    Stepwise     &    AIC   &Random Forest&Gini & {\sc Lasso} \\ \hline
         {\sffamily   checking\_statusA}14& 1112.1730  &    {\sffamily checking\_statusA14}&       30.818197 &               {\sffamily checking\_statusA14} \\
     {\sffamily credit\_amount(4e+03,Inf]}&  1090.3467  &      {\sffamily installment\_rate}&       20.786313 &         {\sffamily credit\_amount(4e+03,Inf]} \\
            {\sffamily credit\_historyA34}&  1071.8062  &       {\sffamily residence\_since}&       19.853029 &                {\sffamily credit\_historyA34} \\
             {\sffamily installment\_rate}&  1056.3428  &             {\sffamily duration(15,36]}&       11.377471 &                 {\sffamily duration(36,Inf]} \\
                   {\sffamily purposeA41}&  1044.1580  &     {\sffamily credit\_historyA34}&       10.966407 &                {\sffamily credit\_historyA31} \\
                   {\sffamily savingsA65}&  1033.7521  &         {\sffamily credit\_amount}&       10.964186 &                       {\sffamily savingsA65} \\
                   {\sffamily purposeA43}&  1023.4673  &      {\sffamily existing\_credits}&       10.482961 &                      {\sffamily housingA152} \\
                  {\sffamily housingA152}&   1015.3619 &{\sffamily other\_payment\_plansA143}&       10.469886 &                  {\sffamily duration(15,36]} \\
      {\sffamily other\_payment\_plansA143}&  1008.8532  &         {\sffamily telephoneA192} &       10.217750 &                       {\sffamily purposeA41} \\
          {\sffamily personal\_statusA93} & 1001.6574  &                  {\sffamily  age}&      10.071736  &                {\sffamily installment\_rate} \\
                  {\sffamily savingsA64 }&  996.0108  &          {\sffamily   savingsA65}&       9.547362  &         {\sffamily  property\_magnitudeA124} \\
          {\sffamily  other\_partiesA103} &  991.0377  &    {\sffamily checking\_statusA12}&       9.502445  &                     {\sffamily age(25,Inf]} \\
          {\sffamily checking\_statusA13} &  985.9720  &           {\sffamily housingA152} &       8.757095  &              {\sffamily checking\_statusA13} \\
         {\sffamily  checking\_statusA12} & 982.9530  &               {\sffamily jobA173}&       8.734460  &                      {\sffamily purposeA43} \\
              {\sffamily  employmentA74} &  980.2228  &    {\sffamily personal\_statusA93}&       8.715932  &               {\sffamily other\_partiesA103} \\
                {\sffamily  age(25,Inf]} &  977.9145  &{\sffamily property\_magnitudeA123}&       8.634527  &                   {\sffamily employmentA72} \\
                 {\sffamily  purposeA42} &  975.2365  &  {\sffamily   personal\_statusA92}&       8.438480  &                      {\sffamily savingsA64} \\
             {\sffamily duration(15,36] }&  972.5094  &          {\sffamily   purposeA43}&       8.362432  &                {\sffamily    employmentA74} \\
            {\sffamily duration(36,Inf]} &  966.7004  &         {\sffamily employmentA73}&       8.225416  &                {\sffamily       purposeA46} \\
                {\sffamily   purposeA49} &  965.1470  &     {\sffamily     employmentA75}&       8.089682  &            {\sffamily   personal\_statusA93} \\
                 {\sffamily purposeA410} &  963.2713  &         {\sffamily     duration(36,Inf]}&       8.029945  &              {\sffamily personal\_statusA92} \\
           {\sffamily credit\_historyA31} &  962.1370  &            {\sffamily purposeA42}&       8.025749  &                      {\sffamily savingsA63} \\
                {\sffamily   purposeA48} & 961.1567   &{\sffamily property\_magnitudeA122}&       7.908813  &                 {\sffamily   telephoneA192}  \\ \hline
\end{tabular}
\caption{Cr\'edit: choix de variables, tri s\'equentiel, par approche {\em stepwise}, par fonction d'importance dans une for\^et al\'eatoire et par {\sc lasso}.}\label{tab:compare}
\end{table}

Avec l'approche {\em stepwise} et l'approche {\sc lasso}, on reste sur des mod\`eles logistiques lin\'eaires. Dans le cas des for\^etes al\'eatoires (et des arbres), des int\'eractions entre variables peuvent \^etre prises en compte, lorsque 2 variables sont pr\'esentes. Par exemple la variable {\sffamily residence\_since} est pr\'esente tr\`es haut parmi les variables pr\'edictives (troisi\`eme variable la plus importante).% sans rester dans une proc\'edure {\em stepwise}, et devenant non nulle tr\`es tard.

\subsection{Les d\'eterminants des salaires (r\'egression)} 
\label{appli4}

Afin d'expliquer les salaires (individuels) en fonction du niveau d'\'etude, de l'exp\'erience de la personne, et son genre, il est classique d'utiliser l'\'equation de salaire de Mincer - d\'ecrite dans \citeNP{Mincer} - tel que le rappelle \citeNP{Lemieux}:
\begin{equation}
\log (\textsf{wage}) = \beta_0+\beta_1 \,\textsf{ed}+\beta_2 \,\textsf{exp}+\beta_3 \,\textsf{exp}^2+\beta_4\,\textsf{fe} +\varepsilon
\label{eq:lm:app1}
\end{equation}
où {\sf ed} est le niveau d'\'etudes, {\sf ex} l'exp\'erience professionnelle et {\sf fe} une variable indicatrice, \'egale \`a 1 si l'individu est une femme et \`a 0 sinon. D'apr\`es la th\'eorie du capital humain, le salaire esp\'er\'e augmente avec l'exp\'erience, de moins en moins vite, pour atteindre un maximum avant de diminuer. L'introduction du carr\'e de {\sf exp} permet de prendre en compte une telle relation. La pr\'esence de la variable  {\sf fe} permet quand \`a elle de mesurer une \'eventuelle discrimination salariale entre les hommes et les femmes.

Le mod\`ele (\ref{eq:lm:app1}) impose une relation lin\'eaire entre le salaire  et le niveau d'\'etude, et une relation quadratique entre le salaire et l'exp\'erience professionnelle. Ces relations peuvent paraître trop restrictives. Plusieurs \'etudes montrent notamment que le salaire ne diminue pas apr\`es un certain age, et qu'une relation quadratique ou un polynôme de degr\'e plus \'elev\'e est plus adapt\'e (comme d\'ecrit dans \citeNP{MurphyWelch} et  \citeNP{Bazen}).

Le mod\`ele (\ref{eq:lm:app1}) impose \'egalement que la diff\'erence salariale entre les hommes et les femmes est ind\'ependente du niveau d'\'etude et de l'exp\'erience. Il est trop restrictif si, par exemple, on suspecte que l'\'ecart de salaire moyen entre les hommes et les femmes est faible pour les postes non-qualifi\'es et fort pour les postes qualifi\'es, ou faible en d\'ebut de carri\`ere et fort en fin de carri\`ere ({\em effets d'int\'eractions}). 

Le mod\`ele le plus flexible est le mod\`ele enti\`erement non-param\'etrique~:
\begin{equation}
\log (\textsf{wage}) = m(\textsf{ed, exp, fe}) +\varepsilon
\label{eq:np:app1}
\end{equation}
où $m(\cdot)$ est une fonction quelconque. Il a l'avantage de pouvoir tenir compte de relations non-lin\'eaires quelconques et d'int\'eractions complexes entre les variables. Mais, sa grande flexibilit\'e se fait au d\'etriment d'une interpr\'etation plus difficile du mod\`ele. En effet, il faudrait un graphique en 4-dimensions pour repr\'esenter la fonction $m$. Une solution consiste \`a repr\'esenter la fonction $m$ en 3 dimensions, en fixant la valeur de l'une des variables, mais la fonction repr\'esent\'ee peut  être tr\`es diff\'erente avec une valeur fix\'ee diff\'erente.

\begin{table}
\begin{center}
\begin{tabular}{lccccc}
\hline\\[-2ex]
 & Mod\`ele (\ref{eq:lm:app1}) & \multicolumn{4}{c}{Mod\`ele (\ref{eq:np:app1})} \\
% in-sample: 0.1951 0.1909 0.1889 0.1050 0.1729 0.2014
\cline{3-6}\\[-2ex]
%$CV_{10}$ 
$\widehat{\mathcal{R}}^{10-{\text{\sf CV}}}$ & {\sc ols} &  {\sc s}plines & {\sc b}agging & {\sc r.f}orest & {\sc b}oosting\\
\hline\\[-2ex]
out-of-sample & 0.2006 &  0.2004 & 0.2762 & 0.2160 & 0.2173 \\
\hline
\end{tabular}
\caption{Salaire: analyse de validation crois\'ee par blocs ($K=10$) : performances de l'estimation des mod\`eles lin\'eaire (\ref{eq:lm:app1}) et enti\`erement non-param\'etrique (\ref{eq:np:app1}).}
\label{tab:wage}
\end{center}
\end{table}

Nous utilisons les donn\'ees d'une enquête de l'US Census Bureau dat\'e de mai 1985, issues de l'ouvrage de \citeNP{Berndt} et disponibles sour R.\footnote{C'est le \revision{jeu de donn\'ees} {\sffamily CPS1985} de la \revision{biblioth\`eque} {\sffamily AER}.} Nous estimons les deux mod\`eles et utilisons une analyse de validation crois\'ees par 10 blocs pour s\'electionner la meilleure approche. Le mod\`ele param\'etrique (\ref{eq:lm:app1}) est estim\'e par Moindres Carr\'es Ordinaires ({\sc ols}). Le mod\`ele enti\`erement non-param\'etrique (\ref{eq:np:app1}) est estim\'e par la m\'ethode des {\em splines},  car il en comprend peu de variables, ainsi que par les m\'ethodes {\em bagging}, {\em random forest} et {\em boosting}.

Le tableau \ref{tab:wage} pr\'esente les r\'esultats de la  validation crois\'ee en 10 blocs ({\em 10-fold cross-validation}). Le meilleur mod\`ele est celui qui minimise le crit\`ere %$CV_{10}$
$\widehat{\mathcal{R}}^{10-{\text{\sf CV}}}$. Les r\'esultats montrent que le mod\`ele (\ref{eq:lm:app1}) est au moins aussi performant que le mod\`ele (\ref{eq:np:app1}), ce qui sugg\`ere que le mod\`ele param\'etrique (\ref{eq:lm:app1}) est correctement sp\'ecifi\'e.

%%%%%%%%%%%%%%%%%
\subsection{Les d\'eterminants des prix des logements \`a Boston (r\'egression)}
\label{appli5}

Nous reprenons ici l'un des exemples utilis\'e dans \citeNP{James}, dont les donn\'ees sont disponibles sous {\sffamily R}. Le \revision{jeu de donn\'ees} contient les valeurs m\'edianes des prix des maisons ({\sf medv}) dans $n=506$ quartiers autour de Boston, ainsi que 13 autres variables, dont le nombre moyen de pi\`eces par maison ({\sf rm}), l'age moyen des maisons ({\sf age}) et le pourcentage de m\'enages dont la cat\'egorie socio-professionnelle est peu \'elev\'ee ({\sf lstat}).\footnote{C'est le \revision{jeu de donn\'ees} \texttt{Boston} de  la librairie \texttt{MASS}. Pour une description compl\`ete des donn\'ees, voir: {\sffamily https://stat.ethz.ch/R-manual/R-devel/library/MASS/html/Boston.html}.} 

Consid\'erons le mod\`ele de r\'egression lin\'eaire suivant :
\begin{equation}
\textsf{medv} = \alpha + \bx\transpose\bbeta +\varepsilon
\label{eq:lm:app2}
\end{equation}
où $\bx=[\textsf{chas,nox,age,tax,indus,rad,dis,lstat,crim,black,rm,zn,ptratio}]$ est un vecteur en dimension 13 et $\bbeta$ est un vecteur de $13$ param\`etres. Ce mod\`ele sp\'ecifie une relation lin\'eaire entre la valeur des maisons et chacune des variable explicatives.

Le mod\`ele le plus flexible est le mod\`ele enti\`erement non-param\'etrique~:
\begin{equation}
\textsf{medv} = m( \bx) +\varepsilon.
\label{eq:np:app2}
\end{equation}
L'estimation de ce mod\`ele avec les m\'ethodes du noyau ou les splines peut être probl\'ematique, car le nombre de variables est relativement \'elev\'e (il y a ici 13 variables), ou au moins trop \'elev\'e pour envisager estimer une surface en dimension 13. %Ce probl\`eme est connu comme \'etant le {\em fl\'eau de la dimension}. Les m\'ethodes d'estimation {\em bagging, random forest} et {\em boosting}, issues de l'apprentissage statistique sont alors plus appropri\'ees.
Nous estimons les deux mod\`eles et utilisons une analyse de validation crois\'ee par 10-blocs pour s\'electionner la meilleure approche. Le mod\`ele param\'etrique (\ref{eq:lm:app2}) est estim\'e par Moindres Carr\'es Ordinaires ({\sc ols}) et le mod\`ele enti\`erement non-param\'etrique (\ref{eq:np:app2}) est estim\'e par trois m\'ethodes diff\'erentes: {\em bagging}, for\^et al\'eatoire et {\em boosting} (nous utilisons ici les valeurs par d\'efaut utilis\'ees dans \citeNP{James}, pp. 328-331).

\begin{table}
\begin{center}
\begin{tabular}{lcccc}
\hline\\[-2ex]
& Mod\`ele (\ref{eq:lm:app2}) &  \multicolumn{3}{c}{Mod\`ele (\ref{eq:np:app2})} \\[1ex]
\cline{3-5}\\[-2ex]
 %$CV_{10}$
 $\widehat{\mathcal{R}}^{10-{\text{\sf CV}}}$& {\sc ols} &  {\sc b}agging & {\sc r}.{\sc f}orest & {\sc b}oosting \\
\hline\\[-2ex]
in-sample & 21.782 &    1.867 & 1.849 & 7.012  \\
out-of-sample & 24.082 &   9.590  & 9.407 & 11.789 \\
\hline
\end{tabular}
\caption{Prix des logements \`a Boston: analyse de validation crois\'ee par blocs ($K=10$): performances de l'estimation des mod\`eles lin\'eaire (\ref{eq:lm:app2}) et enti\`erement non-param\'etrique (\ref{eq:np:app2}).}
\label{tab:boston}
\end{center}
\end{table}

Le tableau \ref{tab:boston} pr\'esente les r\'esultats de la  validation crois\'ee en 10 blocs ({\em 10-fold cross-validation}). La premi\`ere ligne ({\em in-sample}) pr\'esente la qualit\'e de l'ajustement des mod\`eles en utilisant seulement les donn\'ees d'apprentissage, c'est-\`a-dire celles qui ont servi \`a estimer le mod\`ele, pour calculer le {\sc mse}. La deuxi\`eme ligne ({\em out-of-sample}) pr\'esente la qualit\'e de l'ajustement en utilisant d'autres donn\'ees que celles ayant servies \`a estimer le mod\`ele, pour calculer l'erreur quadratique.  \`A partir  des r\'esultats in-sample, les m\'ethodes de {\em bagging} et de {\em random forest} paraissent incroyablement plus performantes que l'estimation {\sc ols} du mod\`ele lin\'eaire (\ref{eq:lm:app2}), le crit\`ere %$CV_{10}$ 
$\widehat{\mathcal{R}}^{10-{\text{\sf CV}}}$ passant de 21.782 \`a 1.867 et 1.849.
Les r\'esultats {\em out-of-sample} vont dans le même sens, mais la diff\'erence est  moins importante, le crit\`ere %$CV_{10}$ 
$\widehat{\mathcal{R}}^{10-{\text{\sf CV}}}$ passant de 24.082 \`a 9.59 et 9.407. Ces r\'esultats illustrent un ph\'enom\`ene classique des m\'ethodes non-lin\'eaires, comme le {\em bagging} et la for\^et al\'eatoire, qui peuvent être tr\`es performantes pour pr\'edire les donn\'ees utilis\'ees pour l'estimation, mais  moins performantes pour pr\'edire des donn\'ees hors-\'echantillon. C'est pourquoi la s\'election de la meilleure estimation est habituellement bas\'ee sur une analyse {\em out-of-sample}, telle que pr\'esent\'ee dans la deuxi\`eme ligne.

La diff\'erence entre l'estimation du mod\`ele lin\'eaire (\ref{eq:lm:app2}) et du mod\`ele enti\`erement non-param\'etrique (\ref{eq:np:app2}) est importante (24.082 vs 9.590, 9.407 et 11.789). Un tel \'ecart sugg\`ere que le mod\`ele lin\'eaire est mal sp\'ecifi\'e, et que des relations non-lin\'eaire et/ou des effets d'int\'eractions sont pr\'esentes dans la relation entre le prix des logements, {\sf medv}, et les variables explicatives $\bx$. Ce r\'esultat nous conduit \`a  chercher une meilleure sp\'ecification param\'etrique.

\`A partir du mod\`ele param\'etrique  (\ref{eq:lm:app2}), et afin de prendre en compte d'\'eventuelles non-lin\'earit\'es, le mod\`ele additif g\'en\'eralis\'e ({\sc gam}) suivant peut être consid\'er\'e :
\begin{equation}
\textsf{medv} = m_1 (x_1)+m_2 (x_2)+\dots+m_{13} (x_{13}) +\varepsilon,
\label{eq:sp:app2}
\end{equation}
où $m_1,  m_2, \dots m_{13}$ sont des fonctions inconnues. L'avantage de ce mod\`ele est qu'il permet de consid\'erer n'importe quelle relation non-lin\'eaire entre la variable d\'ependante et chacune des variables explicatives. De plus, il ne souffre pas du probl\`eme du fl\'eau de la dimension, car chacune des fonction est de dimension 1, et il est facilement interpr\'etable. Toutefois, il ne prend pas en compte d'\'eventuels effets d'int\'eractions. 

L'estimation du mod\`ele additif g\'en\'eralis\'e (\ref{eq:sp:app2}) par la m\'ethode des {\em splines}, dans le cadre d'une analyse de validation crois\'ee par 10-blocs, donne une valeur %$CV_{10}=
$\widehat{\mathcal{R}}^{10-{\text{\sf CV}}}=13.643$. Par rapport au mod\`ele param\'etrique (\ref{eq:lm:app2}), il y a un gain important (13.643 vs. 24.082). Mais la diff\'erence avec le mod\`ele enti\`erement non-param\'etrique (\ref{eq:np:app2}) reste cons\'equente (13.643 vs 9.590, 9.407, 11.789). Une telle diff\'erence sugg\`ere que la prise en compte de relations individuelles pouvant être fortement non-lin\'eaires n'est pas suffisante, et que des effets d'int\'eractions entre les variables sont pr\'esents. Nous pourrions inclure dans le mod\`ele les variables d'int\'eractions les plus simples entre toutes les paires de variables ($x_i\times x_j$), mais cela impliquerait de rajouter un tr\`es grand nombre de variables au mod\`ele initial (78 dans notre cas), qui ne serait pas sans cons\'equence sur la qualit\'e de l'estimation du mod\`ele. Quoi qu'il en soit, nous pouvons dire pour le moment que le mod\`ele lin\'eaire est mal sp\'ecifi\'e et qu'il existe des effets d'int\'eractions pouvant être forts dans la relation entre {\sf medv} et $X$, l'identification de tels effets restant d\'elicat. 

\begin{table}[t]
\small
%\size
\begin{center}
\begin{tabular}{|lrr|}
\hline\\[-2ex]
&           {\sf \%IncMS}E & {\sf IncNodePurity} \\\hline
{\sffamily rm}   &     61.35     & 18345.41 \\
{\sffamily lstat} &    36.20     & 15618.22 \\
{\sffamily dis }   &   29.37      & 2601.72 \\
{\sffamily nox }  &    24.91     &  1034.71 \\
{\sffamily age  } &    17.86      &  554.50 \\
{\sffamily ptratio} &  17.43      &  626.58 \\
{\sffamily tax  }   &  16.60       & 611.37 \\
{\sffamily crim }  &   16.26      & 1701.73 \\
{\sffamily indus}  &    9.45      &  237.35 \\
{\sffamily black}   &   8.72      &  457.58 \\
{\sffamily rad }    &   4.53       & 166.72 \\
{\sffamily zn  }     &  3.10        & 35.73 \\
{\sffamily chas}    &   0.87      &   39.05 \\
\hline
\end{tabular}
\caption{Prix des logements \`a Boston: mesures de l'importance de chacune des variables dans l'estimation {\em random forest} du mod\`ele (\ref{eq:np:app2}), en consid\'erant tout l'\'echantillon.}
\label{tab:importance}
\end{center}
\end{table}

Afin d'aller plus loin, les outils d\'evelopp\'es en apprentissage statistique peuvent être \`a nouveau d'un grand recours. Par exemple, l'estimation {\em random forest} s'accompagne de mesures de l'importance de chacune des variables dans l'estimation du mod\`ele (d\'ecrit dans la section pr\'ec\'edante). Le tableau \ref{tab:importance} pr\'esente ces mesures dans le cadre du mod\`ele (\ref{eq:np:app2}), estim\'e sur  l'\'echantillon complet. Les r\'esultats sugg\`erent que les variables {\sf rm} et {\sf lstat} sont les variables les plus importantes pour expliquer les variations des prix des logements {\sf medv}.
Ce constat nous conduit \`a enrichir la relation initiale, en rajoutant les int\'eractions li\'ees \`a ces deux variables seulement, qui sont les plus importantes. 

\begin{figure}[t]
\begin{center}
\includegraphics[width=.7\textwidth,height=.35\textheight]{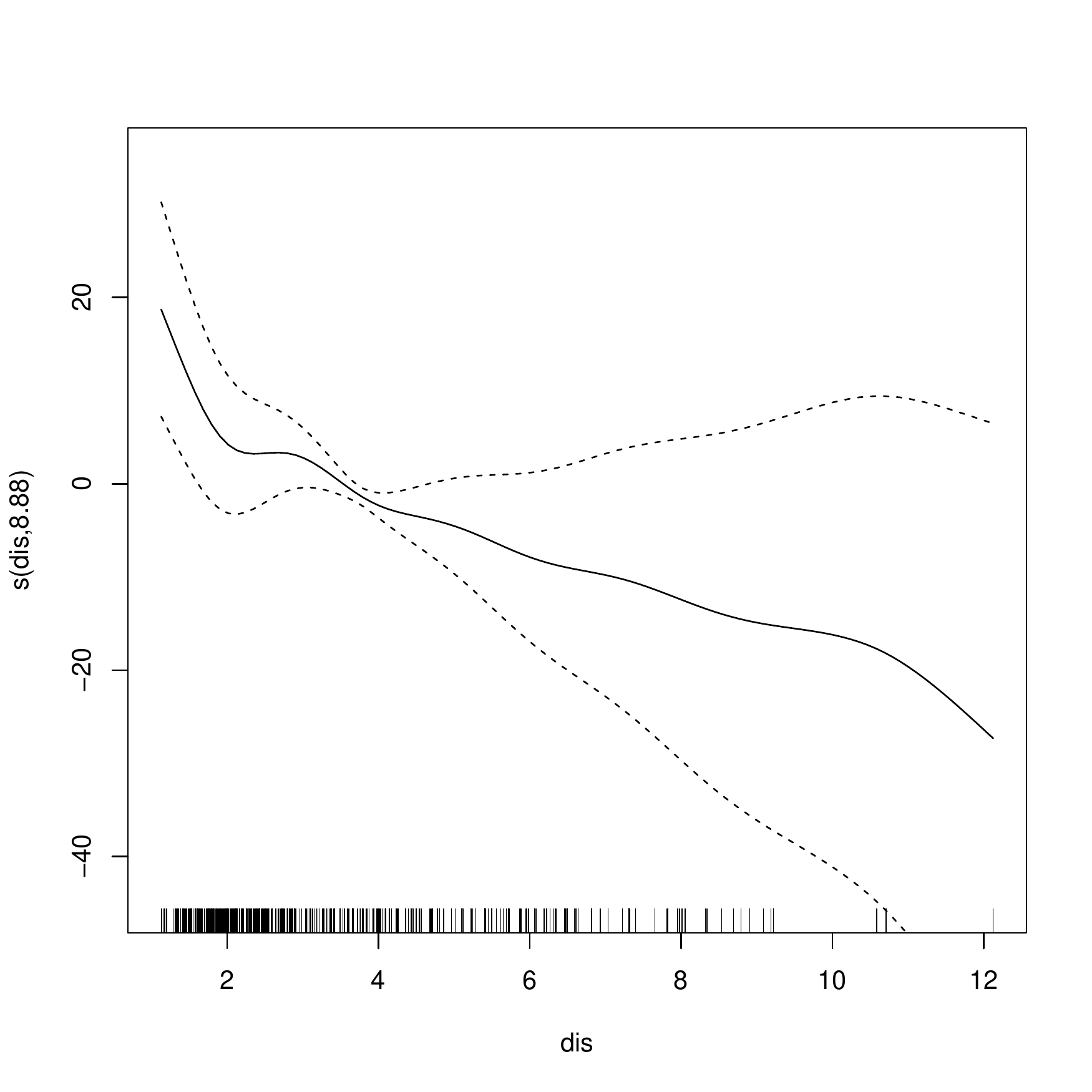}
\end{center}
\caption{Estimation de la relation $m_7(x_7)$ dans le mod\`ele additif g\'en\'eralis\'e (\ref{eq:lm2:app2}), où $x_7={\sf dis}$.}
\label{fig:appli:gam}
\end{figure}

Nous estimons le mod\`ele additif g\'en\'eralis\'e incluant les variables d'int\'eractions, sur l'\'echantillon complet:
\begin{align}
\textsf{medv} = m_1(x_1)+ \cdots + m_{13}(x_{13})+ (\textsf{rm}{:}x) \,\gamma + (\textsf{lstat}{:}x) \,\delta +\varepsilon,
\label{eq:lm2:app2}
\end{align}
où $(\textsf{rm}{:}x)$ repr\'esente les variables d'int\'eractions de {\sf rm} avec toutes les autres variables de $x$ et $(\textsf{lstat}{:}x)$ repr\'esente les variables d'int\'eractions de {\sf lstat} avec toutes les autres variables de $x$.\footnote{On a 
$(\textsf{rm}{:}x)=[${\sf rm}$\times${\sf chas}, {\sf rm}$\times${\sf nox},  {\sf rm}$\times${\sf age}, {\sf rm}$\times${\sf tax}, {\sf rm}$\times${\sf indus}, 
{\sf rm}$\times${\sf rad}, {\sf rm}$\times${\sf dis}, {\sf rm}$\times${\sf lstat}, {\sf rm}$\times${\sf crim}, {\sf rm}$\times${\sf black}, {\sf rm}$\times${\sf zn}, {\sf rm}$\times${\sf ptratio}$]$ et 
$(\textsf{lstat}{:}x)=[${\sf lstat}$\times${\sf chas}, {\sf lstat}$\times${\sf nox}, {\sf lstat}$\times${\sf age}, {\sf lstat}$\times${\sf tax}, {\sf lstat}$\times${\sf indus}, 
{\sf lstat}$\times${\sf rad}, {\sf lstat}$\times${\sf dis}, {\sf lstat}$\times${\sf crim}, {\sf lstat}$\times${\sf black}, {\sf lstat}$\times${\sf zn}, {\sf lstat}$\times${\sf ptratio}$]$.
}
L'analyse des r\'esultats de cette estimation sugg\`ere que les fonctions $\hat m_i$ sont lin\'eaires pour toutes les variables, sauf pour la variable {\sf dis}, dont la relation estim\'ee est pr\'esent\'ee dans la figure \ref{fig:appli:gam}.  Cette variable mesure la distance moyenne \`a cinq centres d'emplois de la r\'egion. L'effet semble diminuer plus rapidement avec la distance, lorsque celle-ci n'est pas tr\`es \'elev\'ee. Au del\`a d'une certaine distance (au del\`a de 2, en $\log$), l'effet est r\'eduit, il continue \`a diminuer mais plus doucement. Cette relation non-lin\'eaire peut  être approch\'ee par une r\'egression lin\'eaire par morceaux en consid\'erant un n\oe{}ud.

Finalement, l'analyse pr\'ec\'edente nous conduit \`a consid\'erer  le mod\`ele lin\'eaire suivant:
\begin{equation}
\textsf{medv} = \alpha + \bx\transpose\bbeta+ (\textsf{dis}-2)_+ \, \theta + (\textsf{rm}{:} x) \, \gamma + (\textsf{lstat}{:}x)\, \delta +\varepsilon
\label{eq:my:app2}
\end{equation}
où $(\textsf{dis}-2)_+$ est \'egal \`a  la valeur de son argument si ce dernier est positif\revision{, et à 0 sinon}. Par rapport au mod\`ele lin\'eaire initial, ce mod\`ele inclut une relation lin\'eaire par morceaux avec la variable {\sf dis}, ainsi que des effets d'int\'eractions entre {\sf rm}, {\sf lstat} et chacune des autres variables de $\bx$. 

Le tableau \ref{tab:boston:cv2} pr\'esente les r\'esultats de la  validation crois\'ee en 10 blocs ({\em 10-fold cross-validation}) de l'estimation des mod\`eles param\'etriques (\ref{eq:lm:app2}) et (\ref{eq:my:app2}),  estim\'es par Moindres Carr\'es Ordinaires ({\sc ols}), et du mod\`ele additif g\'en\'eralis\'e (\ref{eq:sp:app2}) estim\'e par les splines. Il montre que l'ajout des variables d'int\'eractions et de la relation lin\'eaire par morceaux dans le mod\`ele (\ref{eq:my:app2}) donne des r\'esultat beaucoup plus performants que le mod\`ele initial (\ref{eq:lm:app2}): le crit\`ere %$CV_{10}$ 
$\widehat{\mathcal{R}}^{10-{\text{\sf CV}}}$ est divis\'e par plus de deux, il passe de 24.082 \`a 11.759. En comparant ces r\'esultats avec ceux du tableau  \ref{tab:boston}, on constate \'egalement que le mod\`ele param\'etrique (\ref{eq:my:app2}), estim\'e par {\sc ols}, est aussi performant que le mod\`ele g\'en\'eral (\ref{eq:np:app2}) estim\'e par {\em boosting} (%$CV_{10}
$\widehat{\mathcal{R}}^{10-{\text{\sf CV}}}=11.789$). La diff\'erence avec les m\'ethodes {\em bagging} et for\^et al\'eatoire n'est quant \`a elle pas tr\`es importante (%$CV_{10}
$\widehat{\mathcal{R}}^{10-{\text{\sf CV}}}=9.59, 9.407$)

Finalement, les m\'ethodes {\em bagging}, for\^et al\'eatoire et {\em boosting} ont permis de mettre en \'evidence une mauvaise sp\'ecification du mod\`ele param\'etrique initial, puis de trouver un mod\`ele param\'etrique beaucoup plus performant, en prenant compte des effets de non-lin\'earit\'es et d'int\'eractions appropri\'ees.

\begin{table}[t]
\begin{center}
\begin{tabular}{lccc}
\hline\\[-2ex]
 & Mod\`ele (\ref{eq:lm:app2}) & Mod\`ele (\ref{eq:sp:app2}) & Mod\`ele (\ref{eq:my:app2})  \\
%$CV_{10}$ 
$\widehat{\mathcal{R}}^{10-{\text{\sf CV}}}$ & {\sc ols} & {\sc S}plines & {\sc ols}\\
\hline\\[-2ex]
out-of-sample & 24.082 & 13.643 & 11.759  \\
\hline
\end{tabular}
\caption{Prix des logements \`a Boston: analyse de validation crois\'ee par blocs ($K=10$) : performances de l'estimation du mod\`ele lin\'eaire (\ref{eq:lm:app2}) et du mod\`ele lin\'eaire (\ref{eq:my:app2}) incluant les effets d'int\'eractions et une non-lin\'earit\'e par morceaux.}
\label{tab:boston:cv2}
\end{center}
\end{table}

\section{Conclusion}

Si les « deux cultures » (ou les deux communaut\'es) de l'\'econom\'etrie et du {\em machine learning} se sont d\'evelopp\'ees en parall\`ele, le nombre de passerelles entre les deux ne cesse d'augmenter. Alors que \citeNP{Varian} pr\'esentait les apports importants de l'\'econom\'etrie \`a la communaut\'e du {\em machine learning}, nous avons tent\'e ici de pr\'esenter des concepts et des outils d\'evelopp\'es au fil du temps par ces derniers, qui pourraient \^etre utiles aux \'econom\`etres, dans un contexte d'explosion du volume de donn\'ees. Si nous avons commenc\'e par opposer ces deux mondes, c'est aussi pour mieux comprendre leurs forces et leurs faiblesses. Les fondements probabilistes de l'\'econom\'etrie sont incontestablement sa force, avec non seulement une interpr\'etabilit\'e des mod\`eles, mais aussi une quantification de l'incertitude. N\'eanmoins, nous l'avons vu \`a plusieurs reprises sur des donn\'ees r\'eelles, les performances pr\'edictives des mod\`eles de {\em machine learning} sont int\'eressantes, car elles permettent de mettre en avant une mauvaise sp\'ecification d'un mod\`ele \'econom\'etrique. De la m\^eme mani\`ere que les techniques non-param\'etriques permettent d'avoir un point de r\'ef\'erence pour juger de la pertinence d'un mod\`ele param\'etrique, les outils de {\em machine learning} permettent d'am\'eliorer un mod\`ele \'econom\'etrique, en d\'etectant un effet non-lin\'eaire ou un effet crois\'e oubli\'e.

Une illustration des interactions possibles entre les deux communaut\'es se trouve par exemple dans \citeNP{Belloni}\citeNP{Belloni2}, dans un contexte de choix d'instrument dans une r\'egression. Reprenant les donn\'ees de  \citeNP{AngristKrueger} sur un probl\`eme de r\'eussite scolaire, ils montrent comment mettre en \oe{}uvre efficacement les techniques d'\'econom\'etrie instrumentale quand on peut choisir parmi 1530 instruments disponibles (probl\`eme qui deviendra r\'ecurrent avec l'augmentation du volume de donn\'ees).
Comme nous l'avons vu tout au long de cet article, m\^eme si les approches peuvent \^etre fondamentalement diff\'erentes dans les deux communaut\'es, bon nombre d'outils d\'evelopp\'es par la communaut\'e du {\em machine learning} m\'eritent d'\^etre utilis\'es par les \'econom\`etres.


\newpage


\begin{thebibliography}{99}
\bibitem[Ahamada \& Flachaire (2011)]{Ahamada} Ahamada, I. \& E. Flachaire (2011). Non-Parametric Econometrics. Oxford
University Press.
\bibitem[Aigner {\em et al.} (1977)]{Aigneretal} Aigner, D., Lovell, C.A.J \& Schmidt, P. (1977). Formulation and estimation of stochastic frontier production function models. {\em Journal of Econometrics}, {\bf 6}, 21--37.
\bibitem[Aldrich (2010)]{Aldrich} Aldrich, J. (2010). The Econometricians' Statisticians, 1895-1945. {\em History of Political Economy}, {\bf 42} 111--154.
\bibitem[Altman {\em et al.} (1994)]{Altman} Altman, E., Marco, G. \& Varetto, F. (1994). Corporate distress diagnosis: Comparisons using linear discriminant analysis and neural networks (the Italian experience).{\em Journal of Banking \& Finance} {\bf 18}, 505--529.
\bibitem[Angrist \& Lavy (1999)]{Angrist} Angrist, J.D. \& Lavy, V. (1999). Using Maimonides' Rule to Estimate the Effect of Class Size on Scholastic Achievement. {\em Quarterly Journal of Economics}, {\bf 114}, 533--575. 
\bibitem[Angrist, J.D. \& Pischke, J.S. (2010)]{AngristPischke10} Angrist, J.D. \& Pischke, J.S. (2010). The Credibility Revolution in Empirical Economics: How Better Research Design Is Taking the Con out of Econometrics.  {\em Journal of Economic Perspective}, {\bf 24},  3--30.
\bibitem[Angrist \& Pischke (2015)]{AngristPischke} Angrist, J.D. \& Pischke, J.S. (2015). Mastering Metrics. Princeton University Press.
\bibitem[Angrist \& Krueger (1991)]{AngristKrueger} Angrist, J.D. \& Krueger, A.B. (1991). Does Compulsory School Attendance Affect Schooling and Earnings? {\em Quarterly Journal of Economics}, {\bf 106}, 979--1014.
\bibitem[Bottou (2010)]{bottou} Bottou, L. (2010) Large-Scale Machine Learning
with Stochastic Gradient Descent {\em Proceedings of the 19th International Conference on Computational Statistics (COMPSTAT'2010)}, 177--187.
\bibitem[Bajari {\em et al.} (2015)]{Bajari} Bajari, P., Nekipelov, D., Ryan, S.P. \& Yang, M. 2015. Machine learning
methods for demand estimation. {\em American Economic Review}, {\bf 105} 481--485.
\bibitem[Bazen \& Charni (2015)]{Bazen} Bazen, S. \& K. Charni (2015). Do earnings really decline for older workers?
AMSE 2015-11 Discussion Paper, Aix-Marseille University.
\bibitem[Bellman (1957)]{Bellman} Bellman, R.E. (1957). Dynamic programming. Princeton University Press.
\bibitem[Belloni {\em et al.} (2010]{Belloni} Belloni, A., Chernozhukov, V. \& Hansen, C. (2010). Inference Methods for High-Dimensional Sparse Econometric Models. {\em Advances in Economics and Econometrics}, 245--295
\bibitem[, 2012)]{Belloni2} Belloni, A., Chen, D., Chernozhukov, V. \& Hansen, C. (2012). Sparse Models and Methods for Optimal Instruments With an Application to Eminent Domain. {\em Econometrica}, {\bf 80}, 2369--2429. 
\bibitem[Benjamini \& Hochberg (1995)]{Benjamini} Benjamini, Y. \& Hochberg, Y. (1995). Controlling the false discovery rate: a practical and powerful approach to multiple testing. {\em Journal of the Royal Statistical Society, Series B}, {\bf 57}:289--300.
\bibitem[Berger (1985)]{Berger} Berger, J.O. (1985). Statistical decision theory and Bayesian Analysis (2nd ed.). Springer-Verlag. 
\bibitem[Berk (2008)]{Berk} Berk, R.A. (2008). Statistical Learning from a Regression Perspective. Springer Verlag.
\bibitem[Berkson (1944)]{Berkson1} Berkson, J. (1944). Applications of the logistic function to bioassay. {\em Journal of the American
Statistical Association}, {\bf 9}, 357--365.
\bibitem[Berkson (1951)]{Berkson2} Berkson, J. (1951). Why I prefer logits to probits. {\em Biometrics}, {\bf 7} (4), 327--339.
\bibitem[Bernardo \& Smith (2000)]{Bernardo} Bernardo, J.M. \& Smith, A.F.M. (2000). Bayesian Theory. John Wiley.
\bibitem[Berndt (1990)]{Berndt} Berndt, E. R. (1990). The Practice of Econometrics: Classic and Contemporary. Addison Wesley.
\bibitem[Bickel {\em et al.} (1997)]{Bickel} Bickel, P.J., Gotze, F. \& van Zwet, W. (1997). Resampling fewer than $n$ observations: gains, losses and remedies for losses. {\em Statistica Sinica}, 7, 1-31.
\bibitem[Bishop (2006)]{Bishop} Bishop, C. (2006). Pattern Recognition and Machine Learning. Springer Verlag.
\bibitem[Blanco {\em et al.} (2013)]{Blanco} Blanco, A. Pino-Mejias, M., Lara, J. \& Rayo, S. (2013). Credit scoring models for the microfinance industry using neural networks: Evidence from peru. {\em Expert Systems with Applications}, {\bf 40}, 356--364.
\bibitem[Bliss (1934)]{Bliss} Bliss, C.I. (1934). The method of probits. {\em Science}, {\bf 79}, 38--39.
\bibitem[Breiman (2001a)]{Breiman} Breiman, L. (2001a). Statistical Modeling: The Two Cultures. {\em Statistical Science}, {\bf 16}:3, 199--231.
\bibitem[B\"uhlmann \& van de Geer (2011)]{Buhlmann} B\"uhlmann, P. \& van de Geer, S. (2011). Statistics for high-dimensional data: methods, theory and applications. Springer Verlag.
\bibitem[Breiman (2001b)]{Breiman2001} Breiman, L. (2001b). Random forests. {\em Machine learning}, {\bf 45}:1, 5--32.
%\bibitem[Cheysson (1887)]{Cheysson}
\bibitem[Brown (1986)]{Brown} Brown, L.D. (1986) Fundamentals  of  statistical  exponential  families:  with  applications  in
statistical decision theory. Institute of Mathematical Statistics, Hayworth, CA, USA.
\bibitem[B\"uhlman \& van de Geer (2011)]{BuhlmanDeGeer} B\"uhlmann, P. \& van de Geer, S. (2011). Statistics for High Dimensional Data: Methods, Theory and Applications. Springer Verlag.
\bibitem[Cand\`es \& Plan (2009)]{Candes} Cand\`es, E. \& Plan, Y. (2009). Near-ideal model selection by $\ell_1$ minimization. {The Annals of Statistics}, 37:5, 2145--2177.
\bibitem[Clarke {\em et al.} (2009)]{ClarkeEtal} Clarke, B.S., Fokou\'e, E. \& Zhang, H.H. (2009). Principles and Theory for Data Mining and Machine Learning. Springer Verlag.
\bibitem[Cortes \& Vapnik (1995)]{Cortes} Cortes, C. \& Vapnik, V. (1995). Support-vector networks. {\em Machine Learning} {\bf 20} 273--297.
\bibitem[Cybenko (1989)]{Cybenko}Cybenko, G. (1989). Approximation by Superpositions of a Sigmoidal Function 1989 {\em Mathematics of Control,  Signals, and Systems}, {\bf 2}, 303--314.
\bibitem[Darmois (1935)]{Darmois} Darmois, G. (1935). Sur les lois de probabilites a estimation exhaustive. {\em Comptes Rendus de l'Acad\'emie des Sciencs, Paris}, {\bf 200} 1265--1266.
\bibitem[Daubechies et al. (2004)]{Daubechies}  Daubechies, I. Defrise, M. \& De Mol, C. (2004). An iterative thresholding algorithm for linear inverse problems with sparsity constraint. {\em Communications on Pure and Applied Mathematics}, 57:11, 1413--1457 
\bibitem[Davison (1997)]{Davison} Davison, A.C. (1997). Bootstrap. Cambridge University Press.
\bibitem[Davidson \& MacKinnon (1993)]{DavidsonMacKinnon1} Davidson, R. \& MacKinnon, J.G. (1993). Estimation and Inference in Econometrics. Oxford University Press.
\bibitem[Davidson \& MacKinnon (2003)]{DavidsonMacKinnon2} Davidson, R. \& MacKinnon, J.G. (2003). Econometric Theory and Methods. Oxford University Press.
\bibitem[Duo (1993)]{Duo} Duo, Q. (1993). The Formation of Econometrics. Oxford University Press.
\bibitem[Debreu (1986)]{Debreu} Debreu, G. 1986. 
Theoretic Models: Mathematical Form and Economic Content. {\em Econometrica}, {\bf 54}, 1259--1270.
\bibitem[Dhillon {\em et al.} (2014)]{Dhilonetal2013} Dhillon, P., Lu, Y. Foster, D.P. \& Ungar, L.H. (2014). New Subsampling Algorithms for Fast Least Squares Regression. {\em in }{Advances in Neural Information Processing Systems 26}, Burges, Bottou, Welling, Ghahramani \& Weinberger Eds., Curran Associates.
\bibitem[Efron \& Tibshirani (1993)]{Efron} Efron, B. \& Tibshirani, R. (1993). Bootstrap. Chapman Hall CRC.
\bibitem[Engel (1857)]{Engel} Engel, E. (1857). Die Productions- und Consumtionsverh\"altnisse des K\"onigreichs Sachsen. {\em Statistisches Bureau des K\"oniglich S\"achsischen Ministeriums des Innern}.
\bibitem[Feldstein \& Horioka (1980)]{FH} Feldstein, M. \& Horioka, C. (1980). Domestic Saving and International Capital Flows. {\em Economic Journal}, {\bf 90}, 314--329.
\bibitem[Flach (2012)]{Flach} Flach, P. (2012). Machine Learning. Cambridge University Press.
\bibitem[Foster \& George (1994)]{FosterGeorge94} Foster, D.P. \& George, E.I. (1994). The Risk Inflation Criterion for Multiple Regression. {\em The Annals of Statistics}, 22:4, 1947--1975.
\bibitem[Friedman (1997)]{Friedman97} Friedman, J.H. (1997). Data Mining and Statistics: What's the Connection. {\em Proceedings of the 29th Symposium on the Interface Between Computer Science and Statistics}.
\bibitem[Frish \& Waugh (1933)]{FrishWaugh} Frisch, R. \& Waugh, F.V. (1933). Partial Time Regressions as Compared with Individual Trends. {\em Econometrica}. {\bf 1}, 387--401.
\bibitem[Gneiting (2011)]{Gneiting} Gneiting, T. (2011). Making and Evaluating Point Forecasts. {\em Journal of the American Statistical Association}, {\bf 106}, 746--762.
\bibitem[Givord (2010)]{Givord} Givord, P. (2010). M\'ethodes \'econom\'etriques pour l'\'evaluation de politiques publiques. {INSEE Document de Travail}, {\bf 08}
\bibitem[Grandvalet {\em et al.} (2005)]{Grandvalet} Grandvalet, Y., Mari\'ethoz, J., \& Bengio, S. 2005. Interpretation of SVMs with an application to unbalanced classification. {\em Advances in Neural Information Processing Systems} {\bf 18}.
\bibitem[Groves \& Rothenberg (1969)]{GrovesRothenberg69} Groves, T. \& Rothenberg, T. (1969). A note on the expected value of an inverse matrix. {\em Biometrika}, 56:3, 690--691.
\bibitem[Haavelmo (1944)]{Haavelmo} Haavelmo, T. (1944). The probability approach in econometrics, {\em Econometrica}, {\bf 12}:iii-vi and 1--115.
\bibitem[Hastie \& Tibshirani (1990)]{Trevor} Hastie, T. \& Tibshirani, R. (1990). Generalized Additive Models. Chapman \& Hall/CRC.
\bibitem[Hastie {\em et al.} (2009)]{HastieEtal} Hastie, T., Tibshirani, R. \& Friedman, J. (2009). The Elements of Statistical Learning. Springer Verlag.
\bibitem[Hastie (2005)]{Hastie} Hastie, T., Tibshirani, W. \& Wainwright, M. (2015). Statistical Learning with Sparsity. Chapman CRC.
\bibitem[Hastie {\em et al.} (2016)]{HTT}  Hastie, T., Tibshiriani, R. \& Tibshiriani, R.J. (2016). Extended comparisons of best subset selection, forward stepwise selection and the Lasso. {\em ArXiV}, {\sffamily https://arxiv.org/abs/1707.08692}.
\bibitem[d'Haultef\oe{}uille \& Givord (2014)]{dHaultefoeuille}  d'Haultef\oe{}uille, X. \& Givord, P. (2014) La r\'egression quantile en pratique. {\em \'Economie \& Statistiques}, {\bf 471}, 85--111.
\bibitem[Hebb (1949)]{Hebb} Hebb, D.O. (1949). The organization of behavior, New York, Wiley.
\bibitem[Heckman (1979)]{Heckman} Heckman, J.J. (1979). Sample selection bias as a specification error. {\em Econometrica}, {\bf 47}, 153--161.
\bibitem[Heckman {\em et al.} (2003)]{Heckman2} Heckman, J.J., Tobias, J.L. \& Vytlacil, E. (2003). Simple Estimators for Treatment Parameters in a Latent-Variable Framework. {\em  The Review of Economics and Statistics}, {\bf 85}, 748--755.
%\bibitem[Hendry (1995)]{Hendry} Hendry, D.F. (1995). Dynamic Econometrics. Oxford University Press, Oxford.
\bibitem[Hendry \& Krolzig (1995)]{HendryK} Hendry, D F. \& Krolzig, H.-M. (2001). Automatic Econometric Model Selection. Timberlake Press.
\bibitem[Herbrich {\em et al.} (1999)]{Herbrich} Herbrich, R., Keilbach, M., Graepel, T. Bollmann-Sdorra, P. \& Obermayer, K. (1999). Neural Networks in Economics. {\em in} Computational Techniques for Modelling in Economics, T. Brenner Eds. Springer Verlag, 169--196.
\bibitem[Hoerl (1962)]{Hoerl} Hoerl, A.E. (1962). Applications of ridge analysis to regression problems. {\em Chemical Engineering Progress}, {\bf 58}:3, 54--59.
\bibitem[Hoerl \& Kennard (1980)]{HoerlKennard} Hoerl, A.E. \& Kennard, R.W. (1981). Ridge regression: biased estimation for nonorthogonal problems {\em This Week's Citation Classic}, ISI, {\sffamily http://bit.ly/2H9LGiD}
\bibitem[Holland (1986)]{Holland} Holland, P. (1986). Statistics and causal inference. {\em Journal of the American Statistical Association}, {\bf 81}, 945--960.
\bibitem[Hyndman {\em et al.} (2009)]{Hyndman} Hyndman, R. , Koehler, A.B., Ord, J.K.	\& Snyder, R.D. (2009). Forecasting with Exponential Smoothing. Springer Verlag.
\bibitem[James {\em et al.} (2013)]{James} James, G., D. Witten, T. Hastie, \& R. Tibshirani (2013). An introduction
to Statistical Learning. Springer Series in Statistics.
\bibitem[Khashman (2011)]{Khashman} Khashman, A. (2011). Credit risk evaluation using neural networks: Emotional versus conven-
tional models. {\em Applied Soft Computing}, {\bf 11}, 5477--5484.
\bibitem[Kean (2010)]{Keen} Kean, M.P. (2010). Structural vs. atheoretic approaches to econometrics. {\em Journal of Econometrics}, {\bf 156}, 3--20.
\bibitem[Kleiner {\em et al.} (2012)]  Kleiner, A., Talwalkar, A., Sarkar , P. \& Jordan, M. (2012). The Big Data Bootstrap. arXiv:1206.6415 .
\bibitem[Koch (2013)]{Koch} Koch, I. (2013). Analysis of Multivariate and High-Dimensional Data. Cambridge University Press. 
\bibitem[Koenker (1998)]{KoenkerGalton} Koenker, R. (1998). Galton, Edgeworth, Frish, and prospects for quantile regression in Econometrics. Conference on Principles of Econometrics, Madison.
\bibitem[Koenker (2003)]{Koenker} Koenker, R. (2003). Quantile Regression. Cambridge University Press.
\bibitem[Koenker \& Machado (1999)]{KoenkerMachado} Koenker, R. \& Machado, J. (1999). Goodness of fit and related inference processes for quantile regression {\em Journal of the American Statistical Association}, 94, 1296-1309.
\bibitem[Kolda \& Bader (2009)]{Kolda} Kolda, T. G. \& Bader, B. W. (2009). Tensor decompositions and applications. {\em SIAM Review} {\bf 51}, 455--500.
\bibitem[Koopmans (1957)]{Koopmans} Koopmans, T.C. (1957). Three   Essays   on   the   State   of   Economic   Science. McGraw-Hill.
\bibitem[Kuhn \& Johnson (2013)]{KuhnJohnson} Kuhn, M. \& Johnson, K. (2013). Applied Predictive Modeling. Springer Verlag.
\bibitem[Landis \& Koch (1977)]{LandisKoch} Landis, J.R. \& Koch, G.G. (1977). The measurement of observer agreement for categorical data. {\em Biometrics}, {\bf 33}, 159--174. 
\bibitem[LeCun {\em et al.} (2015)]{LeCun} LeCun, Y., Bengio, Y. \& Hinton, G. (2015). Deep learning. {\em Nature} {\bf 521} 436--444.
\bibitem[Leeb (2008)]{Leeb} Leeb, H. (2008). Evaluation and selection of models for out-of-sample prediction when the sample size is small relative to the complexity of the data-generating process. {\em Bernoulli} 14:3, 661--690.
\bibitem[Lemieux (2006)]{Lemieux} Lemieux, T. (2006). The « Mincer Equation » Thirty Years After Schooling, Experience, and Earnings. {\em in } Jacob Mincer A Pioneer of Modern Labor Economics, Grossbard Eds, 127--145, Springer Verlag.
\bibitem[Li \& Racine (2006)]{Li} Li, J. \& J. S. Racine (2006). Nonparametric Econometrics. Princeton University Press.
\bibitem[Li {\em et al.} (2017)]{Lili} Li, C., Li, Q., Racine, J. \& Zhang, D. (2017). Optimal Model Averaging Of Varying Coefficient Models. {\em Department of Economics Working Papers 2017-01}, McMaster University.
\bibitem[Lin {\em et al.} (2016)]{Lin} Lin, H.W., Tegmark, M. \& Rolnick, D. (2016). Why does deep and cheap learning work so well?
{\em ArXiv e-prints}.
\bibitem[Lucas (1976)]{Lucas} Lucas, R.E. (1976). Econometric Policy Evaluation: A Critique. {\em Carnegie-Rochester Conference Series on Public Policy}, 19--46.
\bibitem[Mallows (1973)]{Mallows} Mallows, C.L. (1973). Some Comments on $C_p$. {\em Technometrics}, {\bf 15}, 661--675.
\bibitem[McCullogh \& Pitts (1943)]{McCulloch} McCullogh, W.S. \& Pitts, W. (1943). A logical calculus of the ideas immanent in nervous activity. {\em Bulletin of Mathematical Biophysics}, 5:4, 115--133.
\bibitem[Mincer (1974)]{Mincer} Mincer, J. (1974). Schooling, experience and earnings. Columbia University Press.
\bibitem[Mitchell (1997)]{Mitchell} Mitchell, T. (1997). Machine Learning. McGraw-Hill.
\bibitem[Morgan \& Sonquist (1963)]{Morganetal} Morgan, J.N. \& Sonquist, J.A. (1963). Problems in the analysis of survey data, and a proposal. {\em Journal of the American Statistcal Association}, {\bf 58}, 415--434.
\bibitem[Morgan (1990)]{Morgan} Morgan, M.S. (1990). The history of
econometric ideas. Cambridge University Press.
\bibitem[Mohri {\em et al.} (2012)]{MohriEtal} Mohri, M., Rostamizadeh, A. \& Talwalker, A. (2012) Foundations of Machine Learning. MIT Press.
\bibitem[Mullainathan \& Spiess (2017)]{Mullainathan}
Mullainathan, S. \& Spiess, J. (2017). Machine learning: An applied econometric approach. {\em Journal of Economic Perspectives}, {\bf 31} 87--106.
\bibitem[M\"uller (2011)]{Muller} M\"uller, M. (2011). Generalized Linear Models {\em in } Handbook of Computational Statistics, J.E Gentle, W.K. H\"ardle \& Y. Mori Eds. Springer Verlag.
\bibitem[Murphy (2012)]{Murphy} Murphy, K.R. (2012). Machine Learning: a Probabilistic Perspective. MIT Press.
\bibitem[Murphy \& Welch (1990)]{MurphyWelch} Murphy, K. M. \& F. Welch (1990). Empirical age-earnings profiles. {\em Journal of Labor Economics} {\bf 8}, 202--229.
\bibitem[Nadaraya (1964)]{Nadaraya} Nadaraya, E. A. (1964). On Estimating Regression. {\em Theory of Probability and its Applications}, {\bf 9}:1, 141--2.
\bibitem[Natarajan (1995)]{Natarajan}  Natarajan, B. K. (1995). Sparse approximate solutions to linear systems. {\em SIAM Journal on Computing} (SICOMP), 24 227–-234.
\bibitem[Nevo \& Whinston (2010)]{Nevo} Nevo, A. \& Whinston, M.D. (2010). Taking the Dogma out of Econometrics: Structural Modeling and Credible Inference. {\em Journal of Economic Perspective}, {\bf 24},  69--82.
\bibitem[Neyman (1923)]{Neyman} Neyman, J. (1923).  Sur les applications de la th\'eorie des probabilit\'es aux exp\'eriences agricoles : Essai des principes. M\'emoire de master, republib\'e dans {\em Statistical Science}, {\bf 5}, 463--472.
\bibitem[Nisbet, Elder \& Miner (2001)]{Nisbet} Nisbet, R., Elder, J. \& Miner, G. (2011). Handbook of Statistical Analysis and Data Mining
Applications. Academic Press, New York. 
\bibitem[Okun (1962)]{Okun} Okun,  A. (1962). Potential GNP: Its measurement and significance. {\em Proceedings of the Business and Economics Section of the American Statistical Association}, 98--103.
\bibitem[Orcutt (1952)]{Orcutt} Orcutt, G.H. (1952). Toward a partial redirection of econometrics. {\em Review of Economics and Statistics}, {\bf 34} 195--213.
\bibitem[Pagan \& Ullah (1999)]{Pagan} Pagan, A. \& A. Ullah (1999). Nonparametric Econometrics. Themes in Modern Econometrics. Cambridge: Cambridge University Press.
\bibitem[Pearson (1901)]{Pearson} Pearson, K. (1901). On lines and planes of closest fit to systems of points in space. {\em Philosophical Magazine}, {\bf 2}, 559-–572.
\bibitem[Platt (1999)]{Platt} Platt, J. (1999). Probabilistic outputs for support vector machines and comparisons to regularized likelihood methods. {\em  Advances in Large Margin Classifiers}. {\bf 10}, 61--74.
\bibitem[Portnoy (1988)]{Portnoy} Portnoy, S. (1988). Asymptotic Behavior of Likelihood Methods for Exponential Families when the Number of Parameters Tends to Infinity. {\em Annals of Statistics}, {\bf 16}:356--366.
\bibitem[Quenouille (1949)]{Quenouille1} Quenouille, M. H. (1949). Problems in Plane Sampling. {\em The Annals of Mathematical Statistics} {\bf 20}(3):355--375.
\bibitem[Quenouille (1956)]{Quenouille2} Quenouille, M. H. (1956). Notes on Bias in Estimation. {\em Biometrika} {\bf 43}(3-4), 353--360.
\bibitem[Quinlan (1986)]{Quinlan} Quinlan, J.R. (1986). Induction of decision trees. {\em Machine Learning } {\bf 1} 81--106.
\bibitem[Reiers\o{o}l (1945)]{Reiersol} Reiers\o{o}l, O. (1945). Confluence analysis of means of instrumental sets of variables. {\em Arkiv. for 
Mathematik, Astronomi Och Fysik}, {\bf 32}. 
\bibitem[Rosenbaum \& Rubin (1983)]{RosenbaumRubin} Rosenbaum, P. \& Rubin, D. (1983). The Central Role of the Propensity Score in Observational Studies for Causal Effects. {\em Biometrika}, {\bf 70}, 41--55.
\bibitem[Rosenblatt (1958)]{Rosenblatt} Rosenblatt, F. (1958). The perceptron: a probabilistic model for information storage and organization in the brain. {\em Psychological Review}, {\bf 65}, 386--408.
\bibitem[Rubin (1974)]{Rubin} Rubin, D. (1974). Estimating Causal Effects of Treatments in Randomized and Nonrandomized Studies. {\em Journal of Educational Psychology}, {\bf 66}, 688--701.
\bibitem[Ruppert, Wand \& Carroll (2003)]{Ruppert} Ruppert, D., Wand, M. P. \& Carroll, R.J. (2003). Semiparametric Regression. Cambridge University Press.
\bibitem[Samuel (1959)]{Samuel} Samuel, A. (1959). Some Studies in Machine Learning Using the Game of Checkers. {\em IBM Journal of Research and Development}, {\bf 44}:1. 
\bibitem[Schultz (1930)]{Schultz} Schultz, H. (1930). The Meaning of Statistical Demand Curves. University of Chicago.
\bibitem[Shai \& Shai (2014)]{Shaix2} Shai, S.S. \& Shai, B.D. (2014). Understanding Machine Learning
From Theory to Algorithms. Cambridge University Press.
\bibitem[Shao (1993)]{Shao93} Shao, J. (1993). Linear Model Selection by Cross-Validation. {\em Journal of the American Statistical Association} {\bf 88}:(422), 486--494.
\bibitem[Shalev-Shwartz \&  Ben-David (2014)]{Shalev-Shwartz} Shalev-Shwartz, S. \&  Ben-David, S. (2014). Understanding Machine Learning: From Theory to Algorithms. Cambridge University Press.
\bibitem[Shao (1997)]{Shao97} Shao, J. (1997). An Asymptotic Theory for Linear Model Selection. {\em Statistica Sinica}, {\bf 7}, 221--264.
\bibitem[Shapire \& Freund (2012)]{ShapireFreund} Shapire, R.E. \& Freund, Y. (2012). Boosting. MIT Press.
\bibitem[Silverman (1986)]{Silverman} Silverman, B.W. (1986) Density Estimation. Chapman \& Hall.
\bibitem[Simonoff (1996)]{Simonoff} Simonoff, J. S. (1996). Smoothing Methods in Statistics. Springer. 
\bibitem[Stone (1977)]{Stone} Stone, M. (1977). An Asymptotic Equivalence of Choice of Model by Cross-Validation and Akaike's Criterion. {\em Journal of the Royal Statistical Society. Series B }, {\bf 39}:1, 44--47.
\bibitem[Tam \& Kiang (1992)]{Tam} Tam, K.Y. \& Kiang, M.Y. (1992). Managerial applications of neural networks: The case
of bank failure predictions. {\em Management Science}, {\bf 38}, 926--947.
\bibitem[Tan (1995)]{Tan} Tan, H. (1995). Neural-Network model for stock forecasting. MSc Thesis, Texas Tech. University.
\bibitem[Tibshirani (1996)]{LASSO} Tibshirani, R. (1996). Regression shrinkage and selection via the lasso. {\em Journal of the Royal Statistical Society, Series B.}, {\bf 58}, 267--288.
\bibitem[Tibshirani \& Wasserman (2016)]{TibWasserman} Tibshirani, R. \& Wasserman, L. (2016). A Closer Look at Sparse Regression. {\sffamily http://bit.ly/2FrGQ32}
\bibitem[Tikhonov (1963)]{Tikhonov} Tikhonov, A. N. (1963). Solution of incorrectly formulated problems and the regularization method. {\em Soviet Mathematics}, {\bf 4}: 1035--1038.
\bibitem[Tinbergen (1939)]{Tinbergen} Tinbergen, J. (1939). Statistical Testing of Business Cycle Theories. Vol. 1: A Method and its Application to Investment activity; Vol. 2: Business Cycles in the United States of America,
1919—1932. Geneva: League of Nations.
\bibitem[Tobin (1958)]{Tobin} Tobin, J. (1958). Estimation of Relationship for Limited Dependent Variables. {\em Econometrica}, {\bf 26}, 24--36. 
\bibitem[Tropp (2011)]{Tropp2011} Tropp, (2011). Improved analysis of the subsampled randomized Hadamard transform. {\em Advances in Adaptive Data Analysis}, 3:1, 115--126.
\bibitem[Tsen (2001)]{Tsen} Tsen, P. (2001). Convergence of a block coordinate descent for nondifferentiable minization. {\em Journal of Optimization Theory and Applications}, 109:3, 475--494.
\bibitem[Tuff\'ery (2001)]{Tuffery} Tuff\'ery, S. (2001). Data Mining and Statistics for Decision Making. Wiley Interscience.
\bibitem[Tukey (1958)]{Tukey} Tukey, J. W. (1958). Bias and confidence in not quite large samples. {\em The Annals of Mathematical Statistics}, {\bf 29}:614--623.
\bibitem[Vapnik (1998)]{Vapnik98} Vapnik, V. (1998). Statistical Learning Theory. Wiley.
\bibitem[Vapnik \& Chervonenkis (1971)]{VC} Vapnik, C, \& Chervonenkis, A. (1971). On the uniform convergence of relative frequencies of events to their probabilities. {\em Theory of Probability and its Applications}, {\bf 16}:264--280.
\bibitem[Varian (2014)]{Varian} Varian, H.R. (2014). Big Data: New Tricks for Econometrics. {\em Journal of Economic Perspectives},  {\bf 28}(2):3--28. 
\bibitem[Vert (2017)]{Vert} Vert, J.P. (2017). Machine learning in computational biology. ENSAE.
\bibitem[Waltrup {\em et al.} (2014)]{Walltrup} Waltrup, L.S., Sobotka, F., Kneib, T. \& Kauermann, G. (2014). Expectile and quantile regression—David and Goliath? {\em Statistical Modelling}, {\bf 15}, 433 -- 456.
\bibitem[Watson (1964)]{Watson} Watson, G. S. (1964). Smooth regression analysis. {\em Sankhya: The Indian Journal of Statistics, Series A}, {\bf 26}:4, 359--372.
\bibitem[Watt {\em et al.} (2016)]{Watt} Watt, J., Borhani, R. \& Katsaggelos, A. (2016). Machine Learning Refined : Foundations, Algorithms, and Applications. Cambridge University Press.
\bibitem[Widrow \& Hoff (1960)]{Widrow} Widrow, B. \& Hoff,  M.E. Jr. (1960). Adaptive Switching Circuits. IRE WESCON Convention Record, {\bf 4}:96--104.
\bibitem[Wolpert \& Macready (1997)]{Wolpert97} Wolpert, D.H., Macready, W.G. (1997), No Free Lunch Theorems for Optimization, {\em IEEE Transactions on Evolutionary Computation} {\bf 1}, 67.
\bibitem[Wolpert (1996)]{Wolpert96} Wolpert, David (1996), The Lack of A Priori Distinctions between Learning Algorithms, {\em Neural Computation}, 1341-1390.
\bibitem[Working (1927)]{Working} Working, E. J. (1927). What do statistical `demand curves' show? {\em Quarterly Journal of Economics}, {\bf 41}:212--35.
\bibitem[Yu \& Moyeed (2001)]{YuMoyeed} Yu, K. \& Moyeed, R. (2001). Bayesian quantile regression. {\em Statistics \& Probability Letters}, 54, 437--447.

\bibitem[Zinkevich {\em et al.} (2010)]{Zinkevichetal} Zinkevich M.A., Weimer, M., Smola, A. \& Li, L. (2010). Parallelized Stochastic Gradient. {\em Advances in neural information processing systems}, 2595--2603.

\end{thebibliography}
\end{document}